\documentclass[letter]{jsedi_sub}
\usepackage{cancel}
\usepackage{hyperref}
\usepackage{cleveref}
\usepackage{booktabs}
\usepackage[version=4]{mhchem} 

\usepackage{booktabs}
\usepackage{graphicx}
\usepackage{xcolor}
\usepackage{ulem}

\crefname{equation}{Eq.}{Eqs.}
\Crefname{equation}{Eq.}{Eqs.}

\crefname{figure}{Fig.}{Figs.}
\Crefname{figure}{Fig.}{Figs.}

\crefname{table}{Table}{Tables}
\Crefname{table}{Table}{Tables}

\newcommand{\refeq}[1]{\cref{#1}}
\newcommand{\reffig}[1]{\cref{#1}}
\newcommand{\reftab}[1]{\cref{#1}}
\newcommand{\pr}[1]{$^{#1}$}

\title{Core and mantle thermal evolution constraints on the onset of plate tectonics and a long-lived geodynamo}

\shorttitle{Core-mantle coupled thermal evolution} 

\author[1]{Valentin Bonnet Gibet
	\orcid{0000-0003-0348-7347}
	\thanks{Corresponding author: 
\href{mailto:valentin.bonnetgibet@proton.me}{\texttt{valentin.bonnetgibet@proton.me}}}
}
\author[1]{Nicola Tosi
	\orcid{0000-0002-4912-2848}
}

\affil[1]{Institute of Space Research, Deutsches Zentrum für Luft- und Raumfahrt (DLR), Berlin, Germany}

\credit{Conceptualization}{Valentin Bonnet Gibet, Nicola Tosi}
\credit{Methodology}{Valentin Bonnet Gibet, Nicola Tosi}
\credit{Software}{Valentin Bonnet Gibet, Nicola Tosi}
\credit{Validation}{Valentin Bonnet Gibet}
\credit{Formal Analysis}{Valentin Bonnet Gibet, Nicola Tosi}
\credit{Writing - Original draft}{Valentin Bonnet Gibet}
\credit{Writing - Review \& Editing}{Valentin Bonnet Gibet, Nicola Tosi}
\credit{Visualization}{Valentin Bonnet Gibet}
\credit{Supervision}{Nicola Tosi}
\credit{Funding acquisition}{Nicola Tosi}

\begin{document}

\publicationonly{
\dois{}
\handedname{Efirstname Elastname}
\receiveddate{}
\accepteddate{}
\publisheddate{}
\theyear{}
\thevolume{}
\thepaper{}  
}

\makesedititle{
  \begin{summary}{Abstract}
     Earth's long-lived geodynamo is difficult to reconcile with recent high estimates of the core thermal conductivity, a problem known as the ``new core paradox''. At the same time, the long-term thermal evolution of the mantle remains uncertain, largely due to the poorly constrained onset of modern-style plate tectonics, which marks the transition to efficient cooling of the interior through mobile-lid convection.  Because core cooling -- and thus magnetic field generation -- depends on the efficiency with which the mantle extracts heat from the core, these two problems are closely linked. Here, we investigate the coupled thermal evolution of mantle and core using a 1D model that incorporates a parametrized transition from stagnant- to mobile-lid convection, defined by its onset time and with a prescribed duration. This framework allows us to assess how different tectonic histories influence Earth’s thermal and magnetic evolution. We perform Bayesian inversions using constraints from the palaeomagnetic record, mantle cooling history, and present-day thermal state. Our results favour a transition from stagnant- to mobile-lid convection during the Archean, which promotes core cooling and enables a geodynamo throughout Earth's history, even for core thermal conductivities in excess of 100 W m\pr{-1} K\pr{-1}. Reproducing simultaneously the inferred mantle cooling history and present-day surface heat loss remains challenging, highlighting limitations of current 1D mantle convection scaling laws. Nevertheless, our results show that a delayed onset of mobile-lid convection can provide a viable solution to the new core paradox. 
    \vspace{1.3cm} 
  \end{summary}
 \begin{summary}{Non-technical summary}
    Earth has sustained a magnetic field for at least the past 3.4 billion years. This field originates from the motion of liquid iron in the planet’s core, a process that requires the core to cool efficiently. However, recent studies suggest that heat may be conducted through the core more efficiently than previously thought, making it harder to sustain the magnetic field over such long timescales. This apparent contradiction is known as the ``new core paradox''. At the same time, the thermal evolution of Earth’s mantle is still debated, especially regarding when plate tectonics began. Because the mantle controls how effectively heat is extracted from the core, the evolution of the mantle and core are closely linked. In this study, we use a numerical model of Earth’s interior to explore how the timing for the onset of plate tectonics affects core cooling and the longevity of the magnetic field. Our results suggest that plate tectonics likely began during the Archean. We find that a late plate tectonics onset favours a long-lived magnetic field even with a high core thermal conductivity. These findings offer a new perspective on how the ``new core paradox'' can be resolved.
    \vspace{1.3cm} 
 \end{summary}
}

\section{Introduction}

\subsection{Magnetic field generation}
The palaeomagnetic record indicates that Earth’s magnetic field is a long-lived feature, existing since at least 3.4~Ga \citep{Biggin2015,bono2021pint}, or possibly even earlier \citep{tarduno2023hadaean}, a result that remains controversial \citep{fu2024}. Although the temporal resolution of palaeomagnetic data decreases as we look further into the past, current evidence suggests that the magnetic field has been continuously active \citep{Biggin2015,bono2021pint}. Earth's magnetic field is generated by convection within the liquid iron outer core. For a magnetic field to be sustained, specific criteria must be met, including sufficient convective power and a favourable flow geometry \citep{christensen2006scaling}. At present-day, the geodynamo is powered by both thermal and compositional convection. The latter is driven by the crystallization of the inner core, which releases latent heat and light elements at the Inner Core Boundary (ICB). 

To match the Preliminary Reference Earth Model (PREM) density profile \citep{PREM}, the outer core must contain a total light element mass fraction of approximately 5-10 wt\% \citep{hirose2021light}. The light elements most often proposed are sulphur, silicon, oxygen, carbon, and hydrogen. Prior to inner core nucleation, the geodynamo would have been thermally driven only. During this phase, core cooling must have been sufficient to maintain a magnetic field consistent with the palaeomagnetic record. However, recent experimental and theoretical estimates suggest that the thermal conductivity of the core could be as high as $\approx$100-120 W m\pr{-1} K\pr{-1} \citep{koker2012electrical,pozzo2022toward}. These values, which are much higher than earlier theoretical estimates \citep{stacey2007}, imply a strong reduction of the convective power available to drive the dynamo due to the increase in the amount of heat conducted along the core adiabatic gradient. Prior to inner core nucleation, estimated at $\approx$\-500--700 Ma \citep{labrosse2015thermal}, the convective power of a purely thermally-driven dynamo is at its minimum; in the presence of high core thermal conductivity, the resulting convective power is likely insufficient to sustain a dynamo \citep{davies2015cooling, labrosse2015thermal,driscoll2023paradox}. Indeed, this period coincides with the lowest magnetic field intensities recorded \citep{bono2019young, lloyd2021first}. 

The difficulty of reconciling a thermally-driven dynamo in the presence of a high core thermal conductivity with the uninterrupted operation of the magnetic field since at least 3.4 Ga is known as the “new core paradox” \citep{olson2013new}. Proposed solutions to this paradox, such as the exsolution of \ce{MgO} at the Core-Mantle Boundary (CMB) \citep{badro2016early,orourke2016powering}, or precession-driven flows \citep{lin2016precession,andrault2016deep}, have been argued to be insufficient to bridge the necessary gap \citep{du2019experimental,landeau2022sustaining,Insixiengmay2025}. In contrast, the precipitation of \ce{SiO2} has been proposed as a viable mechanism capable of powering an early dynamo \citep{Hirose2017crystallization, Zhang2022free, Wilson2022powering,Davies2023}.  Importantly, some experimental studies suggest a lower thermal conductivity for the core \citep{konopkova2016direct,andrault2025experimental,Hsieh2025}, in line with the classical estimates of \citet{stacey2007}, implying less prohibitive conditions for the operation of a thermally-driven dynamo prior to inner-core nucleation. 

The thermal conductivity is also the key parameter governing the potential formation of a thermally-stratified layer at the top of the core, with large  values at the CMB promoting the growth of stratification beneath it \citep{labrosse1997cooling}. Seismological studies have suggested the existence of a stratified layer up to 450~km thick \citep{kaneshima2018array}, while other studies indicate a thinner layer between 90 and 140~km \citep{tanaka2007possibility}, or even no stratified layer at all \citep{irving2018seismically}. This discrepancy mainly reflects differences in the reference Earth models used to interpret SmKS travel times: while PREM requires the presence of a low-velocity layer at the top of the core to fit the observations, the EPOC model of \citet{irving2018seismically} achieves a comparable fit without invoking such a layer.  This layer could also be compositionally stratified owing to its enrichment in light elements \citep{gubbins2013stratified,bouffard2019chemical,davies2020transfer}. Because core evolution is primarily controlled by heat transfer through the overlying solid mantle, an accurate quantification of the heat extracted from the core via mantle convection is essential to understanding both core evolution and magnetic field generation.

\subsection{Mantle convection regime}
The mantle, being orders of magnitude more viscous than the liquid outer core, cools on much longer timescales and thus exerts a strong control on core cooling. The efficiency of core cooling is determined by the heat flux across the CMB, which is difficult to constrain as it depends on both the temperature contrast between the CMB and the deep mantle and on the local vigour of solid-state convection. The present-day CMB temperature also remains poorly constrained.  The existence of a liquid outer core beneath a solid lower mantle bounds the CMB temperature between the solidus of lowermost mantle rocks and the adiabatic core temperature anchored to the melting temperature at the ICB. However, both depend on the composition, which remains uncertain for both the lowermost mantle and the outer core. As a result, current estimates allow a broad range of CMB temperatures, typically between approximately 3899 K and 4300 K \citep{forst2022multidisciplinary}. 

In contrast, the mantle potential temperature is better constrained by petrological modelling, which relates the composition of primary magmas to their temperature and depth of melt generation. Current estimates indicate a present-day mantle potential temperature of $1625 \pm 25$~K \citep{herzberg2010thermal}. A similar mantle potential temperature is obtain using a half-space cooling model to match present-day sea-floor heat fluxes \citep{MCKENZIE2005}. This approach also provides constraints on the temporal evolution of the mantle potential temperature, revealing a secular cooling trend \citep{herzberg2009petrological,ganne2017mantle,sudholz2025xenolith}. In particular, the record indicates enhanced cooling over the last $\approx 2$ Gyr, with inferred cooling rates between -80~K~Gyr\pr{-1} and -110~K~Gyr\pr{-1} \citep{forst2022multidisciplinary}. 

The present-day surface heat loss is well constrained by geophysical observations, with an estimated total value of $46\pm3$~TW \citep{jaupart2015temperature}. This heat loss originates from two main sources: a continental contribution of approximately $14$~TW, including $7$–$8$~TW due to radiogenic heat production, and an oceanic heat loss of about $32$~TW. 

The mantle viscosity is strongly temperature-dependent and controls convective efficiency, influencing both surface and CMB heat loss. Consequently, a warmer mantle exerts two competing effects on the CMB heat flux. On one hand, a higher mantle temperature reduces the thermal contrast across the CMB, decreasing core heat loss. On the other hand, it reduces lower-mantle viscosity, thinning the lower thermal boundary layer and enhancing heat transport. In the absence of surface mobility, this rheological behaviour promotes a stagnant-lid regime characterized by a thick, rigid, and conductive lithosphere that insulates the interior and leads to inefficient mantle cooling \citep[e.g.,][]{moresi1995}. By contrast, on present-day Earth, plate tectonics enables large-scale surface mobility and continuous lithospheric recycling, maintaining a relatively thin thermal boundary layer and enhancing convective heat transport. This mobile-lid regime provides an efficient mechanism to cool the mantle, and sustain an efficient extraction of heat from the core.

Despite its fundamental importance, the timing of Earth’s transition to a global plate tectonic regime comparable to the present-day system remains one of the major unresolved questions in Earth sciences. The dating of preserved plate tectonic features such as ophiolites or blueschists points to an age of plate tectonic between 1 and 3 Ga \citep{stern2007and,condie2008did,stern2018understanding,condie2018zircon}. However, the lack of preserved Hadean and Archean rocks limits direct constraints on early Earth, motivating the use of indirect proxies to infer ancient subduction,  including geochemical signatures \citep{greber2017titanium,lipp2021composition}, palaeomagnetic apparent polar wander paths \citep{cawood2018geological,brenner2020paleomagnetic}, and evidence from silicate-atmosphere chemical exchange \citep{farquhar2002mass}. These methods imply  different estimates for the onset of plate tectonics ranging from shortly after magma-ocean solidification to as late as $\approx$800~Myr \citep[see][and references therein]{harrison2024review,korenaga2025tectonics}, with the majority of these studies suggesting an onset during the Archean. However, as pointed out by \citet{harrison2024review}, these approaches are inherently affected by a strong preservation bias: continuous lithospheric recycling progressively erases structural and geochemical signatures. Even the persistence of plate tectonic is still debated \citep{oneill2018inception}. Single lid episodes could have alternated with plate tectonics \citep{stern2023orosirian}. Moreover, the tectonic style (or styles) that prevailed prior to the establishment of present-day-like plate tectonics remains uncertain, and it is unclear whether the early Earth experienced a single dominant regime or multiple transitional modes \citep{arndt2013episodic}. Stagnant-lid–like convection is frequently invoked as a plausible  tectonic style during the early Earth, both because it is prevalent among terrestrial bodies in the Solar System and because it is compatible with geochemical observations \citep{debaille2013stagnant}. Alternative regimes, such as sluggish-lid or plutonic-lid convection, have also been proposed; however, these regimes are generally expected to be less efficient at cooling the mantle than a fully developed mobile-lid regime \citep{lenardic2018diversity,lourencco2020plutonic}. As a result, the tectonic history of the Earth likely played a critical role in modulating long-term core cooling and, by extension, the sustainability of the magnetic field. 

\subsection{Earth's thermal evolution}
Due to the prohibitive computational costs of 3D magneto-hydrodynamic simulations over billion-year timescales, the long-term evolution of planetary cores is typically investigated using 1D radially-averaged models. These frameworks have evolved from purely thermal models \citep{labrosse1997cooling} to sophisticated thermo-chemical models that account for light-element partitioning and the formation of stable chemical stratification \citep{labrosse2015thermal,davies2015cooling,greenwood2021stable}. By integrating these models backward in time from the present-day inner core radius of 1221~km \citep{PREM}, thermo-chemical models with a high thermal conductivity suggest a  nucleation age that typically falls between 700 and 500~Ma \citep{davies2015cooling,labrosse2015thermal}.  These models also monitor the viability of the geodynamo through proxies like Joule entropy production, which determines whether the available thermal and compositional buoyancy is sufficient to sustain a magnetic field \citep{davies2015cooling,labrosse2015thermal,greenwood2021stable,driscoll2023paradox}. \citet{labrosse2015thermal} estimates that a core heat flow of at least 16 TW at the time of inner core nucleation is required to sustain the long-lived Earth's magnetic field.

To ensure global consistency, these core models are coupled with mantle thermal histories that allow for the high heat fluxes required to sustain a magnetic field throughout Earth's evolution \citep{driscoll2014thermal,labrosse2016,patocka2020}. Notably, backward-in-time models often encounter the problem of the so-called ``thermal catastrophe'' where extrapolating the present-day surface heat loss backward in time results in unrealistically high mantle temperatures in the past \citep{christensen1985thermal,korenaga2006archean}. \citet{driscoll2014thermal} and \citet{patocka2020} have shown that a large present-day CMB heat flux can actually provide a solution to both the new core paradox and the thermal catastrophe. Furthermore, \citet{driscoll2023paradox} coupled the core model of \citet{davies2015cooling} with the mantle model of \citet{driscoll2014thermal} to explore alternative solutions for both problems, finding that a high initial CMB or a high concentration of radiogenic heat-producing elements (HPEs) within the core allow maintaining a long-lived magnetic field. Nevertheless, these studies assume a mobile-lid regime throughout Earth’s entire history. As discussed above, however, mobile lid convection may be a relatively recent feature of Earth’s evolution, associated with the onset of plate tectonics, and the consequences of a transition between convection regimes remain poorly explored in coupled core–mantle models.

Recently, \citet{alasad2024coupled} developed a coupled core–mantle thermal evolution model in which the style of mantle convection is not prescribed but emerges self-consistently, spanning regimes from sluggish-lid to mobile-lid convection as a function of the evolving mantle viscosity and thermal state. Their simulations suggest that Earth operated predominantly in a sluggish-lid regime during the Precambrian, enabled by a hot mantle and the presence of an asthenosphere, before transitioning to modern plate tectonics. This early sluggish-lid regime regulates CMB heat flow, allowing core convection and a long-lived geodynamo prior to inner core nucleation. They found that only models that undergo a transition from sluggish- to active-lid convection satisfy present-day constraints on mantle temperature, inner core size, and magnetic field persistence. While the model of \citet{alasad2024coupled} represents an important step toward self-consistent core–mantle evolution, its predictive power ultimately depends on the use of very specific scaling laws to parametrize mantle convection and lithospheric dynamics. However, these scaling laws rely on idealized assumptions about heat transport and energy dissipation that have not been systematically validated against fully dynamic convection models \citep{korenaga2025tectonics}. Moreover, the model neglects potentially important physical processes, including the formation of a thermally stratified layer at the top of the core, which can substantially modulate the CMB temperature and heat flow over periods of limited cooling. Finally, by assuming a constant core thermal conductivity, the model overlooks its pressure dependence, a factor that plays a central role in determining the core’s entropy budget and the long-term sustainability of the geodynamo across the parameter space. 


The coupled evolution of mantle and core, including the generation of a magnetic field, provides a framework for investigating the long-term history of the Earth using a variety of geophysical constraints. Here, we adopt a formal inverse modelling strategy in which mantle and core observations are used jointly to infer viable evolutionary pathways for Earth's interior. In particular, we explore a possible, yet under-examined, solution to the new core paradox by investigating the consequences of transitions in the convection regime, which are generally neglected in existing core-mantle evolution models.

Since a robust physical theory describing convection regime transitions remains lacking, we adopt a pragmatic approach in which Earth's evolution is modelled assuming an early stagnant-lid regime followed by a mobile-lid regime. The timing of this transition is treated as a free parameter. This allows us to assess whether global thermal evolution models, when constrained by mantle, core and magnetic field observations, can provide quantitative constraints on the onset of plate tectonics.

\section{Core-mantle thermal evolution modelling}

We developed a framework coupling a one-dimensional thermochemical evolution model of the core and a thermal evolution model of the mantle. The core model is a modified version of the model developed by \citet{greenwood2021stable}, which accounts for core secular cooling, inner core nucleation and growth, light element partitioning between liquid and solid, and allows for the development of a thermally stratified layer at the CMB. The mantle model is based on classical Rayleigh-Nusselt scaling laws and boundary layer theory, enabling it to treat both mobile-lid and stagnant-lid convection regimes \citep[e.g.,][]{foley2020heat}. It tracks the time evolution of the potential temperature $T_p$, considering radiogenic heat production, melt extraction, heat flux from the core, and surface heat loss. The model can transition between different convection styles, switching from a stagnant-lid regime with inefficient cooling to a fully-efficient mobile-lid regime.

\subsection{Core modelling}

The core model allows for three distinct layers: a solid inner core, a liquid convective outer core, and a liquid thermally stratified layer below the CMB. The outer core is convective while both the inner core and the stratified layer are treated as conductive. 

\paragraph{Thermal state.}

\begin{figure}[ht!]
    \centering
    \includegraphics[width=\linewidth]{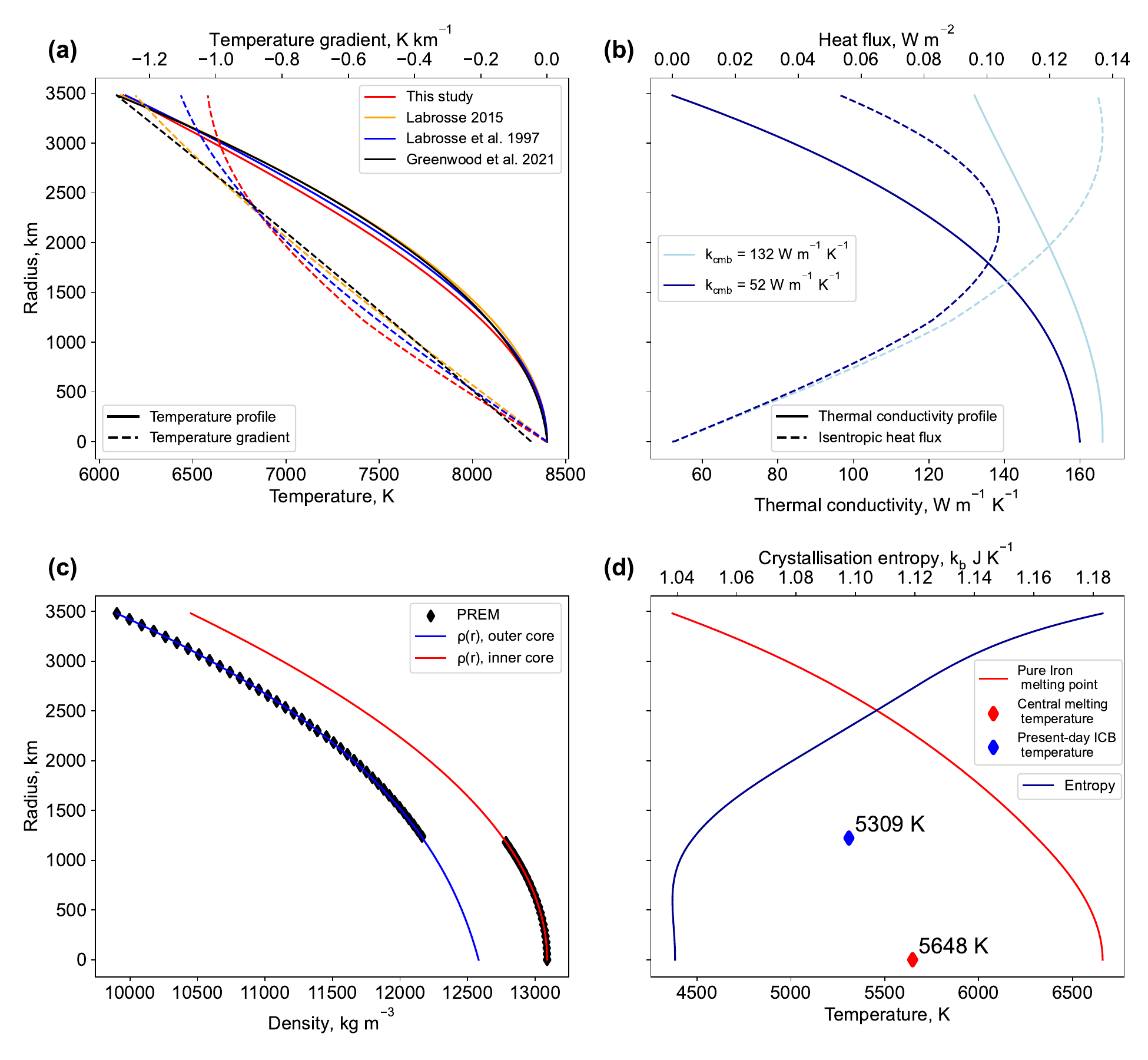}
    \caption{Radial profiles within the core: (a) isentropic temperature ($T_{ct} = 8400$ K) for different numerical computation methods (solid lines) and corresponding temperature gradient (dashed lines); (b) thermal conductivity end-members (solid lines) and resulting isentropic heat flux (dashed lines);  (c) PREM density \citep[][black diamonds]{PREM}, and polynomial-based density profile for the inner (solid red line) and outer core (solid blue line) at the present-day; (d) pure-iron melting curve (red line), crystallisation entropy (blue line), ICB temperature at present-day (blue diamond) and central melting temperature (red diamond).}
    \label{fig:core_fig}
\end{figure}

\citet{labrosse2001age} showed that, below the stratified layer, the convective core can be assumed to follow an isentropic temperature profile:

\begin{equation}
    T_a(r) = T_{ct} \exp\left(-\int_0^r \frac{\alpha_c g(r)}{c_c} dr\right),
    \label{eq:adibat_profil}
\end{equation}

\noindent where $T_{ct}$ is the central temperature of the core, $\alpha_c$ the core coefficient of thermal expansion, $r$ the radius, $g(r)$ the gravitational acceleration, and $c_c$ the heat capacity. Since the exponential term in eq. \eqref{eq:adibat_profil} varies slowly with time, the time derivative of the temperature profile can be approximated as

\begin{equation}
    \frac{ \mathrm{d}T_a}{\mathrm{d}t} = \frac{T_a}{T_{ct}} \frac{ \mathrm{d}T_{ct}}{\mathrm{d}t},
    \label{eq:adibat_approx}
\end{equation}

\noindent indicating that solving for the time evolution of the central temperature provides the complete thermal evolution within the core. The temperature gradient in the the isentropic region is

\begin{equation}
    \frac{ \mathrm{d}T_a}{\mathrm{d}r} = -\frac{\alpha_c g}{c_c} T_a,
    \label{eq:gradiant_adiat}
\end{equation}

\noindent which results in the following heat flux along the profile:

\begin{equation}
    q_{a}(r) = - k(r)\frac{ \mathrm{d}T_a}{\mathrm{d}r},
\end{equation}

\noindent where $k(r)$ is the radially-dependent core thermal conductivity. Following \citet{greenwood2021stable} and \citet{davies2015cooling}, we adopt a polynomial law for the thermal conductivity:

\begin{equation}
    k(r) = k_0 + k_1 r + k_2 r^2 + k_3 r^3,
\end{equation}

\noindent
with coefficients $k_n$ given in Table~\ref{tab:polynom}. Following \citet{greenwood2021stable}, who  proposed three distinct core thermal conductivity profiles based on different compositional scenarios \citep{davies2015cooling}, we adopt two end-member profiles as shown in \reffig{fig:core_fig}b, regardless of the core composition. These are characterized by significantly different values at the CMB, ranging from 52 to 132~W~m\pr{-1}~K\pr{-1}, but similar values at the core center ($\sim$160~W~m\pr{-1}~K\pr{-1}). Consequently, a lower conductivity at the CMB is associated with a stronger increase in conductivity with depth. The two profiles produce very different isentropic heat fluxes: in the low-CMB-conductivity, strongly depth-dependent case (dark blue lines in \reffig{fig:core_fig}b), the isentropic heat flux increases over the upper $\sim$1000~km of the core before decreasing toward zero, whereas in the high-CMB-conductivity case (light blue lines in \reffig{fig:core_fig}b), it remains nearly constant over the first $\sim$600~km and then decreases with depth. To explore thermal conductivity profiles between these two end-members, we introduce a weighting factor on each polynomials coefficients $f_k \in [0,1]$, where $f_k = 0$ corresponds to the low-CMB-conductivity profile and $f_k = 1$ to the high-CMB-conductivity profile. 

\paragraph{Core density profile.}  We express the density profile within the core as

\begin{align}
    \rho(r,r_{ic})\left\{ \begin{array}{ll}
    \rho_l(r) = \rho_{l,0} + \rho_{l,1} r + \rho_{l,2} r^2 + \rho_{l,3} r^3 ,\quad  \mathrm{if} \; r>r_{ic}\\
    \rho_s(r) = \rho_{s,0} + \rho_{s,1} r + \rho_{s,2} r^2 + \rho_{s,3} r^3,\quad \mathrm{if} \;   r<r_{ic}\end{array} \label{eq/density_profile}  \right. \\ 
\end{align}

\noindent where $\rho_s(r)$ is the density profile in the inner core and $\rho_l(r)$ the density profile in the liquid region. The polynomial coefficients are derived from PREM \citep[\reffig{fig:core_fig}c and Table~\ref{tab:polynom}, ][]{PREM}. The profiles of hydrostatic pressure ($P$), gravity acceleration ($g$), and gravity potential ($\psi$) are computed using these density profiles.

\paragraph{Light element partitioning.} Each light element has a different partitioning behaviour during the crystallisation of the inner core. We use the model of \citet{alfe2002composition} of the Fe-Si-O-S system to determine the concentration of each light element in the solid phase during inner core growth. First, the mass fraction $ c_x^{l/s}$, and the molar fraction $c_x^{M,l/s}$ are linked by 

\begin{equation}
 c_x^{l/s} = \frac{M_x}{M}   c_x^{M,l/s},
 \label{eq:mol2mass}
\end{equation}

\noindent where $M_x$ is the molar mass of the element $x$ and $M$ is the molar mass of the alloy. For this study, we choose to fix the core composition to reduce the number of free parameters. We adopt typical initial core concentrations of 5.23~wt\% \ce{O} (16.0~mol\%) and 1.14~wt\% \ce{Si} (2.0~mol\%) that, after the formation of a 1221~km radius inner core, result in concentrations that can fit the PREM density profile \citep[5.50~wt\% and 1.14~wt\% respectively, ][]{davies2015cooling,badro2015core}. Partitioning of light elements begins with the onset of inner-core crystallization. The concentration of light elements in the solid phase is computed assuming chemical equilibrium at the ICB and neglecting pressure effect, which yields the following relation:
\begin{equation}
    \mu_0^l+\lambda^l_x c_x^{M,l} + k_b T_m \ln(c_x^{M,l}) =   \mu_0^s+\lambda^s_x c_x^{M,s} + k_b T_m \ln(c_x^{M,s}),
\end{equation}
\noindent where $\mu_0^{l/s}$ is the reference chemical potential, $\lambda^{l/s}$ is a linear correction to the chemical potential from the ideal solution (see \reftab{tab:polynom}), $k_b$ is the Boltzmann constant, and $T_m$ is the melting temperature at the ICB. The concentration of a given element varies as the inner core grows: 

\begin{equation}
    \frac{\mathrm{d}c^{l}_{x}}{\mathrm{d}t} = C_x\frac{\mathrm{d}r_{ic}}{\mathrm{d}t},
     \label{eq:Cx}
 \end{equation}

\noindent where $C_x$ is obtained assuming mass conservation of the element and the total volume conservation of the core, which results in the following expression:
\begin{equation}
C_x   =  \frac{4\pi r_{ic}^2 \left(\rho_s^{ic} c^{s}_x   - \rho_l^{ic} c^{l}_x \right)}{M^{OC}},
\end{equation}
\noindent where $\rho_s^{ic}$ and $\rho_l^{ic}$ are the density at the inner core boundary respectively in the solid and liquid, and the $M^{OC}$ the mass of the outer core. 

\paragraph{Melting temperature.} Light elements do not only affect core density but also the melting temperature of the alloy. We consider the influence of composition following the model of \citet{alfe2002composition} in which the effect of each light element on the melting curve of pure iron is computed separately. This model links the effect on the melting curve of a given light element to its partitioning behaviour between solid and liquid as follows:
\begin{equation}
    T_m =  T_m^{Fe} \left( 1 + k_b\frac{ \displaystyle\sum_x c_x^{M,s}-c_x^{M,l}}{ \Delta S^{Fe}}  \right),
\end{equation}
\noindent where $T^{Fe}_m$ is the melting temperature of pure iron in Kelvin and $\Delta S^{Fe}$ is the entropy of crystallization expressed in multiples of the Boltzmann constant $k_b$ (\reftab{tab:polynom} and \reffig{fig:core_fig}d). Both $T^{Fe}_m$ are pressure-dependent: 
\begin{align}
    T^{Fe}_m(P) &= T^{Fe}_{m,0} + T^{Fe}_{m,1} P + T^{Fe}_{m,2} P^2 + T^{Fe}_{m,3} P^3,  \\
    \Delta S^{Fe}(P) &= S^{Fe}_{0} + S^{Fe}_{1} P + S^{Fe}_{2} P^2 + S^{Fe}_{3} P^3 + S^{Fe}_{4} P^4.
\end{align}
\noindent Since the gradient of the melting curve is larger than the isentropic gradient, the  crystallisation of the Earth's core starts at the center and proceeds upward. Owing to its compositional dependency, the melting point decreases upon inner core growth as
\begin{equation}
    \frac{\mathrm{d}T_m}{\mathrm{d}r_{ic}} = k_b\frac{T_m^{Fe}(r_{ic})}{ \Delta S^{Fe}(r_{ic})} \sum_x\frac{\mathrm{d}c^{M,l}_x}{\mathrm{d}r_{ic}}   ,
    \label{eq:Tmdri_lightform}
\end{equation}
\noindent where $\frac{\mathrm{d}c^{M,l}_x}{\mathrm{d}r_{ic}}$ is the increase of the molar concentration of an element $x$ as the inner core grows. 

\begin{figure}[ht!]
    \centering
    \includegraphics[width=0.6\linewidth]{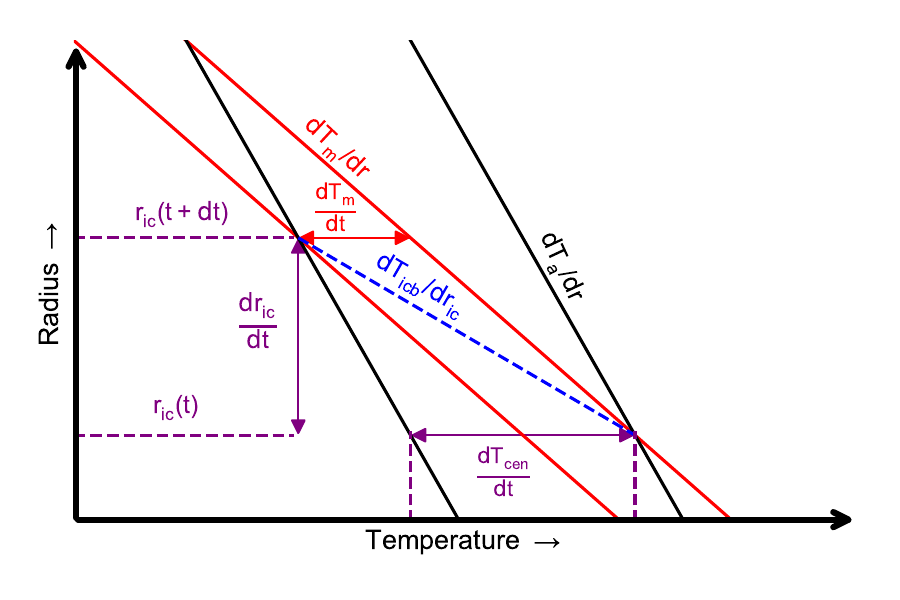}
    \caption{Scheme of inner core growth during secular cooling. The inner core size is given by the intersection of the temperature profile (black lines) and the melting curve (red lines). The melting temperature decreases as the inner core grows because of the enrichment of the outer core in light elements. The inner core radius growth rate ($\frac{\mathrm{d}r_{ic}}{\mathrm{d}t}$) depends on the difference between the isentropic gradient ($\frac{\mathrm{d}T_a}{\mathrm{d}r}$) and the melting temperature gradient ($\frac{\mathrm{d}T_m}{\mathrm{d}r}$), and on the drop in melting temperature ($\frac{\mathrm{d}T_m}{\mathrm{d}t}$).}
    \label{fig:dric_schem}
\end{figure}

\begin{table}[]
\centering
\caption{Polynomial coefficients for radius- or pressure-dependent core profiles from \cite{PREM,alfe2002iron,davies2015cooling}.}
\label{tab:polynom}
\begin{tabular}{@{}llllll@{}}

\toprule
Profile                                                            & Polynomials  &  values            &              &             &          \\ \midrule
                                                     & n = 0       & n=1         & n=2          & n=3         & n=4      \\ \midrule
Thermal conductivity, $k(r) = \displaystyle\sum^n k_n r^n$        &   W m\pr{-1} K\pr{-1}          &    W m\pr{-2} K\pr{-1}           &  W m\pr{-3} K\pr{-1}            &     W m\pr{-4} K\pr{-1} &     \\
low-CMB-conductivity, $k_n$:                                                             & 160         & -2.41$\times10^{-6}$   & -4.04$\times10^{-12}$    & -1.20$\times10^{-18}$   &       \\
high-CMB-conductivity,  $k_n$:                                                          & 166         & 5.90$\times10^{-7}$   & -5.25$\times10^{-12}$    & 6.55$\times10^{-19}$    &       \\ \midrule
Density,    $\rho(r) = \displaystyle\sum^n \rho_n r^n$  &  kg m\pr{-3}            &    kg m\pr{-4}         &      kg m\pr{-5}         &  kg m\pr{-6}           &          \\
PREM Outer core, $\rho_n$:  & 12581  & -1.9836$\times10^{-4}$ & -0.89742$\times10^{-10}$  & -2.1377$\times10^{-17}$  &       \\
PREM inner core, $\rho_n$: & 13089 & 0 & -2.1774$\times10^{-10}$ & 0 &      \\ \midrule
Pure-iron melting temperature       & K & K Pa\pr{-1} &   K Pa\pr{-2}            &   K Pa\pr{-3}           &          \\
 $T_m^{{Fe}}(P) = \displaystyle\sum^n {{T}_n} P^n$, $T_n$:                                                               & 1700        & 2.73$\times10^{-8}$   & -6.65$\times10^{-20}$    & 7.95$\times10^{-32}$    &       \\
Crystallisation entropy           & $k_b$ &   $k_b$ Pa\pr{-1}        &       $k_b$ Pa\pr{-2}         &     $k_b$ Pa\pr{-3}         &      $k_b$ Pa\pr{-4}     \\
$\Delta S^{{Fe}}(P) = \displaystyle\sum^n {S}_n P^n$ , $S_n$:    & 1.91        & -1.19$\times10^{-11}$   & 7.09 $\times10^{-23}$     & -1.94 $\times10^{-34}$  & 1.95  $\times10^{-46}$\\ \bottomrule
\end{tabular}

\end{table}

\paragraph{Inner core growth.} The inner core nucleation occurs when the central temperature drops below the melting temperature $T_m(r=0)$ (\reffig{fig:core_fig}d, red diamond). As the temperature profile does not depend on the inner core size, the inner core radius ($r_{ic}$) is simply defined as the radius at which the melting curve intersects the isentropic profile, $T_m(r_{ic}) = T_a(r_{ic})$. The development of this condition yields the growth rate of the inner core as a function of the central temperature cooling rate:

\begin{equation}
    \frac{\mathrm{d}r_{ic}}{\mathrm{d}t} = \frac{T_{icb}}{T_{ct}} \left(  \left.\frac{\mathrm{d}T_m}{\mathrm{d}r}\right|_{r_{ic} } -\left.\frac{\mathrm{d}T_a}{\mathrm{d}r}\right|_{r_{ic}} +\left.\frac{\mathrm{d}T_m}{\mathrm{d}r_{ic}}\right|_{r_{ic} }  \right)^{-1}  \frac{\mathrm{d}T_{ct}}{\mathrm{d}t} \equiv C_r \frac{\mathrm{d}T_{ct}}{\mathrm{d}t},
    \label{eq:Cr}
\end{equation}

\noindent where $T_{icb}$ is the temperature at the inner core boundary, $\left.\frac{\mathrm{d}T_m}{\mathrm{d}r} \right|_{r_{ic}}$ is the melting curve gradient at the inner core boundary, $\left.\frac{\mathrm{d}T_a}{\mathrm{d}r} \right|_{r_{ic}}$ is the adiabatic gradient at the inner core boundary, and $\frac{\mathrm{d}T_{m}}{\mathrm{d}r_{ic}} $ is the variation in melting temperature upon inner core growth (\reffig{fig:dric_schem}). 

\paragraph{Core heat balance.} The energy budget of the core below the stratified layer (see section below) is given by:

\begin{equation}
    Q_{top} =  Q_{s} + Q_{g} + Q_{L},
\end{equation}

\noindent where $Q_{top}$ is the heat loss at the convective core outer boundary. In absence of a stratified layer, this corresponds to  the heat flow across the mantle bottom thermal boundary layer $Q_{CMB}$, while in the presence of a stratified layer it corresponds to the isentropic $q_{ad}(r_{sl})A(r_{sl})$, where $A(r_{sl})$ is the surface area of the core at the radius $r_{sl}$ corresponding to the base of the stratified layer.  $Q_{s}$ is the secular cooling of the core, $Q_{L}$ is the latent heat released during the crystallization of the inner core, and $Q_{g}$ is the gravitational energy due to light elements release upon inner core freezing. No internal heating is considered here, in agreement with a recent study on potassium partitioning showing that only a negligible amount can be incorporated within the core during iron-silicate differentiation \citep{xiong2018ab}.  Following \citet{greenwood2021stable}, the heat balance of the core below the stratified layer can be expressed as:

\begin{equation}
    Q_{top} = - \frac{c_c}{T_{ct}}\frac{ \mathrm{d}T_{ct}}{\mathrm{d}t} \int_0^{r_{sl}} \rho T_a \mathrm{d}V + \sum_x \alpha_x^l C_r C_x \frac{ \mathrm{d}T_{ct}} {\mathrm{d}t} \int_{r_{ic}}^{r_{sl}} \rho \psi \mathrm{d}V + 4\pi r_{ic}^2 \rho^s_{icb} L C_r \frac{ \mathrm{d}T_{ct}}{\mathrm{d}t},
    \label{eq:heat_balance}
\end{equation}

\noindent where, in the sum over each light element $x$, $\alpha_x^l$ is the chemical expansion coefficient of a light element, and $C_x$ and $C_r$ are defined in eqs. \eqref{eq:Cx} and \eqref{eq:Cr}, respectively. $\rho^s_{icb}$ is the density of the inner core at the radius of the inner core boundary and $L = \Delta S^{Fe}T_{m} N_a(1000/M_{Fe})$ is the latent heat of crystallisation in J kg\pr{-1}. We can then factor $\frac{ \mathrm{d}T_{ct}}{\mathrm{d}t}$ in eq. \eqref{eq:heat_balance} and express it as:

\begin{equation}
    \frac{\mathrm{d}T_{ct}}{\mathrm{d}t} =\frac{Q_{top}}{Q^{'}_{s} + Q^{'}_{g} + Q^{'}_{L}},
    \label{eq:dTcendt}
\end{equation}

\noindent where $Q^{'}_{s}$, $Q^{'}_{g}$ and $Q^{'}_{L}$ are the respective heat flows divided by  $\frac{ \mathrm{d}T_{ct}}{\mathrm{d}t}$. From eq. \eqref{eq:dTcendt}, which we solve with an explicit Euler method, we can then readily obtain the inner core size and light element partitioning from the change in central temperature. 

\begin{figure}[ht!]
    \centering
    \includegraphics[width=0.6\linewidth]{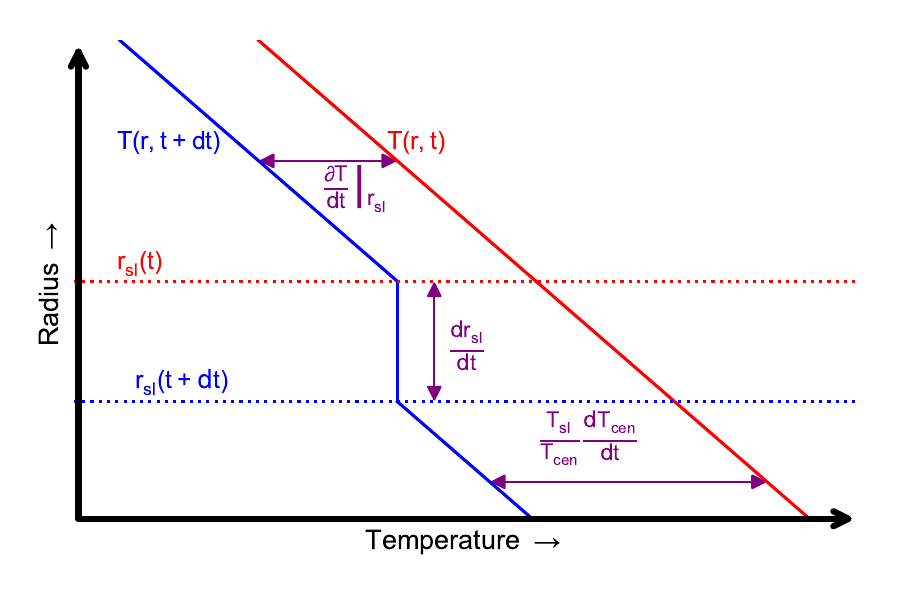}
    \caption{Stratified layer growth method from \citet{greenwood2021stable}. The thermally stratified layer growth rate depends only on the difference between the cooling rate of the isentropic temperature profile and the cooling rate at the base of the conductive profile in the thermally stratified layer. This method preserves the continuity of the temperature profile but not of the heat flux. }
    \label{fig:drsl_schem} 
\end{figure}

\paragraph{Thermally stratified layer.} We assume that heat is transported by conduction within the thermally stratified layer whose temperature profile is obtained by solving a time-dependent heat equation with spherical symmetry:

\begin{equation}
    \rho c_c \frac{\partial T}{\partial t} = \frac{1}{r^2}\frac{\partial}{\partial r}\left( k r^2 \left(\frac{\partial T}{\partial r}\right)\right),
    \label{eq:heateq_sl}
\end{equation}

\noindent where the density $\rho$ and thermal conductivity $k$ are radially-dependent. The boundaries conditions are $ \left. \frac{\partial T}{\partial r}\right|_{r_c} = - \frac{q_{b}}{k(r_b)}$ at the CMB, and $\left. \frac{\partial T}{\partial r}\right|_{r_{sl}} = \left.\frac{\mathrm{d}T_a}{\mathrm{d}r}\right|_{r_{sl}}$ at the base of the layer. Equation \eqref{eq:heateq_sl} is solved numerically using a finite volume method combined with a fully implicit scheme to ensure numerical stability regardless of the time step.

A thermally stratified stratified layer at the top of the convective core forms when the heat extracted at the core-mantle boundary is lower than the heat conducted along the adiabatic gradient:
\begin{equation}
    Q_{cmb} < Q_a,
\end{equation}
\noindent where $Q_a =q_a(r_b)A(r_b)$. To determine the growth of this layer over time, we follow the algorithm proposed by \citet{greenwood2021stable}. The growth of the stratified layer arises from the difference in cooling rate between the stratified and convective regions. As the two layers cool at different rates, a temperature discontinuity forms at the boundary. To ensure temperature continuity, the new stratified layer radius is obtained through a linear extrapolation of the temperature $T_{sl}$ at the base of the stable layer, which is extended until it intersects the adiabatic profile (\reffig{fig:drsl_schem}). From geometric considerations, the stratified layer growth rate is expressed as follow
\begin{equation}
    \frac{\mathrm{d}r_{sl}}{\mathrm{d}t} =\left( \frac{T_{sl}}{T_{ct}} \frac{\mathrm{d}T_{ct}}{\mathrm{d}t} - \left. \frac{\partial T}{\partial t}\right|_{r^+_{sl}} \right) \left. \frac{\mathrm{d}T_a}{\mathrm{d}r}  \right|_{r^-_{sl}}^{-1},
    \label{eq:sl_growth}
\end{equation}
\noindent where $ \left. \frac{\partial T}{\partial t}\right|_{r^+_{sl}}$ is the cooling rate at the base of the stratified layer and $\left. \frac{\mathrm{d}T_a}{\mathrm{d}r}  \right|_{r^-_{sl}}^{-1}$ is the adiabatic profile just below the stratified layer. The growth rate depends on the difference in cooling rates between the isentropic core and the stratified layer. \citet{greenwood2021stable} have shown that this method can well reproduce the results of \citet{labrosse1997cooling} that used a more sophisticated method to track the growth of the stratified layer. 

\paragraph{Entropy generation as magnetic field proxy.} In 1D thermal evolution models, the dynamo process is assessed using proxies that indicate whether magnetic field generation is possible under given core conditions \citep{christensen2006scaling}. A commonly used proxy is the magnetic Reynolds number, which must exceed a critical value ($\sim$45) for dynamo action to be expected  \citep{christensen2006scaling}. However, calculating the magnetic Reynolds number requires a velocity scale, which is difficult to specify without making strong assumptions about the core flow dynamics \citep{schaeffer2025energetically}. For this reason, alternative proxies have been proposed, such as the magnetic dipole moment or the Joule entropy production \citep{gubbins2004gross,olson2006dipole}. Here, we adopt the Joule entropy as our main proxy. Because magnetic field generation acts as an entropy sink, the entropy budget must allow for positive Joule entropy production in order for a dynamo to be sustained. By analogy with the energy budget, we formulate the entropy balance as follows:

\begin{equation}
    E_{k} + E_{J} = E_{s} + E_{L} + E_{g},
\end{equation}

\noindent where the left hand sides lists the different entropy sinks, namely thermal conduction $E_{k}$, ohmic and viscous dissipation, i.e. Joule entropy $E_{J}$, and the right hand side the entropy sources, namely secular cooling $E_{s}$, latent heat released upon inner core crystallisation $E_{L}$, and compositional convection $E_{g}$. The available Joule entropy is

\begin{equation}
    E_{J} = - \int_{r_{ic}}^{r_{sl}} k \left(\frac{\nabla T}{T}\right)^2 \mathrm{d}V -\int_{r_{ic}}^{r_{sl}} \left(\frac{1}{T_{cmb}} - \frac{1}{T} \right)\rho c_c \frac{\mathrm{d}T}{\mathrm{d}t}\mathrm{d}V  + \left( \frac{1}{T_{cmb}} - \frac{1}{T_{icb}} \right) Q_{L} + \frac{Q_{g}}{T_{cmb}}
    \label{eq:entropy_dev}
\end{equation}

\noindent where $T_{cmb}$ is the temperature at the CMB. In order to sustain a magnetic field, $E_J$ must be positive and can be thus calculated from the other terms $E_{k}$, $E_{s}$, $E_{L}$ and $E_{g}$, which are readily computed based on our core model. As in \citet{greenwood2021stable}, we evaluate $E_{s}$ and $E_{k}$ across the inner core, the convective outer core and the stratified layer. Although, fingering convection could occur within the stratified layer \citep{rosenthal2025finger}, we neglect the associated entropy generation as in \citet{knibbe2021modelling}.
\subsection{Core model benchmark}

To validate our implementation, we performed a benchmark against the model of \citet{greenwood2021stable}, hereafter G21. We adopted the same core parameters reported in Tables~\ref{tab:param} and~\ref{tab:polynom} and, for this test, set a constant thermal conductivity of 80~W~m\pr{-1}~K\pr{-1}. Entropy was computed over the entire core and the  term $\frac{\mathrm{d}T_m}{\mathrm{d}r_{ic}}$ was omitted from eq. \eqref{eq:Cr}, following G21. The imposed CMB heat flow follows a linear decrease in time, with a present-day value of 3~TW and a slope of $-11$~TW~Gyr$^{-1}$. Both models start from the same central temperature, $T_{ct}^0 = 8550$~K at 4.5~Ga.

\begin{figure}[ht!]
    \centering
    \includegraphics[width=\linewidth]{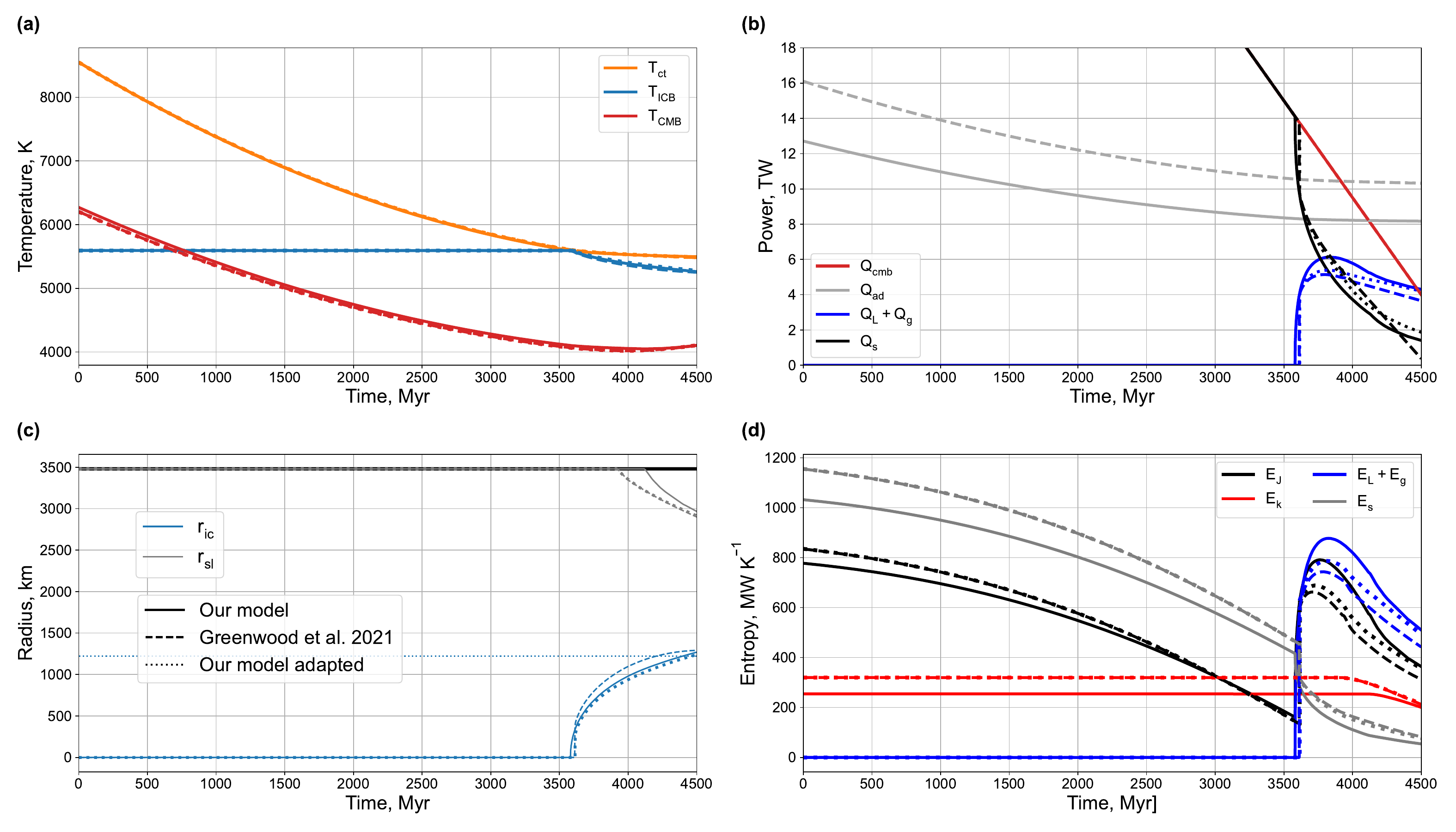}
    \caption{Benchmark for core thermal evolution for three cases: our model  (solid lines), \citet{greenwood2021stable} model (dashed lines), and our model modified to use the polynomial fit for the isentropic temperature profile from G21 (dotted lines). Time evolution of (a) central core temperature (orange), CMB temperature (red), and ICB temperature (blue); (b) CMB heat flow (red), isentropic heat flow (grey), secular cooling (black), and heat flows due to latent heat and gravitational energy release upon inner core freezing (blue); (c) inner-core radius (blue) and thickness of the thermally stratified layer (grey); and (d) Joule entropy (black), entropy sink trough conduction (red), and entropy sources due to inner core freezing (blue) and secular cooling (grey).}
    \label{fig:bench}
\end{figure}

The thermal evolutions match well, with only slight differences. The CMB temperatures exhibit a small difference from the start (67 K, \reffig{fig:bench}a, red lines), while central temperatures are initially identical and gradually separate (up to 16 K, \reffig{fig:bench}a, orange lines). These differences translate into distinct inner-core nucleation ages, with G21 predicting a slightly slower core cooling and hence a slightly more recent onset of inner core growth (\reffig{fig:bench}c, blue lines). The main difference is observed in the two times marking the onset of a thermally stratified layer; the G21 case exhibits earlier stratification (\reffig{fig:bench}c, grey lines), associated with a $\sim$3.5~TW larger isentropic at the CMB (\reffig{fig:bench}b, grey lines). We attribute these discrepancies to the method used to obtain the core temperature profile. Our model computes the profile by direct numerical integration of the governing equation, whereas G21 evaluate the profile using a polynomial fit derived from a numerically-integrated reference profile. The two approaches yield similar absolute temperatures but noticeably different temperature gradients (\reffig{fig:core_fig}a), key factors controlling respectively the growth of the inner core and of the thermal stratification.

To verify this hypothesis, we ran a third simulation in which we replaced our isentropic profile obtained via integration by the same polynomial parametrization used in G21 (\reffig{fig:bench}, dotted line), employing identical polynomial coefficients. This new thermal evolution closely reproduces that of G21, confirming that the computation of the temperature profile was the main cause of the observed differences. For comparison, we also shown in \reffig{fig:core_fig}a the analytical solution of \citet{labrosse1997cooling}, which is closer to our numerically integrated profile than the polynomial fit of G21, particularly in the temperature gradient. The isentropic profile obtained by assuming constant Gr\"uneisen parameters, as in \citet{labrosse2015thermal} and show in \reffig{fig:core_fig}a, lies between the two: it features a gradient at the CMB greater than that of our model but smaller than that of GW21, while having a zero gradient at the center of the core, as in our model. The differences that we observe could be explained by the polynomial fits introducing non-negligible errors near domain boundaries (core center and CMB), precisely where accurate values of the temperature and its gradient are most important for predicting inner core nucleation and stratified layer growth. We therefore favour the numerical integration of the temperature profile.

\subsection{Mantle modelling}

In order to simulate the transition from a low- to high-efficiency cooling regime, we propose a model that allows switching from stagnant-lid to mobile-lid mantle convection at an arbitrarily chosen time during thermal evolution. The model also accounts for cooling via melt extraction and heating from radiogenic decay. This two-regimes approach has the advantage of being based on well-established scaling laws to parametrize convective heat fluxes. 

\paragraph{Parametrization.}  The Rayleigh number is the ratio of buoyancy to viscous forces multiplied by the ratio of momentum and thermal diffusivities. It quantifies the vigour of convection and is expressed as

\begin{equation}
    \mathcal{R}a = \frac{\alpha_m\rho_m g\Delta T D_m^3 }{\kappa_m \eta},
\end{equation}

\noindent where $\alpha_m$ is the mantle coefficient of thermal expansion,  $\rho_m$ is the average mantle density, $g$ is the gravity acceleration, $\Delta T$ is the superadiabatic temperature jump, $D_m$ is the mantle thickness, $\kappa_m$ is the mantle thermal diffusivity, and $\eta$ is the mantle viscosity. The latter depends on temperature and pressure following an Arrhenius law: 

\begin{equation}
 \eta\left(T,P\right) = \eta_{0} \exp\left( \frac{\Delta_a + PV}{RT}-\frac{\Delta_a+ P_{0}V}{RT_{0}}\right),
\end{equation}

\noindent where $\eta_{0}$ is the reference viscosity at reference pressure $P_{0}$ and reference temperature $T_{0}$, $\Delta_a$ is the activation energy, and $V$ is the activation volume. The Nusselt number $\mathcal{N}u$ expresses the ratio of convective to conductive heat fluxes:

\begin{equation}
    \mathcal{N}u = \frac{Q_{conv} D_m}{A k_m \Delta T},
    \label{eq:nusselt}
\end{equation}

\noindent where $Q_{conv}$ is the convective heat flow, $A$ is the area, and $k_m$ is the mantle thermal conductivity. The Nusselt number scales with the Rayleigh number as 

\begin{equation}
 \mathcal{N}u   = C \mathcal{R}a^\beta\theta^{-(\beta+1)},
    \label{eq:NuRa_scaling}
\end{equation}

\noindent where $\beta$ is a constant exponent of $\approx1/3$ for $\mathcal{R}a < 10^{11}$ \citep{solomatov1995scaling}, $\theta=\log(\Delta\eta)$, the Frank—Kamenetskii parameter, $\Delta\eta$ is the viscosity contrast, and $C$ a constant that can also be expressed in terms of a reference Rayleigh number, i.e.  $C = \mathcal{R}a_{ref}^{-\beta}$ \citep{solomatov1995scaling,turcotte2002geodynamics}. Thus, this scaling law is applied within the frame of boundary layer theory \citep{turcotte2002geodynamics}, which assumes that the mantle interior is well mixed and that temperature jump and associated heat transfer are concentrated in thin thermal boundary layers, one  with thickness $\delta_u$ below the surface or below the stagnant lid, and the other, with thickness $\delta_c$, atop of the CMB. 

\paragraph{Surface heat flow.} In a mobile-lid regime where the viscosity contrast $\Delta\eta =0$, meaning that the whole mantle volume undergoes convection, the surface heat flow is derived directly from eqs. \eqref{eq:nusselt} and \eqref{eq:NuRa_scaling} and expressed as

\begin{equation}
     Q_{surf}^{ML} =   A_\mathrm{p}\frac{k_{um} \Delta T_{um}}{D_{m}} \left( \frac{{\mathcal{R}a}^{um}}{\mathcal{R}a_\mathrm{ref}^{um}} \right)^{\frac{1}{3}},
     \label{eq:QML_surf}
\end{equation}

\noindent where $A_p$ the planet's surface, $k_{um}$ is the upper mantle thermal conductivity, $\Delta T_{um}$ is the temperature jump between the upper mantle temperature $T_m$ and the surface temperature, ${\mathcal{R}a}^{um} = \frac{\alpha_m\rho_m g\Delta T_{um} D_m^3 }{\kappa_m \eta_{um}}$ is the upper mantle Rayleigh number with $\eta_{um}$ the viscosity of the upper mantle at the pressure of the base of the cold boundary layer $\delta_u$ and temperature $T_m$, and $\mathcal{R}a_\mathrm{crit}^{um}$ is the critical Rayleigh number for the upper mantle. 

In a stagnant-lid regime, the rigid lithosphere acts as an insulating layer that does not participate in convection. Therefore, the classical scaling laws apply only to the convecting part of the mantle, below the stagnant lid. Some studies addressed this issue by treating separately the convecting mantle via scaling laws and solving a full heat conduction equation in the stagnant lid \citep[e.g.,][]{schubert1979subsolidus,spohn1991,morschhauser2011,tosi2017,Thiriet2019}. Here, we adopt the same approach as for the mobile-lid regime, except that the Frank–Kamenetskii parameter is defined as $\theta = (\Delta_a \Delta T_{um})/(RT_m^2)$ to account for the strong temperature dependence of viscosity, which leads to the development of a stagnant lid. The surface heat flow in this case is given by  \citep{Davaille1993,foley2020heat}

\begin{equation}
    Q_{surf}^\mathrm{SL} = A_{p} \frac{k_{um} \Delta T_{um}}{D_m} \left( \frac{{\mathcal{R}a}_{um}}{\mathcal{R}a_\mathrm{ref}^{um}} \right)^{\frac{1}{3}}  \theta^{-\frac{4}{3}}.
\end{equation} 

\begin{figure}[ht!]
    \centering
    \includegraphics[width=\linewidth]{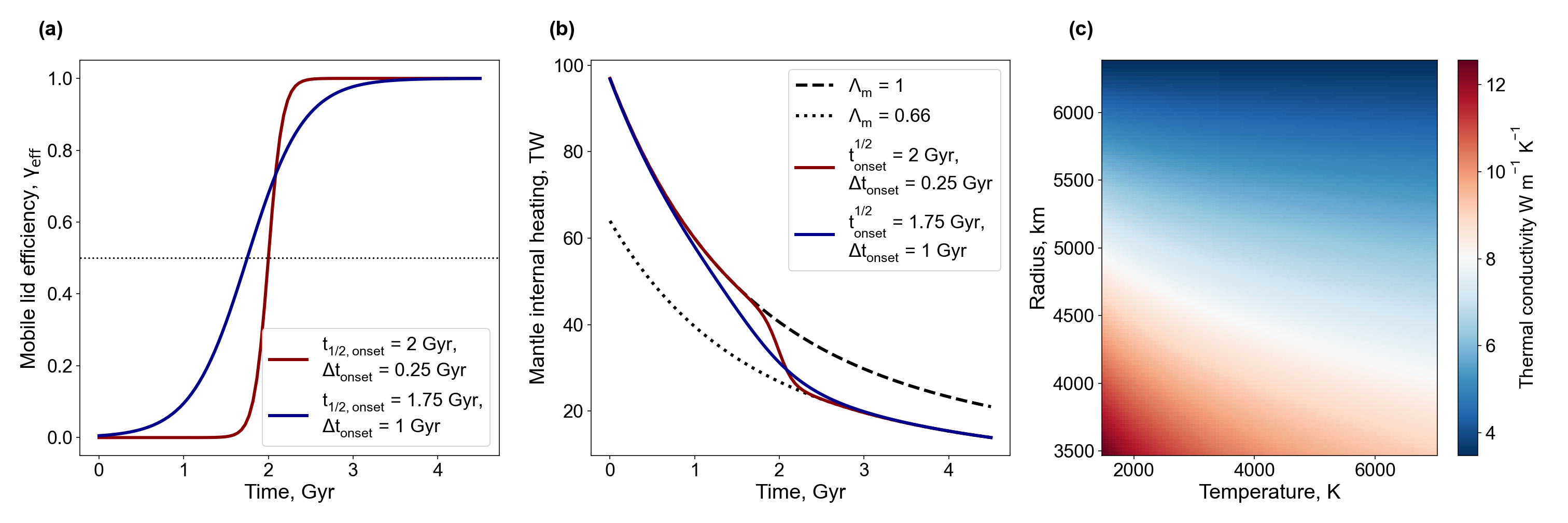}
    \caption{(a) Mobile-lid efficiency factor $\gamma_{eff}$ as a function of time (eq. \eqref{eq:gamma_eff}), for two different times and durations of the transition. (b) Mantle internal heat production (eq. \eqref{eq:Qradt}) as a function of time for two different enrichment factors: $\Lambda_m = 1$ and 0.66 over 4.5 Gyr (dashed and dotted lines, respectively, transition from $\Lambda_m = 1$ to 0.66 following the two mobile-lid efficiency evolutions of panel a (blue and red solid lines). Mantle thermal conductivity as a function of depth and temperature based on eq. \eqref{eq:thermal_cond} \citep{tosi2013mantle}.}
    \label{fig:mantle_fig}
\end{figure}

\paragraph{Stagnant-lid to mobile-lid transition.} Because no fully established model allows for a self-consistent transition from stagnant-lid to mobile-lid mantle convection, we impose this transition parametrically. The onset of plate tectonic is unlikely to occur as an abrupt global event in which the entire surface becomes mobile simultaneously \citep{stern2018evolution}. Therefore, we assume a progressive increase in the efficiency of mobile-lid convection, which we parametrize through the coefficient $\gamma_{eff}$ that ranges from 0 (pure stagnant lid) to 1 (fully-developed mobile lid). To describe this temporal evolution, we adopt a logistic function (\reffig{fig:mantle_fig}a):

\begin{equation}
   \gamma_{eff}(t) = \left( 1+ \exp\left(\frac{3(t-t_{1/2,onset})}{\Delta t_{onset}}\right)\right)^{-1},
   \label{eq:gamma_eff}
\end{equation}

\noindent where $ t_{1/2,onset}$ is the time at which the transition heat flow has reached half of the heat flow associated with the mobile lid regime, called ``transition time'' in the following, and $\Delta t_{onset}$ is the time interval across which $\gamma_{eff}$ increases from $\approx4.7\%$ to 50\%. The time of the onset of the transition is then defined as:

\begin{equation}
    t_{onset} = t_{1/2,onset} - \Delta t_{onset}.
    \label{eq:onset_time}
\end{equation}

The total surface flow is calculated by weighting the two regimes:

\begin{equation}
    Q_{surf} = \gamma_{eff}  Q_{surf}^{ML} + (1 - \gamma_{eff})Q_{surf}^{SL}.
    \label{eq:Qsurf}
\end{equation}

The thickness of the cold boundary layer corresponding to the surface heat flow \eqref{eq:Qsurf} is

\begin{equation}
    \delta_u = \frac{k_{um} \Delta T_{um}}{Q_{surf}},
\end{equation}

\noindent where $k_{um}$ is the thermal conductivity within the boundary layer. The pressure and temperature for determining the Rayleigh number $\mathcal{R}a_{um}$ are taken at the base of the thermal boundary layer, which in turn depends on these pressure and temperature, and the resulting non-linear problem is solved with the Brent’s method as implemented in the SciPy package \citep{scipy2025}.

\paragraph{Core-mantle boundary heat flow.}  The CMB heat flow is derived from eq. \eqref{eq:NuRa_scaling} using $\Delta\eta = 0$ and local parameters:

\begin{equation}
    Q_{cmb}=   A_{c}\frac{k_{lm}(T_{cmb},D_{m}) \Delta T_{lm}}{D_{m}} \left( \frac{{\mathcal{R}a}_{lm}}{\mathcal{R}a_{ref}^{c}} \right)^{\frac{1}{3}},
    \label{eq:Qcmb}
\end{equation}

\noindent where  $A_{c}$ is the CMB area, $k_{lm}(T,z)$ is the temperature- and depth-dependent thermal conductivity of the lower mantle \citep{tosi2013mantle}, $\Delta T_{lm} = T_c - T_b$ is the temperature jump across $\delta_{c}$, $T_b$ the adiabatic temperature at the top of $\delta_c$, ${\mathcal{R}a}_{lm}= \frac{\alpha_m\rho_m g\Delta T_{lm} D_m^3 }{\kappa_m \eta_{lm}}$ is the lower mantle Rayleigh number, $\eta_{lm}$ is the lower mantle viscosity calculated at the temperature $T_{lm} = (T_c + T_b)/2$ and at the pressure of the top of the bottom boundary layer, $\mathcal{R}a_{ref}^{c}$ is the reference Rayleigh number of the lower mantle \citep{deschamps2001thermal,Thiriet2019}, which is given by:

\begin{equation}
    \mathcal{R}a_{ref}^{c} = 0.28 Ra_i^{0.21}, 
\end{equation}

\noindent where $Ra_i = \frac{\alpha_m\rho_m g\Delta T_{i} D_m^3 }{\kappa_m \eta_{um}}$ is the internal Rayleigh number in which $\Delta T_{i} = (T_m -T_s) + (T_c -T_b)$ \citep{thiriet2018hemispheric}. We choose a thermal conductivity of the lower mantle that evolves as a function of temperature \citep{tosi2013mantle}, assuming a lower mantle mineralogy with iron-bearing bridgmanite (80\%) and Periclase (20\%). \citet{tosi2013mantle} propose the following expression for the temperature and depth-dependence of the thermal conductivity of the lower mantle:

\begin{equation}
    k_{m}(T,z) =  3.48 + z\times5.17\times10^{-6} \times \left(\frac{300}{T}\right)^{0.31},
    \label{eq:thermal_cond}
\end{equation}

\noindent where we use $z = D_m$ for the depth (in km) and $T = T_{cmb}$ for the temperature (in K). This parametrization gives  a thermal conductivity between 9 and 10 W m\pr{-1} K\pr{-1} for $z=2890$ km and $T_{CMB} \in[3700-6000]$ K (\reffig{fig:mantle_fig}c), in good agreement with recent experimental results on iron-bearing bridgmanite \citep{edmund2024thermal}. For the upper mantle thermal conductivity $k_{um}$, we use a constant value as the expected temperature variations are small and would not affect $k_{m}$ significantly. Finally, the bottom thermal boundary layer thickness is given by 

\begin{align}
    \delta_{c} = \frac{k_{m}(T_{cmb},D_{m}) \Delta T_{lm}}{Q_{cmb}}.
\end{align}

\paragraph{Melt extraction.} To account for the cooling effect of volcanism, we follow the approach of \citet{driscoll2014thermal} and include a mantle melting heat flow term, $Q_{melt}$, into the global energy balance. While sophisticated models exist for stagnant lid planets that explicitly treat melt extraction and crust formation \citep{morschhauser2011,bonnet2025martian}, such frameworks have not been developed for plate tectonics, which would require a consistent treatment of crustal recycling. The approach of \citet{driscoll2014thermal} assumes a crust that is thin and constant in time -- an assumption that is reasonable for Earth, where the present-day crust represents approximately 1\% of the bulk silicate volume \citep{mooney2023Earth}. Under this assumption, the mass of melt extracted to form crust is compensated by an equal mass of crust recycled back into the mantle. The associated magmatic heat loss is defined as:

\begin{equation}
    Q_{melt} = \dot{M}_{melt} (L_{melt} + c_m \Delta T_{melt}),
\end{equation}

\noindent where $L_{melt}$ is the latent heat of fusion, $c_m$ is the specific heat of the mantle, and $\Delta T_{melt}$ is the temperature difference between the mantle temperature $T_m$ and the solidus temperature $T_{sol}$. The melt production rate, $\dot{M}_{melt}$, is derived from the decompression melting of upwelling mantle material. In the context of mid-ocean ridges, it is expressed as:

\begin{equation}
    \dot{M}_{melt} = \rho_m V_{up} \phi,
\end{equation}

\noindent
where $\rho_m$ is the mantle density, $V_{up}$ is the volumetric upwelling rate, and $\phi$ is the melt fraction calculated as in \citet{driscoll2014thermal}. Finally, the melt heat flow can be simply written as \citep{driscoll2014thermal}: 

\begin{equation}
  Q_{melt} =  A_p  \frac{ 1.16\kappa\rho_{sol}\phi}{\delta_u^\mathrm{ML}} \left( L_{melt} + c_m \Delta T_{melt} \right),
  \label{eq:Qmelt}
\end{equation}

\noindent where $\delta_u^\mathrm{ML}$ is the hot boundary layer of the mobile case. 

\paragraph{Internal heating.} The internal heat production resulting from the radioactive decay of $^{40}$\ce{K}, $^{232}$\ce{Th}, $^{235}$\ce{U} and $^{238}$\ce{U}  as a function of time is given by:

\begin{equation}
    Q_{rad}(t) = M_m \Lambda_{m} \sum_i^n H_i X_i^0 \exp\left(- \frac{\ln(2)}{\tau_i^{1/2}}t\right),
    \label{eq:Qradt}
\end{equation}

\noindent where $M_m$ is the mantle mass, $n=4$ is the number of radioactive isotopes, ${H}_i$ is the heating rate due to the decay of the isotope $i$, $X_i^0$ is the bulk silicate concentration at $t=0$, $\tau_{1/2,i}$ the half-life, and $\Lambda_m$ is the amount of heat-producing elements in the mantle in proportion to the bulk silicate \citep{McDonough1995,ruedas2017radioactive,foley2020heat}. On Earth, heat-producing elements are primarily divided between the continental crust and the mantle. The heat generated in the continental crust does not affect the global mantle heat budget. At the present-day, 8 TW are associated with continental heat production, corresponding to 34\% of the total \citep{jaupart2015temperature}. The evolution of this value in time is uncertain and is linked to the formation, differentiation and recycling of the continents \citep{condie2014growth}. Here, we do not consider the formation of continents and, for simplicity, we assume an enrichment factor that decreases according to the mobile lid efficiency:

\begin{equation}
    \Lambda_m = \Lambda_m^\mathrm{pd} \times \gamma_{eff} \ + (1 - \gamma_{eff})\Lambda_{bulk},
    \label{eq:lambda_m}
\end{equation}

\noindent where $\Lambda_m$ varies from $\Lambda_{bulk} = 1$ during the stagnant lid phase ($\gamma_{eff} = 0$) to 0.66 during the mobile phase ($\gamma_{eff} = 1$). This assumption effectively implies that continents form concomitantly with the onset of plate tectonics. Although this scenario is unlikely, it represents a reasonable first-order approximation and it is more realistic than prescribing a constant $\Lambda_m$ (0.66 or 1) throughout the evolution (\reffig{fig:mantle_fig}b). 

\paragraph{Thermal evolution of the mantle.}

The Earth's mantle thermal history is determined by a heat balance expressed as a function of the secular cooling of the mantle potential temperature $T_p$: 

\begin{equation}
    M_m  c_m \epsilon_m \frac{\mathrm{d}T_p}{\mathrm{d}t} = - Q_{surf}  + Q_{rad} - Q_{melt} + Q_{cmb},
    \label{eq:energy_balance}
\end{equation}

\noindent where $c_m$ is the mantle heat capacity, $\epsilon_m$ is the mantle potential temperature divided by the volume-averaged temperature of the entire mantle, and the four heat flows are those described above, namely the convective heat flow  ($Q_{surf}$, eq. \eqref{eq:Qsurf}),  the heat flow due to internal heating  ($Q_{rad}$, eq. \eqref{eq:Qradt}), the heat flow due to melt extraction ($Q_{melt}$, eq. \eqref{eq:Qmelt}), and the CMB heat flow ($Q_{CMB}$, eq. \eqref{eq:Qcmb}). Between the upper and lower thermal boundary layers, the mantle is assumed to follow an adiabatic temperature profile. The temperatures at the top and bottom of the adiabatic mantle, $T_m$ and $T_b$, are therefore given by
\begin{equation}
    T_{m/b} = T_p \exp\left(\frac{\alpha_m g}{c_m}z_{m/b}\right),
\end{equation}
\noindent where $z_m=\delta_u$ and $z_b=D_m-\delta_c$ denote the depths of the top and bottom of the convecting mantle, respectively.

\begin{table}[h!]
\centering
\caption{Mantle and core parameters.}
\label{tab:param}
\begin{tabular}{@{}lllll@{}}
\toprule
Mantle    &                                   &            &        &     \\ \midrule
$R_p$        & Earth radius                       & 6370       & km        & \citet{PREM} \\ 
$R_c$        & Core radius                       & 3480       & km          & \citet{PREM} \\ 
$g_{um}$     & Surface gravity                   & 9.8        & m s\pr{-2}    & \citet{PREM}   \\
$g_{lm}$    & CMB gravity                       & 10.6       & m s\pr{-2}      & \citet{PREM} \\
$\rho_m $   & Average mantle density            & 4460       & kg m\pr{-3}     & \citet{PREM} \\
$c_m$        & Mantle heat capacity              & 1200       & J kg\pr{-1} K\pr{-1}  &\citet{driscoll2014thermal} \\
$\alpha_m$  & Mantle thermal expansion coefficient       & $3 \times 10^{-5}$   &      K\pr{-1}    & \citet{tosi2013mantle}   \\
$k_{um}$     & Upper mantle thermal conductivity  & 4.2        & W m\pr{-1} K\pr{-1}     & \citet{tosi2013mantle}  \\
$R$         & Gas constant              & 8.314      & J mol\pr{-1} K\pr{-1} \\
$T_{0}$    & Reference temperature             & 1700       & K       &    \\
$P_{0} $   & Reference pressure                & 4.5        & GPa      &  \\
$\mathcal{R}a^{um}_{ref}$  & Reference Rayleigh number          & 450       &        --  & \citet{Thiriet2019}  \\
$\rho_{sol}$  & Solid density for melt extraction & 3300       & kg m\pr{-3}   & \citet{driscoll2014thermal}   \\
$L_{melt} $  & Latent heat of crystallisation     & 320   & kJ kg\pr{-1}    & \citet{driscoll2014thermal}  \\
$T_s$        & Surface temperature               & 288        & K       & \citet{foley2020heat}    \\
$T_p^0$      & Initial mantle potential temperature & 2100 & K & \\ 
$X_{\ce{U}}$     & Present-day BSE Uranium content               & 20.0    & ppb       & \citet{McDonough1995} \\
$X_{\ce{Th}}$     & Present-day BSE Thorium content               & 80.0    & ppb       & \citet{McDonough1995} \\
$X_{\ce{K}}$     & Present-day BSE Potassium content             & 240.0   & ppm        & \citet{McDonough1995} \\ \midrule
Core      &                                   &            &          &   \\ \midrule 
$\alpha_c$  & Core thermal expansion coefficient       & $10^{-5}$   &     K\pr{-1}      & \citet{davies2015cooling}  \\
$c_c$     & Core heat capacity                & 800        & J kg\pr{-1} K\pr{-1}  & \citet{davies2015cooling} \\
$k_b$        & Boltzmann constant                & $1.38\times10^{-23}$   & J K\pr{-1}    &   \\
$N_a$        & Avogadro constant                 & $6.022\times 10^{23}$  & mol\pr{-1}   &   \\
$M_c$        & Core mass                         & $1.94\times10^{24}$   & kg        & \citet{PREM}  \\
$M(\ce{Fe})$  & Iron molar mass                   & 55.845     & g mol\pr{-1}      &  \\
$M(\ce{x}) $   & O -- Si molar mass                   & 15.99  -- 28.0    &   g mol\pr{-1}     &  \\
$c_l$        & O -- Si initial core concentration        & 0.16 -- 0.02       &   mol\%  & \citet{badro2015core,davies2015cooling} \\
$\Delta \mu_0^{l-s} $            & O -- Si change in chemical potential & -4.17 --  -0.08 & $10^{-19}$ J mol\pr{-1} & \citet{alfe2002composition}\\
$\lambda_x^s $                 & O -- Si linear correction in liquid  & 0 -- 4.33            & $10^{-19}$  J mol\pr{-1} & \citet{alfe2002composition} \\
$\lambda_x^s$             & O -- Si linear correction in solid    & 5.21 --  5.77     &   $10^{-19}$  J mol\pr{-1} & \citet{alfe2002composition}  \\
$\alpha_x^l$ & O -- Si chemical expansion coefficient       & 1.1 --  0.86 &    --      & \citet{davies2015cooling}   \\ \bottomrule
\end{tabular}

\end{table}

\subsection{Three examples of thermal evolution}

\begin{figure}[h!]
    \includegraphics[width=\linewidth]{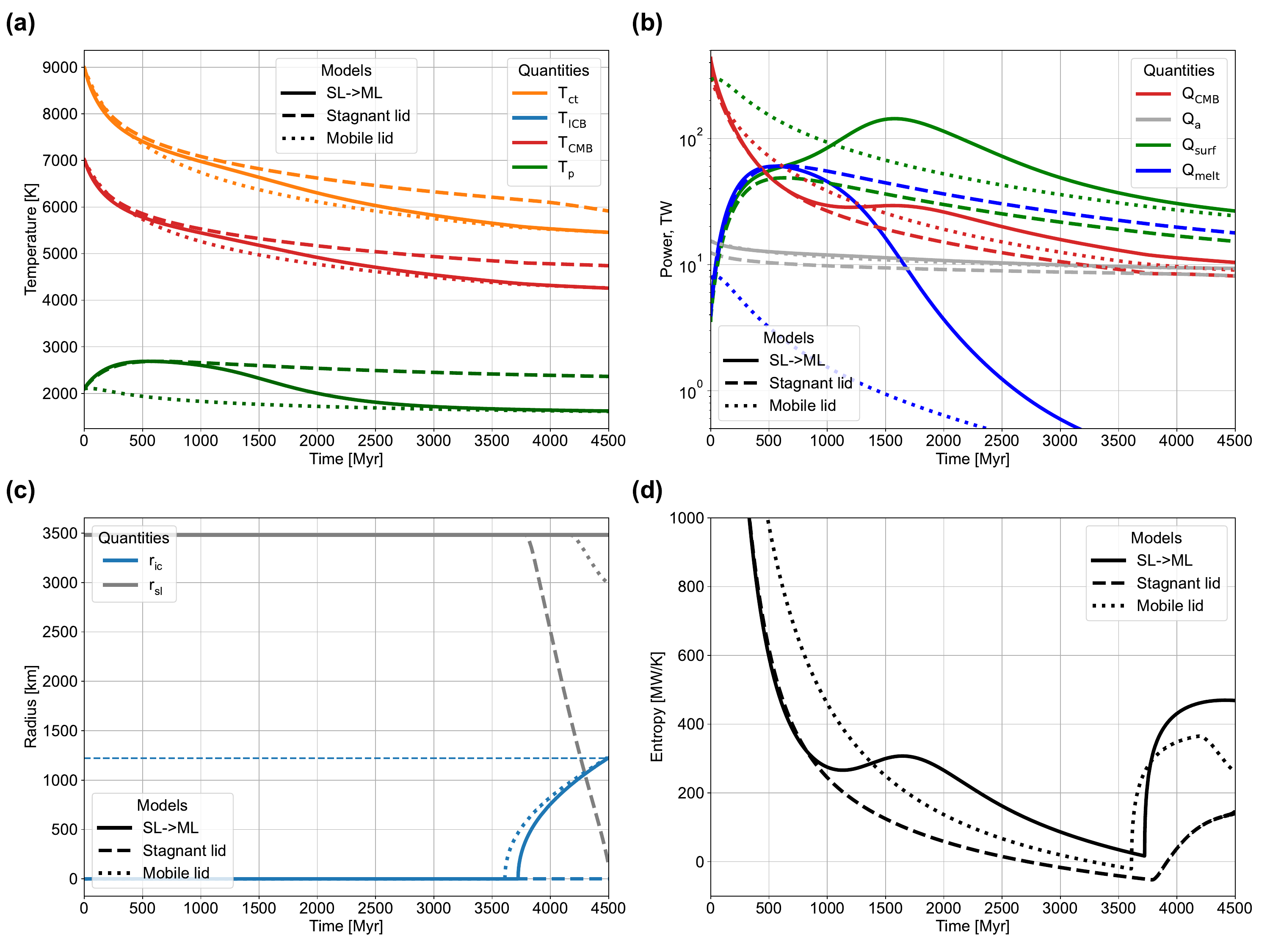}
    \caption{Examples of coupled mantle–core thermal evolution for three tectonic histories: a mobile-lid end-member (dotted lines), a stagnant-lid end-member (dashed lines), and a case transitioning from stagnant- to mobile-lid (solid line; $t^{1/2}_{{onset}} = 1.7$~Gyr and $\Delta t_{{onset}} = 700$~Myr). The four panels shown the evolution of (a) central core temperature (orange), CMB temperature (red), and mantle potential temperature (green); (b) CMB (red), isentropic (grey), surface (green), and melt-related (blue) heat flows; (c) inner-core radius (blue) and thickness of the thermally stratified layer (grey); and (d) entropy available for Joule dissipation. Model parameters are listed in Table~\ref{tab:param}. In addition, we used $\eta_0 = 2.5 \times 10^{21}$~Pa~s, $\Delta_a = 300$ kJ mol\pr{-1}, $V = 4.54$~cm\pr{-1}~mol\pr{-1}, $k_{{cmb}} = 105$~W~m\pr{-1}~K\pr{-1}, and $T_{ct}^0 = 9000$~K. The core thermal conductivity differs in the stagnant-lid case, where $k_{{cmb}} = 88$~W~m\pr{-1}~K\pr{-1} is adopted to prevent complete thermal stratification of the core before the end of the thermal evolution. The activation volume also differs slightly in the transition case, with $V = 4.375$~cm\pr{3}~mol\pr{-1}, to obtain similar final inner-core radius as in the mobile-lid case.}
    \label{fig:1D_examples}
\end{figure}

To illustrate our model and the influence of different mantle convection regimes, we computed three thermal-evolution scenarios: a mobile-lid end-member, a stagnant-lid end-member, and a case that transitions from a stagnant to a mobile lid regime (\reffig{fig:1D_examples}). All simulations use an identical set of parameters (see \reffig{fig:1D_examples} and Tables \ref{tab:param} and \ref{tab:polynom}), except for the activation volume and the core thermal conductivity. The activation volume is the same for the two end-member scenarios, but  slightly lower in the case with an imposed regime transition in order to obtain the same final inner-core radius as in the mobile-lid scenario. We use a high core thermal conductivity, with $f_k = 0.66$, corresponding to $k_{cmb} =$108~W~m\pr{-1}~K\pr{-1} at the CMB to illustrate the new core paradox. For the stagnant-lid case, however, we adopted a lower value of $k_{cmb} = 89$~W~m\pr{-1}~K\pr{-1} to prevent complete thermal stratification of the core before the end of the simulation.

\begin{table}[]
\centering
\caption{Present-day values for our three thermal evolutions examples.}
\label{tab:1D}
\begin{tabular}{@{}llllll@{}}
\toprule
Output & & Stagnant-lid & Mobile-lid & Transition & Unit \\ \midrule
Mantle potential temperature &$T_p$    &      2363        &     1612       &      1621      & K  \\
Mantle average cooling rate since 2 Ga &$\left. \frac{\mathrm{d}T_p}{\mathrm{d}t}\right|_\mathrm{2 \ Gyr}$&     -63         &      -38      &       -94  & Gyr K\pr{-1}     \\
Core central temperature &$T_{ct}$    &       5915       &       5456     &    5458  & K      \\
Core-mantle boundary temperature &$T_{cmb}$   &        4741      &        4263    &       4259  & K   \\
Inner core size &$r_{ic}$   &      0        &    1237        &      1226    & km  \\
Stable layer thickness &$R_c - r_{sl}$    &         3355    &        486    &      0     & km  \\
Isentropic heat flow at the CMB&$Q_a$ ,     &            8.2  &      9.3      &       9.3  & TW    \\
CMB heat flow at the CMB &$Q_{cmb}$   &          8.1    &       9.1    &       10.4   & TW    \\
Surface heat loss &$Q_{surf} $    &           15.3   &        24.4    &   26.6  & TW      \\
Melt extration heat loss &$Q_{melt}$ &            17.8  &         0.2   &      0.2   & TW    \\
Joule entropy &$E_J$     &             149 &     263       &        469   & MW K\pr{-1} \\ 
Minimum Joule entropy &$E_J^{min}$    &      -53       &   -21         &   17      & MW K\pr{-1}   \\\bottomrule
\end{tabular}

\end{table}

The differences among the three scenarios are mainly due to mantle cooling, which in turn affects core cooling. For the two end-member cases, the mantle temperatures rapidly diverge (\reffig{fig:1D_examples}a, green lines). In the stagnant-lid case, about half of the cooling is due to melt extraction, and the other half is due to surface heat loss, whereas in the mobile-lid case it is controlled by surface heat loss mainly (\reffig{fig:1D_examples}b). These two cooling modes yield contrasting mantle thermal states: the mobile-lid regime produces a relatively cold mantle, while the stagnant-lid regime produces a warm mantle (\reftab{tab:1D}), partly due to a higher internal heating ($\gamma_{eff}=0$ in \refeq{eq:lambda_m}). However, both share a low average mantle cooling rate in the last 2 Gyr: respectively $\approx$-63 K Gyr\pr{-1} and $\approx$-38 K Gyr\pr{-1} (\reftab{tab:1D}). 

On the core side, the difference in temperature between regimes grows slowly with time, with the stagnant-lid case leading to a systematically warmer core at present-day (\reffig{fig:1D_examples}a). In our model, the bottom thermal boundary layer is insensitive to the tectonic style; therefore, this behaviour is only due to differences in the lower mantle temperature. Because the stagnant lid cools less efficiently, the lower mantle is warmer and the temperature jump across the bottom boundary layer smaller, leading to a slightly reduced CMB heat flow. Although this reduction is small, when integrated over 4.5~Gyr, it has a large impact on the core evolution and its final structure (\reffig{fig:1D_examples}c).

In the mobile-lid scenario, inner-core nucleation starts after $\approx$3.7~Gyr, with the inner core growing to a radius of about 1237~km, close to its present-day value (\reftab{tab:1D}). After 4.2~Gyr, the CMB heat flow falls below the isentropic heat flow at the CMB, and a stratified layer forms that grows to a thickness of about 486~km (\reffig{fig:1D_examples}c and \reftab{tab:1D}). In the stagnant-lid scenario, the warmer core and lower CMB heat flow cause a stratified layer to form after approximately 3.8~Gyr, with the core becoming fully stratified within 700~Myr (\reffig{fig:1D_examples}c). In this case, the absence of an inner core accelerates the growth of the stratification because the core experiences a larger temperature decrease (eq.\ref{eq:sl_growth}).

In the stagnant-lid scenario, due to the reduced core cooling, the available entropy for Joule dissipation becomes negative at around 2.7 Gyr (\reffig{fig:1D_examples}d and \reftab{tab:1D}). It increases slightly during the subsequent stratification of the core. In the mobile-lid scenario, the available entropy drops below zero later, at approximately 3.3~Gyr, and remains negative until the inner-core starts to grow at approximately 3.7~Gyr, after which it increases again until the present day (\reftab{tab:1D}). Therefore, under the assumption of a high core thermal conductivity, the mobile-lid model cannot sustain a continuous, thermally-driven geodynamo until inner core nucleation (\reffig{fig:1D_examples}d). This behaviour illustrates the new core paradox, whereby a high core thermal conductivity results in insufficient cooling before inner core nucleation to power the geodynamo \citep{labrosse2015thermal,driscoll2023paradox}.

We observe that at the onset of stratified layer growth, the available entropy for Joule dissipation increases continuously until the core is fully stratified in absence of an inner core. This behaviour arises because the entropy sink due to conduction along the isentrope ($E_k$) is defined as the volume integral of $k \left(\nabla T / T\right)^2$ (first term on the right-hand-side of eq. \eqref{eq:entropy_dev}), which peaks near the top of the core in absence of a stratified layer (\reffig{fig:core_fig}a-b). As the stratified layer develops from the CMB downward, it progressively decreases the temperature gradient in the region that contributes both the strongest dissipation and the largest volume, leading to a marked reduction in $E_k$ (\reffig{fig:bench}d). By contrast, $E_s$ primarily depends on the integral of the temperature and density profiles, as well as on the cooling rate of the central temperature (second term on the right-hand-side of eq. \eqref{eq:entropy_dev} ). As the temperature in the stratified layer tends to increase compared to the isentropic profile and the density does not change, $E_s$ decreases more slowly than $E_k$. In addition, the growth of the stratified layer enhances the cooling rate of the central temperature. This occurs because, in absence of inner core growth, the denominator of eq. \eqref{eq:dTcendt} decreases faster than the nominator, reflecting a reduction in the volume-to-area ratio and thus a more efficient cooling of the convective core. Since the adiabatic heat flux changes only slightly at the top of the core (\reffig{fig:core_fig}a-b), the convective core cools increasingly more rapidly as the stratified layer grows (\reffig{fig:1D_examples}a, orange dashed line). As a consequence, the entropy budget predicts positive available entropy even for a fully stratified core, calling into question the validity of the formulation used here, especially the integration domain chosen for $E_s$. Indeed, even if the stratified layer domain is excluded from the integration of $E_s$, the available entropy remains negative. Nevertheless, this suggests that a moderate thermal stratification may actually favour the long-term sustainability of a geodynamo on geological timescales before inner core nucleation \citep{labrosse2015thermal}.  

The case involving a transition from stagnant-lid to mobile-lid initially evolves similarly to the pure stagnant-lid regime, with slow mantle cooling (\reffig{fig:1D_examples}a). Core cooling is, however, slightly more efficient than in the pure stagnant-lid case due to the lower activation volume, resulting in a cooler core and higher surface and CMB heat fluxes before the transition (\reffig{fig:1D_examples}b). The transition begins at approximately 1.0~Gyr, reaches its midpoint at 1.7~Gyr, and is completed by 2.4~Gyr. During this time interval, the surface heat flux increases while melt extraction progressively decreases (\reffig{fig:1D_examples}b). Since both processes remain significant during the transition, mantle cooling is particularly efficient, leading to a rapid drop in potential temperature to values similar to the mobile-lid regime before 3.5~Gyr (\reffig{fig:1D_examples}a).

At the end of the evolution, the inner-core radius reaches a radius of 1226~km (\reffig{fig:1D_examples}c), the mantle potential temperature is 1621~K (\reffig{fig:1D_examples}a), and the mean mantle cooling rate over the last 2~Ga is $-94$~K~Gyr$^{-1}$, in agreement with observational constraints. During the transition, both surface and CMB heat fluxes temporarily exceed those of the fully mobile-lid case, reflecting the warmer mantle and correspondingly lower viscosity prevailing at the onset of plate tectonics. Consequently, the CMB heat flux remains above the adiabatic heat flux throughout the evolution, inhibiting the development of a thermally stratified layer at the top of the core (\reffig{fig:1D_examples}c). The resulting enhancement of core cooling after the transition maintains a positive available entropy even prior to inner-core nucleation, thereby allowing a long-lived geodynamo (\reffig{fig:1D_examples}d). This transitional scenario simultaneously satisfies constraints on the present-day thermal state, mantle cooling rate, and magnetic field history, owing to the combined effects of reduced mantle viscosity and delayed mantle cooling associated with the late onset of mobile-lid tectonics.

Taken together, these illustrative cases highlight the strong sensitivity of coupled mantle–core evolution to a limited number of poorly constrained parameters, including mantle viscosity profile, core thermal conductivity, and the timing and duration of the transition between convection regimes. While the end-member and transitional scenarios reproduce distinct thermal evolutions, several combinations of parameters are capable of satisfying subsets of the observational constraints. This non-uniqueness undermines the ability of forward modelling to deliver definitive conclusions and motivates a systematic exploration of the parameter space. To quantify parameter trade-offs, assess the relative influence of each control parameter, and identify families of solutions consistent with present-day mantle and core constraints, we therefore perform a Bayesian inversion of the observational constraints using a Markov Chain Monte Carlo (MCMC) approach.

\section{Bayesian inversion.}

We perform a Bayesian inversion to estimate the parameters of our coupled 1D core-mantle thermal evolution model that best reconcile model predictions with constraints on both the core and the mantle. The inversion methodology for the parametrized thermal evolution model follows the approach established by \citet{bonnet2025martian}. The inversion is based on a MCMC sampling algorithm, which explores the parameter space via a random walk and generates a collection of models (each model being defined by a set of parameters) that approximates the posterior probability distribution of the model parameters conditioned on the observational constraints.

\paragraph{Bayesian problem.}
A Bayesian inference is used to formulate our inverse problem following the formalism proposed by \citet{gallagher2009markov} and based on Bayes' equation \citep{bayes1763lii}: 

\begin{equation}
    p( \textbf{m}  |  \textbf{d})  \propto p(\textbf{d}  |  \textbf{m}) \ p(\textbf{m}),
    \label{eq:bayes}
\end{equation}

\noindent where $\textbf{m}$ represents the parameters of the thermal evolution model and $\textbf{d}$ represents our ``data'', which consist of observational constraints. $p(\textbf{m})$ is the prior distribution of the model parameters, and $p(\textbf{d}  |  \textbf{m})$ is the likelihood of the observational constraints, i.e. the probability of the data $\textbf{d}$ conditional on the model parameters $p(\textbf{m})$.  $p( \textbf{m}  |  \textbf{d})$ is the posterior probability distribution of the model parameters that we aim to infer. 

\paragraph{Prior constraints.}

The prior probability $p(\textbf{m})$ encodes our a-priori information on the model parameters.  Here, we infer the 7 parameters listed in \reftab{tab:Inversion_MCMC}; the other parameters, held constant, are listed in Table~\ref{tab:param}. In the absence of more informative prior constraints except for possible ranges of values, we adopt a uniform probability distribution for all parameters with bounds listed in \reftab{tab:Inversion_MCMC}. The prior probability is therefore constant within the admissible parameter space and 0 outside of it. 

In addition, we exclude models that develop a full stratification of the core by assigning them a zero prior probability. Indeed, these scenarios differ significantly from the solutions we are looking for as they inevitably lead to the cessation of the geodynamo. 

\paragraph{Observational constraints and likelihood.}

A thermal evolution scenario generates outputs, obtained during the evolution and/or at its end. Some of these outputs can be compared with Earth observations and therefore constitute model predictions, hereafter referred to as forecasts. We use four observational constraints, which form our "data". The data vector \textbf{d} is composed of the present-day inner core radius $r_{ic}$, the age of the magnetic field $t_\mathrm{mag}$, the present-day mantle potential temperature $T_{p}^\mathrm{pd}$, the average mantle cooling rate over the last 2 Gyr $\left. \frac{\mathrm{d}T_p}{\mathrm{d}t}\right|_\mathrm{2 \ Gyr}$, and the present-day total surface heat flow $Q_s^{tot} = Q_{surf}+Q_\mathrm{melt}$. In our case, the observations are assumed to be uncorrelated, so the covariance matrix is diagonal, with entries given by the error variances associated with each observation. In the case of a diagonal covariance matrix, the likelihood can be written as the product of four Gaussian probability distributions:

\begin{equation}
    p(\textbf{d}  |  \textbf{m}) = p(r_{ic}|  \textbf{m}) \times p(t_\mathrm{mag}|  \textbf{m}) \times p(T_{p}^\mathrm{pd}|  \textbf{m}) \times p\left( \left. \frac{\mathrm{d}T_p}{\mathrm{d}t}\right|_\mathrm{2 \ Gyr} |  \textbf{m} \right)\times p(Q_s^{tot}|  \textbf{m}).
    \label{eq:likelihood_gaussien_4v}
\end{equation}

\noindent The probability $p(X| \textbf{m})$ follows then a Gaussian law: 

\begin{equation}
    p(X|  \textbf{m}) = \frac{1}{\sigma_{X} \sqrt{2 \pi}} \exp{-\frac{1}{2} \left( \frac{\bar{X}-X(\textbf{m})}{\sigma_{X}} \right)^2},
    \label{eq:proba_gauss}
\end{equation}

\noindent where $\bar{X}$ and $\sigma_{X}$ are the mean and standard deviation of the data, respectively. 

The likelihood function is constructed based on observational constraints derived from the literature. The present-day mantle potential temperature is constrained to $1625\pm25$~K \citep{herzberg2010thermal}. While seismology provides a very precise measurement of the inner core radius at 1221~km \citep{PREM} with an accuracy at the kilometre  \citep{masters1990summary,irving2018seismically}, the inherent sensitivities of 1D parametrized models makes such precision difficult to achieve in practice. Therefore, we adopt a standard deviation of 20~km to slightly relax the constraint on this output. The average mantle cooling rate over the last 2 Gyr is constrained between -80 and -110~K~Gyr\pr{-1} \citep{forst2022multidisciplinary}. We adopt the median value of -95~K~Gyr\pr{-1} as the mean. By defining the observed range as the standard deviation, we obtain a standard deviation of $\pm15$~K~Gyr\pr{-1}. The present-day surface heat loss  is $46\pm3$~TW \citep{jaupart2015temperature} but since our energy balance (\refeq{eq:lambda_m} and \refeq{eq:energy_balance}) does not include continental heat production, the target surface heat flow for our model corresponds to the total surface heat loss after subtracting the continental radiogenic contribution ($7$ TW in our model), yielding a value of $39$~TW. As the uncertainty reported by \citet{jaupart2015temperature} is $\pm3$~TW, we adopt a $\pm1$~TW uncertainty for the inversion, assuming that this estimate represents approximately a $3\sigma$ interval. Regarding the magnetic field, we define the age of the geodynamo as the time before present during which the Joule entropy production exceeds 1~MW~K\pr{-1} \citep{landeau2022sustaining}. To be consistent with the palaeomagnetic record, this age must be older than 3.4 Ga. The first four constraints follow a Gaussian distribution, while the magnetic field age follows a uniform distribution between 4.5 and 3.4 Ga, i.e the interval from the formation of the Earth to the minimum age of the geodynamo.

\subsection{Markov Chain Monte Carlo algorithm.}

To approximate the posterior distribution, we implement a Markov Chain and a Monte Carlo sampling algorithm \citep{sambridge2013transdimensional}. The latter performs a random walk through the parameter space, starting with a model $\textbf{m}_1$ to which we apply a small, random, perturbation to obtain a model $\textbf{m}_2$:

\begin{equation}
    \textbf{m}_2 =   \textbf{r}_a^\mathrm{gauss}(\textbf{m}_1,\boldsymbol{\sigma}_\mathrm{pertub}),
\end{equation}

\noindent where $\textbf{r}_a^\mathrm{gauss}$ is a vector of 7 random numbers following a Gaussian distribution centred around the $\textbf{m}_1$-parameters, with a standard deviation $\boldsymbol{\sigma}_\mathrm{pertub}$ defined for each model parameter (Table~\ref{tab:Inversion_MCMC}). We calculate the prior probability $p(\textbf{m}_2)$ as described above. We then calculate the posterior probability of our two sets of model parameters, which depends on the likelihood and the prior. We accept the new model $\textbf{m}_2$ in our Markov chain if the following condition is verified:

\begin{equation}
    \frac{p(\textbf{m}_2  |  \textbf{d})}{p(\textbf{m}_1  |  \textbf{d})} > \mathrm{r}_b,
\end{equation}

\noindent where $\mathrm{r}_b$ is a random number chosen uniformly between 0 and 1. If this condition is met, the model $\textbf{m}_2$ is accepted, otherwise the model $\textbf{m}_1$ is retained. The algorithm then restarts from the accepted model ($\textbf{m}_1$ or $\textbf{m}_2$) to which we again apply a perturbation. This operation is repeated a number of times sufficient to sample the entire parameter space. The accepted models form a Markov chain with the following properties: the distribution of the sampled model in the chain is equivalent to the model posterior distribution $p( \textbf{m}  |  \textbf{d})$.  We can then approximate the posterior probability function of our model in the parameter space by calculating the density function of this collection of models. The acceptance rate measures the ratio of accepted over proposed models. This rate increases upon reducing the amplitude of the perturbations and decreases upon increasing them. The optimal acceptance rate in our case is around 23.4\% \citep{roberts2009examplesMCMC}.

\begin{table}[tp]
\centering
\begin{tabular}{@{}lllll@{}}
\toprule
Parameter & Minimum      & Maximum      &$ \sigma_\mathrm{perturb} $ & Unit\\ \midrule
$\log_{10}{(\eta_0)} $     & 20 & 22.5 & 0.04   &  $\log_{10}{(\text{Pa s})}$  \\
$\Delta_a$   & 50  & 450 & 1 & kJ mol\pr{-1} \\
$V$        & 3.5  & 5.5 & 0.02 & $\text{cm}^3$ mol\pr{-1} \\
$ t_{1/2, onset}$    & 0        & 3.7      & 0.02   & $\mathrm{Gyr}$   \\
$\Delta t_{onset}$      & 0.02      & 1.2     & 0.01    & $\mathrm{Gyr}$  \\
$f_k $      &0     & 1    & 0.004  &  $-$  \\
$\mathrm{T}_{ct}^0 $      & 7000   & 11000     & 15  &  K \\ \bottomrule
\end{tabular}
\caption{Model parameters of our sampling algorithm with their range of variability perturbation amplitude. $\eta_0$ is the reference viscosity, $t_{1/2,onset}$ is the mid-time of the transition from stagnant- to mobile-lid, $\Delta t_{onset}$ is half the duration of the transition, $V$ is the activation volume, $f_k$ is the weighting coefficient of the core thermal conductivity, ${T}_p^0$ is the initial mantle potential temperature, and ${T}_{ct}^0 $ is the initial core central temperature.}
\label{tab:Inversion_MCMC}
\end{table}

\section{Inversion results}

\subsection{Incompatible mantle cooling and surface heat-flow constraints}

To constrain the model, we first attempted a joint inversion using all five observational constraints, including both the present-day surface heat loss and the mantle cooling rate over the last 2~Gyr. However, no satisfactory solution was found that simultaneously reproduces these two constraints. The posterior likelihood distributions (see Figure S1 in the Supplementary Information (SI), available via the Zenodo repository) reveal a systematic discrepancy: the inversion consistently yields a surface heat loss that is too low ($37.8\pm0.9$ TW) and a mantle cooling rate that is too high ($-129\pm7$K Gyr\pr{-1}) relative to the observational constraints. Consequently, the results of this inversion cannot be considered statistically valid or conclusive on their own.

This incompatibility reflects the well-known thermal catastrophe viewed from a forward-modeling perspective. Because mantle viscosity strongly depends on temperature, increasing the present-day surface heat loss generally requires a hotter mantle, which also leads to faster long-term cooling. The inversion therefore converges toward low temperature and pressure dependencies of viscosity, a late transition to the mobile-lid regime, and a high present-day CMB heat flux (see Figure S2 and S3 of the SI). Although these characteristics are consistent with mechanisms previously proposed to overcome the thermal catastrophe, they remain insufficient to reconcile the two observational constraints within the present model.

To circumvent this limitation and evaluate the impact of these competing constraints, we isolated them into separate inversions. We define as our reference case the inversion including the mantle cooling rate constraint (Sections \ref{sec:iniphase}, \ref{sec:invparam} and \ref{sec:posterior}), and compare it with a second inversion using the present-day surface heat-loss constraint instead (Section \ref{sec:heatflowinv}). Two additional inversions, without the palaeomagnetic constraint, are then used to assess its influence on the inferred model parameters (Section \ref{sec:varconstr}).

\subsection{Initial Phase}\label{sec:iniphase}

\begin{figure}[ht!]
    \centering
    \includegraphics[width=\linewidth]{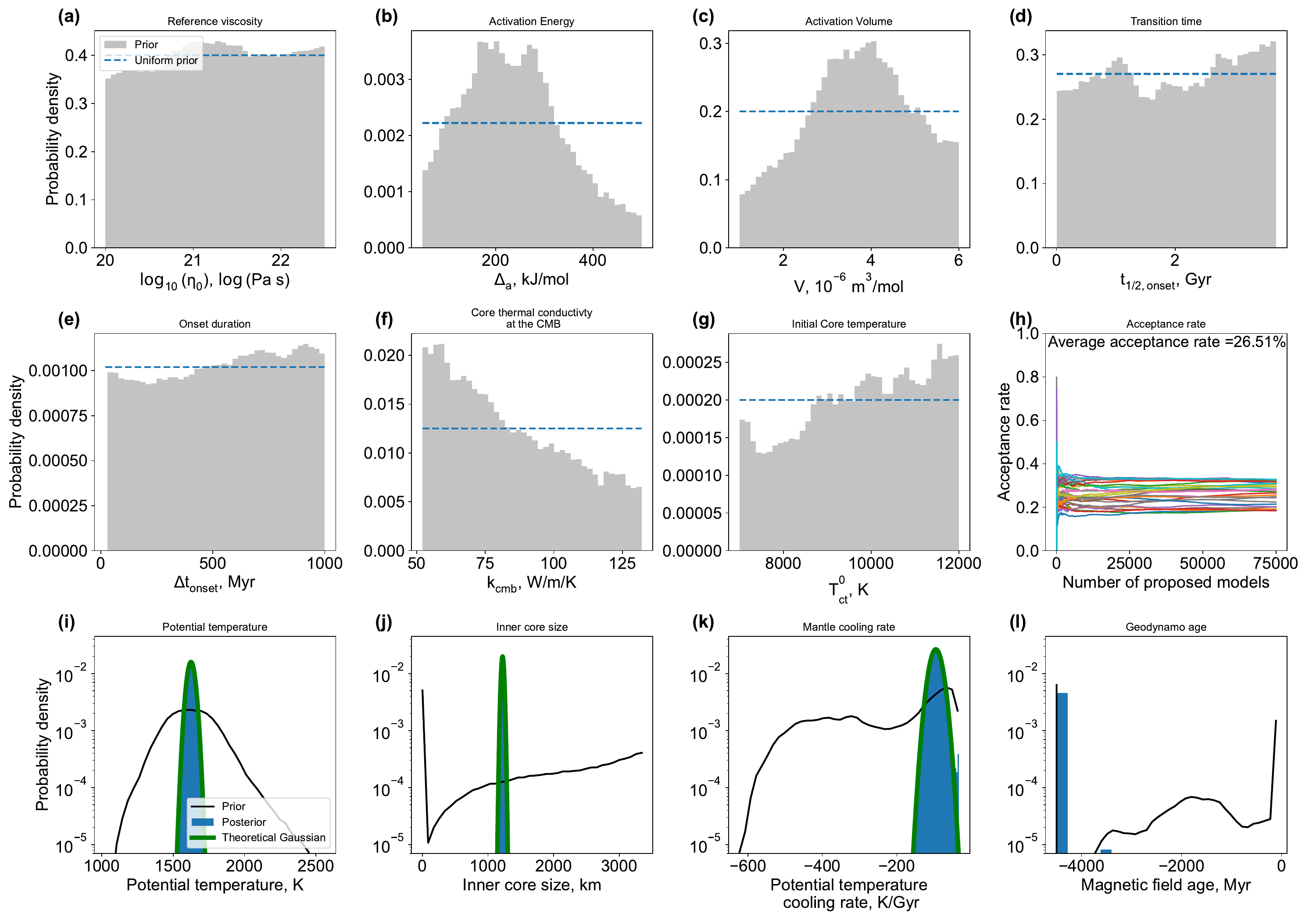}
    \caption{Prior marginal distribution for (\textbf{a}) the reference viscosity, (\textbf{b}) the activation energy, (\textbf{c}) the activation volume, (\textbf{d}) the transition time, (\textbf{e}) the onset duration,  (\textbf{f}) the core thermal conductivity at the CMB, and (\textbf{g}) the initial core temperature. (\textbf{h}) Acceptance rate for the 60 chains as a function of the number of models. Likelihood, prior distributions and posterior likelihood probability for (\textbf{i}) mantle potential temperature, (\textbf{j}) inner core radius, (\textbf{k}) average mantle cooling rate over the last 2 Gyr, and (\textbf{l}) magnetic field age.}
    \label{fig:prior}
\end{figure}

A preliminary exploration phase is required before carrying out the final inversion to properly calibrate the MCMC sampler. We first perform a series of trial inversions using different perturbation amplitudes for each parameter (perturbation vector $\boldsymbol{\sigma}_{\mathrm{perturb}}$, Table~\ref{tab:Inversion_MCMC}). The goal is to identify perturbation amplitudes that yield an acceptance rate as close as possible to the optimal value of 23.4\%. Once the proposal distribution (i.e. the perturbation vector) is specified, we then determine the Markov chain length required to ensure an adequate exploration of the parameter space. To this end, we perform two types of inversions. First, an inversion without constrains -- the prior subset -- is carried out by setting the data likelihood $p(\mathbf{d}\,|\,\mathbf{m}) = 1$, such that the resulting distribution is our prior. Second, we perform an inversion using our four observational constraints -- the reference case. For both cases, 60 independent Markov chains are run in parallel, each starting from a randomly selected initial model. Convergence of the prior case inversion is achieved with 4.5 million sampled models, which we adopt as the total number of proposed models for both the prior and reference inversions. For each chain, the first 8000 accepted models are discarded as burn-in, corresponding to the initial phase during which the algorithm converges toward the high-probability regions of the parameter space. The remaining samples from all chains are combined to form the final ensemble, yielding a total of $N_{\mathrm{model}} = 4.002 \times 10^{6}$ accepted models. For the reference inversion, the resulting acceptance rate is 26.5\%, close to the target optimal value (\reffig{fig:prior}h). For clarity, the parameter $f_k$ is presented in the results in terms of the corresponding thermal conductivity at the CMB.

The prior distribution does not exactly reproduce uniform distributions over the prescribed parameter ranges (\reffig{fig:prior}, grey histograms). This deviation arises because models leading to full thermal stratification of the outer core before 4.5~Gyr are rejected by construction. As a result, the effective prior acceptance rate is approximately 95\%, and the difference between the prior and a uniform distribution highlights the regions of parameter space that systematically lead to complete core stratification. The most prominent effect is observed for the core thermal conductivity at the CMB: the larger $k_{cmb}$, the higher the probability of full stratification (\reffig{fig:prior}f). Activation energies lower than 150 and larger than 350 kJ mol\pr{-1} (\reffig{fig:prior}b) and activation volumes lower than 3 and larger than 5.5 $\text{cm}^3$ mol\pr{-1}  (\reffig{fig:prior}c), and colder initial core temperatures (\reffig{fig:prior}g) also tend to favour full stratification, as all of these factors tend to reduce the CMB heat flux at present-day. For the reference inversion case, the posterior distributions of the four forecast variables closely reproduce the prescribed likelihood functions (\reffig{fig:prior}, blue histograms). A priori, these outputs are highly spread: the mantle potential temperature spans 1163--2500~K (\reffig{fig:prior}i), the inner-core radius ranges from 0 to 3430~km (\reffig{fig:prior}j), the mantle cooling rate varies between $-600$ and $-22$~K~Gyr$^{-1}$ (\reffig{fig:prior}k), and the age of the magnetic field extends from $-4.5$ to 0~Ga (\reffig{fig:prior}l).  The strong contrast between the prior and posterior distributions demonstrates that these forecast are both complementary and highly effective in constraining the model parameters. This specific ensemble of models can be described as points in the 7-dimensional parameter space (one dimension for each inverted parameter, \reftab{tab:Inversion_MCMC}).

\subsection{Posterior distribution of the inverted parameters} \label{sec:invparam}

\begin{figure}[h!]
    \centering
    \includegraphics[width=\linewidth]{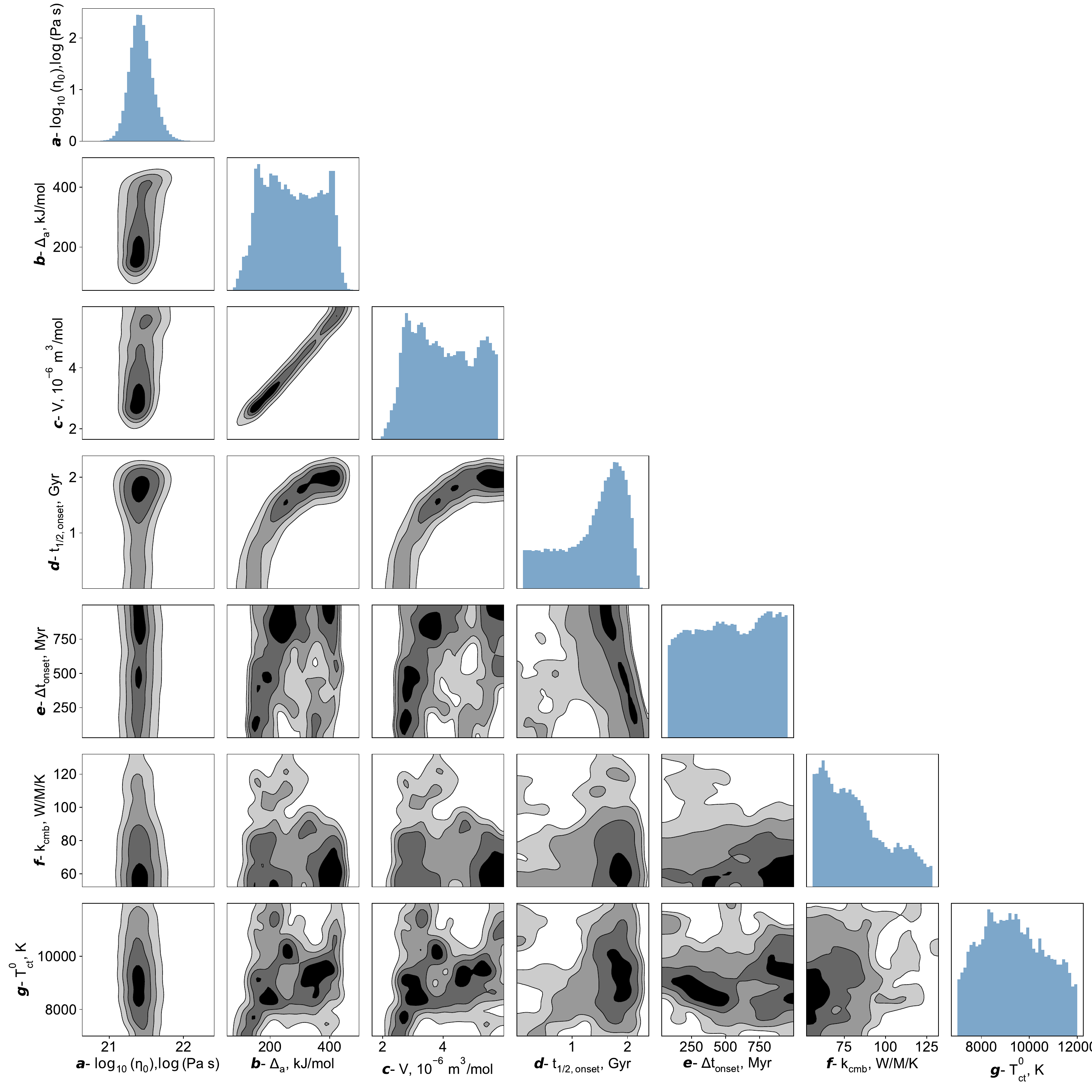}
    \caption{Corner plot representing the projection along our 7 inverted parameters of the posterior probability distribution of our model, i.e. 1D and 2D marginal distributions. For the 1D marginal distributions, the y-axis represents the probability density, while for the 2D marginal distributions, darker colours indicate higher probability density. (\textbf{a})  reference viscosity, (\textbf{b}) activation energy, (\textbf{c}) activation volume, (\textbf{d}) the transition time, (\textbf{d}) onset duration,  (\textbf{e}) thermal conductivity at the CMB, and (\textbf{f})  initial core temperature. The set represent $4.002 \times 10^{6}$ models.}
    \label{fig:corner_plot}
\end{figure}

The inversion can well constrain the reference viscosity. Its posterior distribution can be approximated by a Gaussian, with a mean value of $21.4 \pm 0.17$ in log$_{10}$ space, corresponding to a viscosity of $\approx 2.65 \times 10^{21}$~Pa~s (\reffig{fig:corner_plot}a). The reference viscosity is defined at a pressure of 4.5~GPa, equivalent to a depth of approximately 150~km. This value is in close agreement with viscosity estimates at similar depths derived from glacial isostatic adjustment studies \citep[e.g.,][]{mitrovica1997radial}. For the activation energy, the probability reaches an average of 274 kJ mol\pr{-1}, with a maximum between $\sim200$ and $\sim400$ kJ mol\pr{-1}  (\reffig{fig:corner_plot}b). The activation volume exhibits a maximum probability close to $\sim$$3.2\times10^{-6}$~m$^{3}$~mol$^{-1}$(\reffig{fig:corner_plot}c). These value are consistent with typical experimental and theoretical estimates for the lower mantle, which suggest values close to $\approx 3\times10^{-6}$~m$^{3}$~mol$^{-1}$ and $\sim250$ kJ mol\pr{-1}, but are not in line with upper mantle values \citep{Jain2019,karato2025rheology}. Using the averaged rheological values, it implies a present-day viscosity contrast of approximately a factor 30 between the upper and lower mantle, with a lower-mantle viscosity of the order of $\approx 3.0 \times 10^{22}$~Pa~s, similar to the value commonly inferred for the Earth’s lower mantle \citep{mitrovica1997radial,Lau2016}. Both activation volume and energy show a strong co-variance, the larger the activation energy, the larger the activation volume (\reffig{fig:corner_plot}b-c).


The posterior distribution of the transition time displays a bimodal structure, with a Gaussian-like peak centred at 1.80~Gyr and a ''shoulder'' extending up to the beginning of the evolution (\reffig{fig:corner_plot}d). In contrast, the total duration of the transition remains poorly constrained (\reffig{fig:corner_plot}e), but shows a clear negative correlation with the transition time: late onsets of the transition to mobile lid convection are associated with shorter transition durations. This correlation therefore permits two distinct  end-members scenarios: a relatively early onset with a long-duration transition, or a later onset characterized by a more rapid transition to a mobile-lid regime. Models that show a transition time close to 0 together with a long duration onset are also characterized by  a non-zero $\gamma_{eff}$ 4.5 ago, meaning the absence of a stagnant lid phase (\reffig{fig:corner_plot}d-e). Early transitions are associated with low activation energy and activation volume, whereas more recent transitions require higher values of both parameters (\reffig{fig:corner_plot}d-b/c).  The posterior probability of the core thermal conductivity decreases with increasing conductivity values, again reflecting the signature of the new core paradox (\reffig{fig:corner_plot}f). Nevertheless, the inversion indicates that models with a long-lived geodynamo remain viable for thermal conductivities as high as 110~W~m$^{-1}$~K$^{-1}$, which encompasses approximately 85\% of the accepted solutions, with a non-zero probability until the maximum value of 132~W~m$^{-1}$~K$^{-1}$. The highest thermal conductivity values are preferentially associated with late and rapid tectonic transitions with relatively low activation volume and energy, as these scenarios delay core cooling and thus help sustain sufficient convective power to maintain the geodynamo (\reffig{fig:corner_plot}f-d). 

The inversion also indicates that the initial core temperature is moderately constrained by the inversion (\reffig{fig:corner_plot}g). Relatively warm initial core temperatures are statistically favoured, with an average of 9379 K, whereas colder initial states are disfavoured because they reduce the temperature jump across the core–mantle boundary and thus the convective power available to thermally drive an early geodynamo. This effect further exacerbates the new core paradox by limiting dynamo action before inner core nucleation \citep[e.g.,][]{driscoll2023paradox}.

\subsection{Posterior distribution of selected outputs}\label{sec:posterior}

\begin{figure}[ht!]
    \centering
    \includegraphics[width=0.8\linewidth]{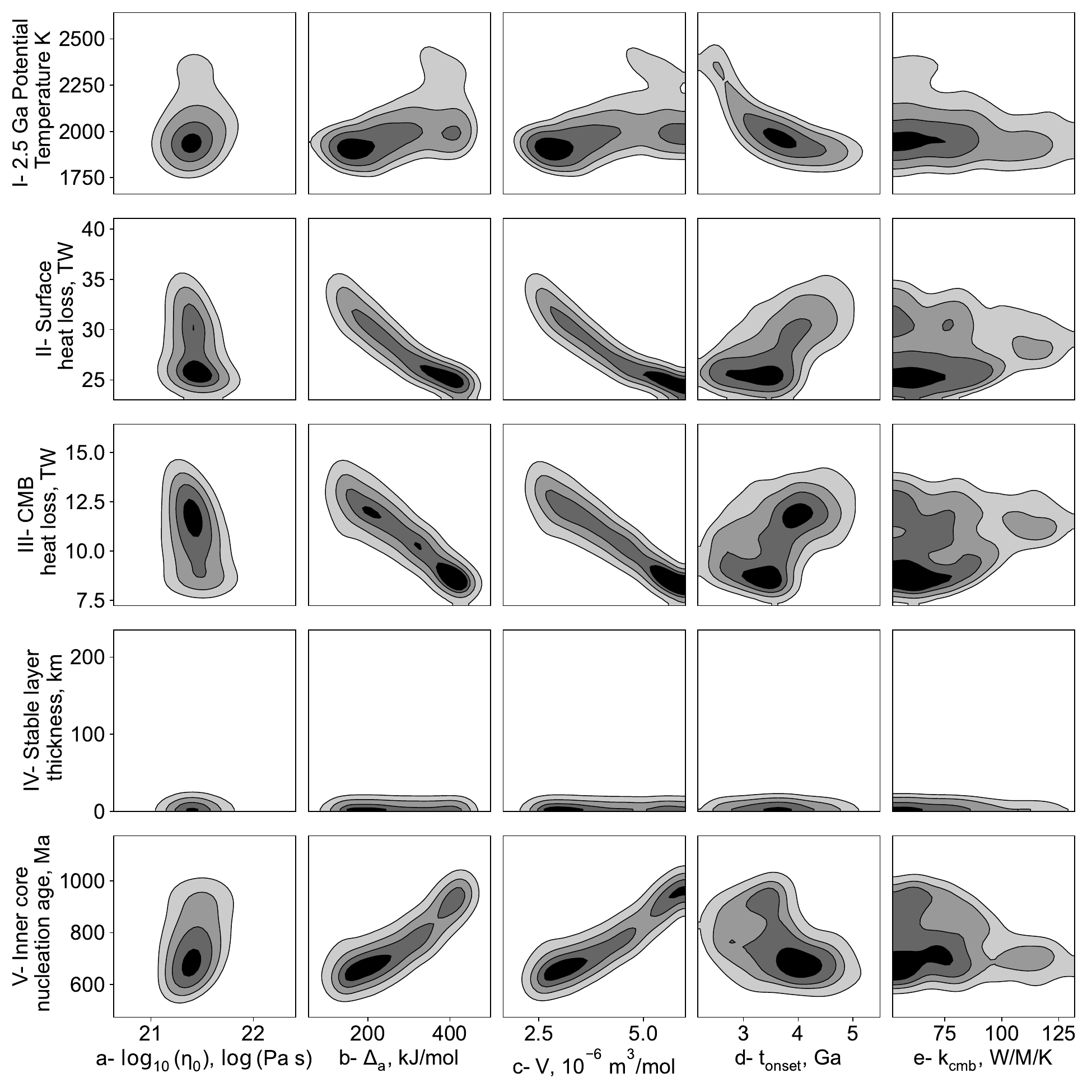}
    \caption{2D marginal posterior distributions of 5 inverted parameters and 5 outputs. Darker colours indicate higher probability density. The columns display (\textbf{a})  the reference viscosity, (\textbf{b}) the activation energy, (\textbf{c}) the activation volume, (\textbf{d}) the onset age, and (\textbf{e}) the thermal conductivity at the CMB. The rows display (\textbf{I}) the mantle potential temperature 2.5 Ga,  (\textbf{II}) the present-day surface heat loss , (\textbf{III}) the present-day CMB heat loss, (\textbf{IV}) the present-day thickness of the thermally stratified layer, and (\textbf{V}) the inner core nucleation age.}
    \label{fig:square_plot}
\end{figure}

The inversion also yields posterior distributions for output quantities, some of which are displayed as functions of the inverted parameters in \reffig{fig:square_plot}. First, we combine the transition time and the transition duration into a single parameter, defined as the onset age of the transition (eq. \eqref{eq:onset_time}). The onset age spreads from about 5.5 to 2.5~Ga, with ages older than 4.5~Ga indicating models that start with a non-zero $\gamma_{eff}$, and a maximum probability at $\sim$4.0~Ga, the start of the Archean (\reffig{fig:corner_plot}d). One way to discriminate between early and late transition scenarios is to consider the mantle potential temperature at 2.5~Ga. This temperature ranges from 1800 to 2400~K (\reffig{fig:square_plot}I). The later the onset, the hotter the mantle is 2.5 Gyr ago (\reffig{fig:square_plot}I-d). Available estimates of mantle potential temperature at 2.5~Ga are consistent with values close to 1955~K \citep{forst2022multidisciplinary}, favouring a relatively early onset to a mobile-lid regime.

The present-day surface heat loss as well as the CMB heat flow, are systematically lower in our inverted model ensemble than commonly inferred from geophysical observations. The ensemble averages are 28.6~TW for surface heat loss (\reffig{fig:square_plot}II) and 11~TW for the CMB heat flow (\reffig{fig:square_plot}III).

$Q_s^{tot}$ is 10.4 TW lower than the target value of 39 TW (see Section 3), which illustrates the difficulty to reconcile mantle cooling rate and surface heat loss. Nevertheless, low reference viscosity, activation energy and activation volume, together with an early transition show more favourable surface and CMB heat losses at present day up to 37 TW and 14 TW, respectively (\reffig{fig:square_plot}II/III--d). High activation volume and activation energy, which are associated with a late transition, favour low surface and CMB heat losses  (\reffig{fig:square_plot}II/III--b/c).

From the core perspective, the inversion indicates that present-day thermal stratification is highly unlikely: fewer than 0.69\% of the accepted models develop a thin stratified layer at the top of the core (\reffig{fig:square_plot}IV). This result reflects the higher present-day CMB heat flow, 11 TW in average, compared with an average isentropic of 6.9~TW (\reffig{fig:square_plot}III). In the absence of such a layer, the CMB temperature, $4260\pm3$ K, is directly determined by the core composition and the present-day inner core size. This value lies within the commonly cited range of CMB temperatures  \citep[3500--4300~K, ][]{andrault2011solidus,davies2015cooling,forst2022multidisciplinary}, but is higher than a recent seismological estimate \citep[3470--3880 K, ][]{deschamps2022estimating}, which suggest a core composition allowing a lower present-day ICB temperature \citep{yuan2023hydrogen}.  The inferred inner-core nucleation age lies between approximately 1000 and 550~Ma. Higher present-day CMB heat flows correspond to more recent inner-core nucleation ages (\reffig{fig:square_plot}V). 

\subsection{Inversion constrained by surface heat flow} \label{sec:heatflowinv}

We performed a second inversion, using the surface heat flow as a constraint in place of the mantle cooling rate. This set contains $2.001\times 10^6$ models with an average acceptance rate of 23.3\%. The parameters posterior co-variation which we obtained are similar to those of the references case (see Figure S4 of the SI). Therefore we report there only the posterior distributions of the output quantities, as in the section above.

\begin{figure}[ht!]
    \centering
    \includegraphics[width=0.8\linewidth]{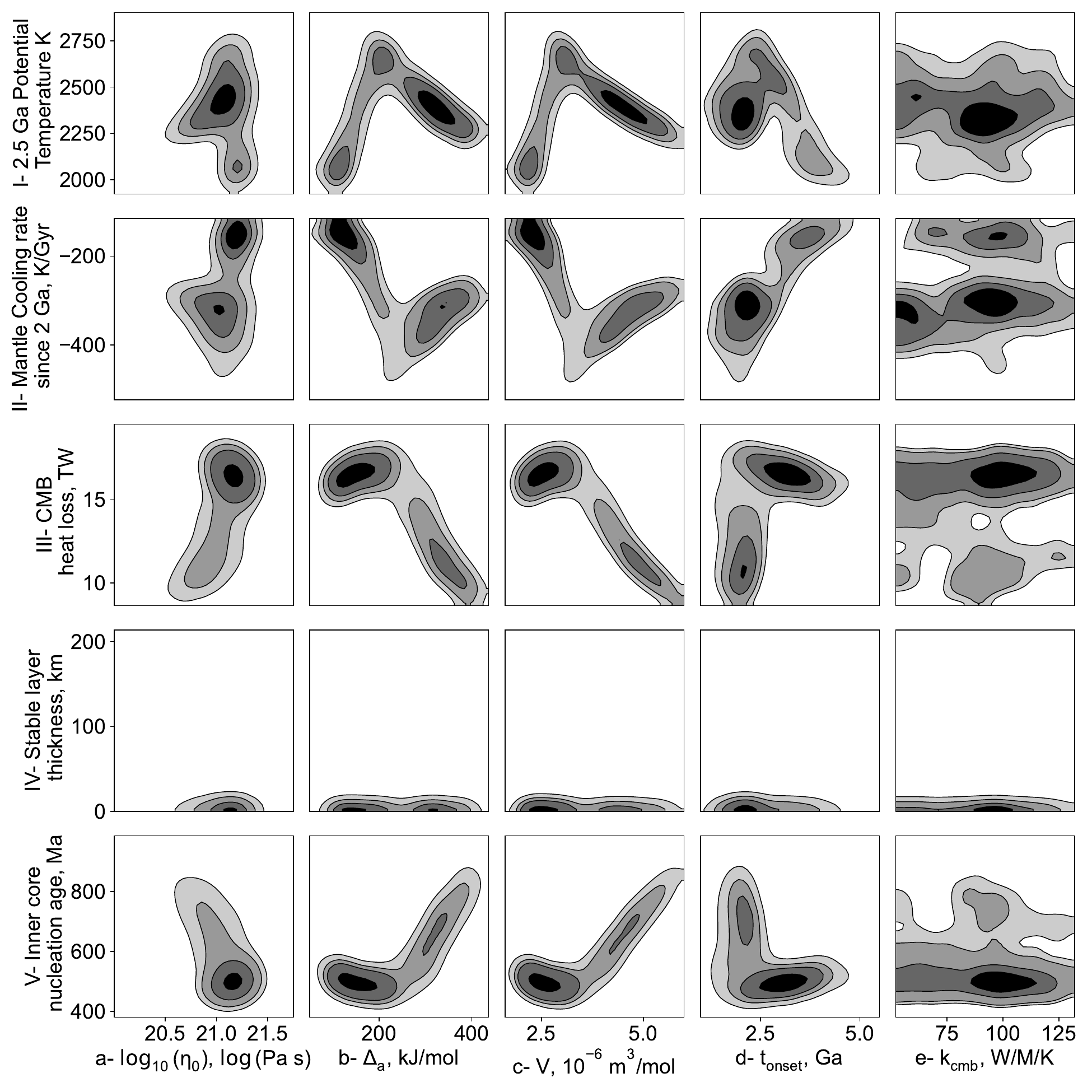}
    \caption{2D marginal posterior distributions of 5 inverted parameters and 5 outputs. Darker colors indicate higher probability density. The columns display (\textbf{a})  the reference viscosity, (\textbf{b}) the activation energy,  (\textbf{c}) the activation volume, (\textbf{d}) the onset age, and (\textbf{e}) the core thermal conductivity at the CMB. The rows display (\textbf{I}) the mantle potential temperature 2.5 Ga, (\textbf{II}) the average mantle cooling rate since 2 Ga,  (\textbf{III}) the present-day CMB heat flow, (\textbf{IV}) the present-day thickness of the thermally stratified layer, and (\textbf{V}) the inner core nucleation age.}
    \label{fig:square_plot_qs}
\end{figure}

The predicted mantle cooling rate is substantially higher than in the reference case, ranging from approximately -130 to -400~K~Gyr$^{-1}$ (\reffig{fig:square_plot_qs}II), and exhibits a bimodal distribution between values lower than $\sim-230$~K~Gyr$^{-1}$ and higher values. The low values are associated with low activation energy, low activation volume, and an early transition to the mobile-lid regime (\reffig{fig:square_plot_qs}II--b,c,d). These models converge toward an almost isoviscous mantle, similar to the solutions identified to get a higher surface heat loss in the reference case. They also predict relatively warm mantle temperatures, between 2000 and 2200 K, prior to 2.5~Ga and a recent inner-core nucleation age of around 500~Ma (\reffig{fig:square_plot_qs}I-V). Higher mantle cooling rates values  are associated with a later tectonic transition, centred near 2.2~Ga, together with larger activation energy and activation volume. These models maintain higher mantle temperatures before 2.5~Ga, approximately 2250--2750~K, and predict inner-core nucleation ages spanning roughly 400 to 900~Ma, with older nucleation ages associated with higher activation energy and activation volume (\reffig{fig:square_plot_qs}I-V).

The present-day CMB heat flow is also shifted toward larger values than in the reference inversion, ranging from approximately 9 to 18~TW (\reffig{fig:square_plot_qs}III). This distribution is similarly bimodal. The highest CMB heat fluxes (up to $\sim$15--17~TW) are obtained for models with low activation energy, low activation volume, and an early mobile-lid transition, whereas models with a later transition predict progressively lower CMB heat fluxes as activation energy and activation volume increase. In all accepted models, these relatively high CMB heat fluxes prevent the development of a thermally stratified layer; representing only 0.7\% of the models (\reffig{fig:square_plot_qs}IV).

Finally, the inversion allows the full range of core thermal conductivity considered here (\reffig{fig:square_plot_qs}e), with the preferred value of $k_{\mathrm{cmb}}$ depending on the combination of activation energy, activation volume, and tectonic transition age. Higher thermal conductivities are associated with lower activation energy and activation volume and an earlier transition to the mobile-lid regime, which together favour higher CMB heat fluxes. Conversely, later transitions and higher activation energy and activation volume favour lower thermal conductivities. Therefore, the new core paradox does not represent a limitation for the solutions satisfying the present-day surface heat-loss constraint.

\subsection{Varying constraints} \label{sec:varconstr}

\begin{figure}[ht!]
    \centering
    \includegraphics[width=\linewidth]{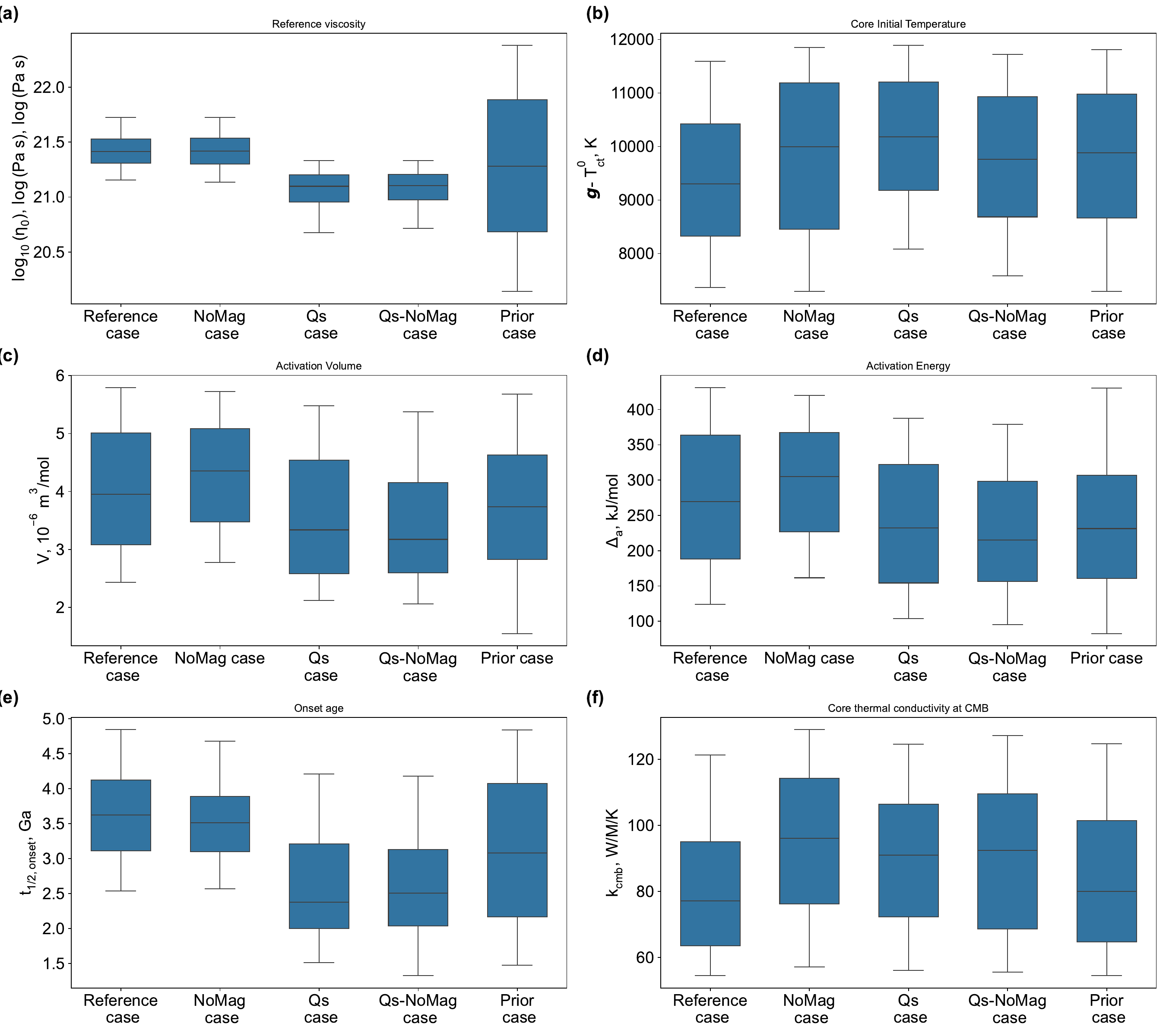}
    \caption{Box-plots representing posterior distribution of a) the reference viscosity, b) the core initial temperature (K), c) the activation volume (m\pr{3} mol\pr{-1}), d) the activation energy (kJ mol\pr{-1}), e)  the transition time (Ga) , f) the core thermal conductivity at the CMB (W m\pr{-1} K\pr{-1}) for different inversions (see text for the details).  Each box-plot displays the median (horizontal line), the interquartile range (blue box), and the 5th and 95th percentiles (whiskers) of the sampled distributions.
}
    \label{fig:boxplot}
\end{figure}

To investigate how the magnetic constraint influences the inferred model parameters, we performed two additional inversions: one using the same constraints as the reference case but without the magnetic field age constraint (NoMag case), and another following the surface heat flux case but also without the magnetic field age constraint (Qs-NoMag case). Both inversions used the same MCMC setup as the previous runs (Reference case), resulting in slightly higher acceptance rates (28.6\% for NoMag, 23.1\% for Qs-NoMag).

We summarize the results in \reffig{fig:boxplot} using box plots. First, the reference viscosity is largely insensitive to the magnetic constraint, with the posterior distributions obtained without this constraint remaining similar to those of the corresponding  inversion that includes it (\reffig{fig:boxplot}a). This is because the reference viscosity is mainly constrained by present-day observations. The initial core temperature is slightly affected by the magnetic constraint, as a warmer initial core is more favourable for sustaining a long-lived magnetic field (\reffig{fig:boxplot}b). While this parameter is only weakly constrained in the reference inversion, the Qs inversion favours higher initial core temperatures. Removing the magnetic constraint shifts the posterior distribution toward lower initial temperatures in the Qs-NoMag case, whereas the NoMag inversion favour favours comparatively higher values, although its posterior remains closer to the prior distribution.

The activation energy and activation volume are only weakly affected by the magnetic constraint in the reference inversion (\reffig{fig:boxplot}c--d). In contrast, for the surface heat-loss inversions, removing the magnetic constraint shifts both parameters toward higher values. A similar trend is observed for the tectonic transition age (\reffig{fig:boxplot}e): the reference inversion favours an earlier transition, whereas the Qs inversion yields a broader distribution extending to later transition ages. Without the magnetic constraint, the Qs-NoMag inversion shows a clear preference for a late transition.

Finally, the magnetic constraint has the strongest influence on the inferred core thermal conductivity (\reffig{fig:boxplot}f). In the reference inversion, the posterior remains close to the prior distribution. The Qs inversion shifts the posterior toward higher thermal conductivities, and this tendency becomes even more pronounced when the magnetic constraint is removed. In the Qs-NoMag case, the posterior is dominated by high thermal conductivity values, reflecting the fact that the high CMB heat fluxes predicted by these models inhibit the formation of a thermally stratified layer and therefore remove the main limitation on large values of $k_{\mathrm{cmb}}$.

\section{Discussion}

\subsection{Constraining Earth's thermal evolution}

Our coupled mantle–core thermal evolution model, which includes a transition in convection regime, successfully reproduces a wide range of observational constraints on Earth's thermal state. However, a key limitation is that the model cannot simultaneously match present-day surface heat loss and the inferred mantle cooling rate. Treating these two constraints separately demonstrates that both can be reproduced while satisfying the other constraints. Our MCMC inversions show that some parameters, in particular the reference viscosity, activation energy and activation volume, are relatively well determined by the present-day constraints (\reffig{fig:corner_plot} and \reffig{fig:boxplot}a-b-c). However, we emphasize that thermal constraints alone are not sufficient to restrict the parameter space. Because of trade-offs between model parameters, the present-day constraints --surface heat loss, potential temperature and core size-- can be reproduced over a wide range of parameter values. Constraints on the thermal evolution of the mantle and core --mantle cooling rate and palaeomagnetic record-- allow us to better define the parameters related to convection regimes and the Earth's magnetic history (\reffig{fig:boxplot}).

This impossibility of simultaneously fitting the mantle cooling rate and surface heat loss stems directly from the strongly temperature-dependent viscosity of the mantle. Models that reproduce the observed surface heat loss require a sufficiently warm and rapidly cooling mantle, whereas models that better reproduce the mantle cooling history tend to predict a lower surface heat loss since 2 Ga (\reffig{fig:square_plot}II and \reffig{fig:square_plot_qs}II). Within this framework, the decrease in mantle temperature inevitably increases viscosity and reduces convective efficiency, making it difficult to simultaneously maintain a high present-day surface heat loss and reproduce the cooling rate. This represents a "forward-in-time" manifestation of the "thermal catastrophe" encountered in backward-in-time models: a strongly temperature-dependent viscosity cannot simultaneously reconcile the Earth's thermal history with its present-day thermal state \citep{davies1980thermal,christensen1985thermal}. To resolve this "thermal catastrophe", the strong covariance between mantle temperature and cooling rate must be reduced. This is highlighted by our inversions, in which low activation energy and low activation volume help reconcile the two constraints but are not sufficient to reproduce them simultaneously (\reffig{fig:square_plot}II and \reffig{fig:square_plot_qs}II). Several physical mechanisms may counter-balance the effect of temperature on viscosity.

Mantle rheology is not only a function of pressure and temperature; it is also sensitive to composition, water content, and grain size \citep{karato2010rheology}. At shallow depths beneath oceanic ridges, melt extraction removes water and iron from the mantle, increasing the viscosity and decreasing the density of the residue, a mechanism called dehydration stiffening \citep{parmentier1992chemical,hirth1996water}. It creates a high-viscosity, low-buoyancy layer within the lithosphere that can significantly reduce surface heat flow \citep{korenaga2009scaling}. Because higher mantle temperatures in the past would have promoted greater melt production, this stiffening effect was likely more pronounced in the past. Conversely, the deep water cycle may act as a buffer; plate tectonics trough subduction gradually increases the mantle water content over time, lowering its viscosity and counteracting the viscosity increase caused by secular cooling \citep{korenaga2011thermal,Seales2022}. Furthermore, smaller grain-sizes decrease the viscosity \citep{karato2010rheology}. Grain size is governed by the competition between mechanical grain reduction, i.e. damage, and grain growth by dynamic recrystallization, i.e. healing \citep{bercovici2014plate,foley2014scaling}. As the mantle cools, the reduction in healing rates may favour smaller grain sizes, thereby reducing the effective viscosity and offsetting the temperature dependence of the viscosity.

Another way to resolve the thermal catastrophe was proposed by \citet{driscoll2014thermal}, who suggested that the core cools faster than the mantle, with a high present-day CMB heat flux ($Q_{\mathrm{cmb}} \approx 15$ TW). Indeed, our inversions constrained by surface heat loss or mantle cooling rate yield similar results in cases with an early transition and low activation parameters (i.e. $\Delta_a, V$). In these cases, $Q_{\mathrm{cmb}}$ reaches 15-17 TW for the surface heat flow case and 12.5-14 TW for the reference case. Combined with $\sim13$ TW of mantle radiogenic heating, this results in only 9-13.5 TW of secular cooling, thereby slowing the decline of mantle potential temperature over time. However, \citet{driscoll2014thermal} obtained these high CMB heat fluxes without considering a low activation energy, but rather relying on a very small viscosity contrast between the upper and lower mantle. This divergence from our results stems from two critical assumptions. First, they incorporated substantial amount of potassium in the core that results in a core radioactive heat production of $\sim85\ \text{TW}$ at $4.5\ \text{Ga}$, although such high potassium concentrations conflict with geochemical estimates \citep[e.g.][]{Blanchard2017,xiong2018ab,Chidester2022}. Second, their lowermost mantle viscosity depends solely on the upper mantle temperature. The core temperature does not affect the lower mantle viscosity and thus the boundary layer thickness; it increases only the temperature jump across the boundary layer. This second difference leads to drastically different CMB heat flows at 4.5 Ga between our model ($\sim300$ TW; \reffig{fig:1D_examples}b) and that of  \citet{driscoll2014thermal} ($\sim40$ TW). Unlike the formulation of \citet{driscoll2014thermal}, our CMB heat flux parameterization has been explicitly benchmarked against 2D and 3D mantle convection simulations \citep{deschamps2001thermal,Breuer2003,Thiriet2019}. Consequently, we argue that their formulation underestimates early heat extraction, which, combined with high core internal heating, artificially delays core cooling until the present day to yield a larger present-day CMB heat flux. Thus, while a rapidly cooling core remains a viable solution, the formulation used by \citet{driscoll2014thermal}, \citet{driscoll2023paradox}, and \citet{patocka2020} is physically unsatisfactory, highlighting the need for alternative mechanisms to sustain high present-day $Q_{\mathrm{cmb}}$

Alternatively, the efficiency of heat transport may be regulated by the convection regime itself. Earth's tectonic history may not have been a simple progression but rather characterized by episodic efficiency, such as Wilson Cycles \citep{burke2011wilson}, intermittent stagnant-lid phases \citep{arndt2013episodic}, or the relative tectonic stability of the ``boring billion'' \citep{stern2018evolution}. In such scenarios, mantle cooling would be non-monotonic, consisting of phases of rapid cooling punctuated by periods of thermal insulation \citep{labrosse2007thermal}. If these phases occur in rapid succession, they may appear as a constant cooling rate in the current, low-resolution $T_p$ record \citep{herzberg2009petrological,forst2022multidisciplinary}. Testing this hypothesis requires higher time resolution data and greater accuracy in paleo-temperature reconstructions. Furthermore, an improved potential temperature record, particularly for the period prior to 2~Ga, remains essential for refining our understanding of the thermal and convective history of the Earth. Conversely, recent constraints from xenolith thermobarometry suggest a significantly lower cooling rate (<50~K~Gyr\pr{-1}) since 1.4~Ga \citep{sudholz2025xenolith}, yet indicate constant equilibration pressures. This implies a stable lithospheric thickness and surface heat flow beneath the continents, suggesting a thermal state near equilibrium characterized by a high Urey ratio, where internal heat production and CMB flux effectively balance surface losses.

Regarding the core, our results highlight that a delayed onset of the transition to a mobile-lid regime helps sustain the geodynamo throughout Earth's thermal history (\reffig{fig:corner_plot} and \reffig{fig:square_plot_qs}). In particular, a late and rapid transition leads to high core cooling rates that are also compatible with high values of the core thermal conductivity. This suggests that non-monotonic mantle cooling is a viable candidate for resolving the new core paradox, as previously proposed also by \citet{alasad2024coupled}. While the palaeomagnetic record provides a less strong constraint on the mantle thermal evolution compared to the potential temperature record, it remains essential for constraining the range of admissible core thermal conductivities. Our reference inversion shows that core thermal conductivities up to 100~W~m\pr{-1}~K\pr{-1} are compatible with a late transition to mobile-lid convection, while values up to 132~W~m\pr{-1}~K\pr{-1} are possible for an early transition combined with low activation energy and activation volume (\reffig{fig:square_plot}e). These high-conductivity solutions are also obtained in the inversion using the surface heat loss as a constraint, where the core thermal conductivity can reach the upper limit of the prior across the inferred range of transition onset ages (\reffig{fig:square_plot_qs}e).

Models that would allow to fit both the mantle cooling rate and the present-day surface heat loss as discussed above, would be even more favourable to high core thermal conductivity, as they would imply larger cooling rates at the time of inner core nucleation. Such scenarios could potentially permit even higher core conductivities than those identified in our inversion, as well as a more recent time of inner core nucleation. While the palaeomagnetic record is compatible with a very young inner core, the data for this period are poorly constrained \citep{bono2019young,lloyd2021first}. The low resolution and high uncertainty of palaeointensity measurements make it difficult to point the age of the  minimum magnetic field intensity, typically associated with a time closely preceding the onset of inner core nucleation \citep{bono2019young,lloyd2021first}. Furthermore, the palaeomagnetic record does not unequivocally imply that the geodynamo may have ceased for a short time interval prior to inner core nucleation. Ultimately, more precise determinations of core thermal conductivity, core composition, and the age of inner core nucleation will provide the necessary constraints to refine our understanding of the coupled evolution between the core and the mantle's tectonic history.


\subsection{Insight on Plate Tectonics onset age}

Our results highlight two scenarios for the Earth's tectonic history: an early onset characterized by a long and gradual increase of cooling efficiency, or a late, rapid transition to a mobile-lid regime. The late-transition scenario is particularly compelling for reconciling the new core paradox. However, this model yields too high mantle temperatures before the transition onset (\reffig{fig:square_plot}I). In contrast, the early-onset scenario favours a more realistic mantle temperature evolution. Regardless of the duration, both scenarios suggest an onset of the transition occurring during the late Hadean or the Archean, with a high probability in the Eoarchean.

Translating the onset age and duration of these transitions into a coherent geodynamic history remains challenging. If modern plate tectonics emerged late, the transition must have been relatively rapid, likely less than 500~Ma, to align with geological observations of plate-like features appearing in the rock record \citep{bedard2024gradual}. In this context, the transition duration represents the time required for tectonic activity to evolve from localized, intermittent lithospheric failures into a globally integrated and fully efficient mobile-lid system.

An alternative explanation for the observed cooling trends is the existence of one or more intermediate convection regimes. Rather than a binary shift between stagnant- and mobile-lid convection, the Earth may have experienced regimes of intermediate cooling efficiency, such as a sluggish lid, plutonic lid, or squishy lid \citep{lenardic2018diversity,gunawardana2024correlating,lyu2025dissecting}. These regimes would likely manifest as a step-like progression in cooling efficiency rather than the single transition assumed in our current model.

\citet{lyu2025dissecting} have recently proposed a detailed regime diagram to distinguish among these convection regimes in dependence of the surface yield stress and mantle internal heating rate. During Earth’s secular cooling, both the internal heating and the lithospheric yield stress decrease, driving the mantle to progressively greater surface mobility and increased localization of surface deformation, until modern-style plate tectonics is established. However, the cooling efficiencies of these intermediate regimes remain poorly characterized and lack suitable scaling laws for heat transport to be used in coupled core-mantle evolution models. 

\subsection{Basal magma ocean and mantle heterogeneities}

Our current model does not explicitly incorporate a basal magma ocean (BMO). Yet, such a feature is widely considered a robust consequence of early Earth history, arising either from the fractional crystallization of a primordial magma ocean or from high initial core temperatures that would have maintained the lowermost mantle molten \citep{labrosse2007crystallizing,boukare2025solidification,lark2026coupled}. In the context of our results, the main effect of an early stagnant-lid regime is to limit mantle cooling during the early Earth. The release of latent heat during BMO crystallization would act as a thermal buffer, significantly slowing the secular cooling of the mantle potential temperature. Furthermore, the BMO effectively insulates the core, reducing the heat flux across the core-mantle boundary and potentially suppressing a thermally-driven geodynamo in the core during the first billion years of Earth's history \citep{lherm2024thermal}. If a BMO were integrated into our framework, satisfying the continuous $3.4\text{ Ga}$ palaeomagnetic record could become more challenging, potentially necessitating an even earlier onset of the mobile-lid regime. A more rapid transition to efficient mantle cooling would be required to accelerate the solidification of the BMO, thereby ending the period of core insulation and allowing the thermally-driven dynamo to initiate and/or persist. Furthermore, core insulation by a BMO and the resulting delay in core cooling could provide a viable mechanism to sustain a higher CMB heat flux at the present day.

As proposed by previous studies, a BMO could have sustained an early geodynamo for sufficiently high values of the silicate electrical conductivity \citep{ziegler2013implications,stixrude2020silicate,lherm2024thermal}. However, a recent study investigating thin-shell dynamo models indicates that the high-aspect-ratio geometry of a basal magma ocean increases the critical magnetic Reynolds number required for dynamo action to approximately 70-80 \citep{schaeffer2025energetically}. This higher threshold suggests that a BMO-driven dynamo is more difficult to sustain without extreme iron enrichment or convective vigour \citep{dragulet2025electrical,schaeffer2025energetically}.

A key limitation of our model is the assumption of a well-mixed mantle, whereas Earth's mantle exhibits strong lateral and radial heterogeneities associated with mantle convection, plate tectonics, and the crystallization of the BMO. Seismic tomography reveals significant lateral thermal and/or compositional heterogeneities near the CMB, the LLSVPs, which can produce large spatial variations in CMB heat flux \citep{Ritsema2020,Frasson2024}. While such lateral variations may influence core dynamics and the development of a thermally stratified layer \citep{TerraNova2024}; it remains unclear to what extent 1D scaling laws can capture the resulting complexity of lower-mantle heat transport. In particular, our parameterization has been benchmarked only for purely thermal convection \citep{Thiriet2019} and does not account for potenteial compositional heterogeneities associated with the LLSVPs, which may affect the mean CMB heat flux. At the surface, the strong contrast between the insulating continental lithosphere and the higher heat flux through oceanic lithosphere may likewise lead to a different thermal evolution than that predicted by our homogeneous 1D model.

Radial heterogeneities may also be important. Variations in mineralogy with depth affect thermal properties such as thermal expansivity and, more importantly, the dominant deformation mechanisms, resulting in a rheological contrast between the upper and lower mantle \citep{tosi2013mantle,Jain2019,karato2025rheology}. Because our inversion is strongly constrained by core evolution, the inferred viscosity parameters primarily reflect the effective rheology of the lower mantle \citep{karato2025rheology}. Introducing distinct rheological laws for the upper and lower mantle could therefore provide additional flexibility and potentially help reconcile the present-day surface heat loss with the long-term mantle cooling rate.

\section{Conclusion}

Earth’s thermal evolution remains subject to several challenges, such as the thermal catastrophe, the new core paradox, and the timing of the onset of plate tectonics. These issues are fundamentally coupled as they are all related to the cooling efficiency of mantle convection. 

We showed that a transition in the convection regime, from stagnant-lid to a fully-developed mobile-lid, can play a key role in reconciling these constraints. In particular, two distinct tectonic pathways that can jointly lead to a long-lived geodynamo and satisfy the mantle cooling rate, even  for high core thermal conductivities:  (1) an early transition beginning at $\approx 4\text{ Ga}$ with an gradual increase in efficiency over an extended interval, or (2) a later transition at $\approx 2.5\text{ Ga}$ characterized by a more rapid shift to a mobile-lid regime. 

At the same time, the purely temperature-dependent viscosity law adopted here limits our ability to reproduce the observed cooling rates over the last billion years and the present-day surface heat loss, highlighting clear priorities for future model developments. In particular, incorporating more realistic rheological laws should help reduce the strong covariance between temperature and surface heat flow, for example by accounting for the effects of grain-size evolution, water content, and dehydration stiffening. In addition, allowing for more complex and potentially non-monotonic histories of cooling efficiency may provide further insight into Earth's tectonic and thermal evolution. Finally, future models should also integrate the evolution and crystallization of a basal magma ocean, which likely played a critical role in regulating both core and mantle cooling.

\begin{acknowledgements}
We thank Stephane Labrosse and an anonymous referee for their thorough and constructive reviews that helped improve a previous version of the paper. This work was supported by the Deutsche Forschungsgemeinschaft (DFG, German Research Foundation) – Project number TO 704/6-1, under the Priority Programme SPP2404 ``Reconstructing the deep dynamics of planet Earth over geologic time (DeepDyn)''.
\end{acknowledgements}

\section*{Data availability}

The source code, inversion raw data and postprocessing code are available in the associated Zenodo's repository \citep{bonnet2026zenodo}. 




\section*{Competing interests}  
Valentin Bonnet Gibet and Nicola Tosi declare no competing interests.    



\printbibliography

@String{AN = "Astronomische Nachrichten"}

@string{PEPI = {Phys. Earth Planet. Inter.}}

@String{SAM = "Stud. App. Math."}

@String{SA = "Sci. Adv."}

@article{PREM,
	Author = {Dziewonski, A. M. and Anderson, D. L.},
	Journal = PEPI,
	Pages = {297--356},
	Title = {Preliminary reference {Earth} model},
	doi = {10.1016/0031-9201(81)90046-7}, 
	Volume = 25,
	Year = 1981}

@article{labrosse2001age,
title = {The age of the inner core},
journal = {Earth Planet. Sci. Lett.},
volume = {190},
number = {3},
pages = {111-123},
year = {2001},
issn = {0012-821X},
doi = {https://doi.org/10.1016/S0012-821X(01)00387-9},
url = {https://www.sciencedirect.com/science/article/pii/S0012821X01003879},
author = {Stéphane Labrosse and Jean-Paul Poirier and Jean-Louis {Le Mouël}}
}

@article{Biggin2015,
  title = {Palaeomagnetic field intensity variations suggest Mesoproterozoic inner-core nucleation},
  volume = {526},
  ISSN = {1476-4687},
  url = {http://dx.doi.org/10.1038/nature15523},
  DOI = {10.1038/nature15523},
  number = {7572},
  journal = {Nature},
  publisher = {Springer Science and Business Media LLC},
  author = {Biggin,  A. J. and Piispa,  E. J. and Pesonen,  L. J. and Holme,  R. and Paterson,  G. A. and Veikkolainen,  T. and Tauxe,  L.},
  year = {2015},
  month = Oct,
  pages = {245–248}
}

@article{Hsieh2025,
  title = {Moderate Thermal Conductivity of Fe‐Ni‐Si Alloy at Earth’s Core Conditions: Implications for Core Thermal Evolution and Geodynamo},
  volume = {52},
  ISSN = {1944-8007},
  url = {http://dx.doi.org/10.1029/2025GL117576},
  DOI = {10.1029/2025gl117576},
  number = {21},
  journal = {Geophys. Res. Lett.},
  publisher = {American Geophysical Union (AGU)},
  author = {Hsieh,  Wen‐Pin and Chiang,  Yi‐Ting and Deschamps,  Frédéric and Chen,  Chao‐Chih and Chang,  Jen‐Wei and Ikuta,  Daijo and Ohtani,  Eiji},
  year = {2025},
  month = Nov 
}

@article{Seales2022,
  title = {Plate tectonics,  mixed heating convection,  and the divergence of mantle and plume temperatures},
  volume = {50},
  ISSN = {1943-2682},
  url = {http://dx.doi.org/10.1130/G50309.1},
  DOI = {10.1130/g50309.1},
  number = {12},
  journal = {Geology},
  publisher = {Geol. Soc. Am. Bull.},
  author = {Seales,  Johnny and Lenardic,  Adrian and Tomasini,  Julian Garrido},
  year = {2022},
  month = Sept,
  pages = {1377–1381}
}

@article{Davaille1993,
  title = {Transient high-Rayleigh-number thermal convection with large viscosity variations},
  volume = {253},
  ISSN = {1469-7645},
  url = {http://dx.doi.org/10.1017/S0022112093001740},
  DOI = {10.1017/s0022112093001740},
  journal = {J. Fluid Mech.},
  publisher = {Cambridge University Press (CUP)},
  author = {Davaille,  Anne and Jaupart,  Claude},
  year = {1993},
  month = Aug,
  pages = {141–166}
}

@article{mitrovica1997radial,
author = {Mitrovica, Jerry X. and Forte, Alessandro M.},
title = {Radial profile of mantle viscosity: Results from the joint inversion of convection and postglacial rebound observables},
journal = {J. Geophys. Res. Solid Earth},
volume = {102},
number = {B2},
pages = {2751-2769},
doi = {https://doi.org/10.1029/96JB03175},
url = {https://agupubs.onlinelibrary.wiley.com/doi/abs/10.1029/96JB03175},
eprint = {https://agupubs.onlinelibrary.wiley.com/doi/pdf/10.1029/96JB03175},
year = {1997}
}

@article{bonnet2025martian,
  title={Martian highlands differentiation concomitant to dichotomy formation},
  author={Bonnet Gibet, Valentin and Michaut, Chlo{\'e} and Bodin, Thomas and Wieczorek, Mark and Dubuffet, Fabien},
  journal={J. Geophys. Res. Planets},
  volume={130},
  number={3},
  pages={e2024JE008486},
  doi = {http://doi.org/10.1029/2024JE008486},
  year={2025},
  publisher={Wiley Online Library}
}

@article{bayes1763lii,
  title={{LII}. {A}n essay towards solving a problem in the doctrine of chances. By the late Rev. Mr. Bayes, FRS communicated by Mr. Price, in a letter to John Canton, AMFR S},
  author={Bayes, Thomas},
  journal={Philos. Trans. R. Soc.},
  volume={},
  number={53},
  pages={370--418},
  year={1763},
  publisher={The Royal Society London}
}

@article{gallagher2009markov,
  title={Markov chain {M}onte {C}arlo ({MCMC}) sampling methods to determine optimal models, model resolution and model choice for {E}arth {S}cience problems},
  author={Gallagher, Kerry and Charvin, Karl and Nielsen, Soren and Sambridge, Malcolm and Stephenson, John},
  journal={Marine and Petroleum Geology},
  volume={26},
  number={4},
  pages={525--535},
  year={2009},
  publisher={Elsevier}
}

@article{karato2025rheology,
  title={Rheology of the lower mantle: a review},
  author={Karato, Shun-ichiro and Girard, Jennifer and Cho, Heechen E},
  journal={Progress in Earth and Planetary Science},
  volume={12},
  number={1},
  pages={1--20},
  year={2025},
  doi = {https://doi.org/10.1186/s40645-025-00695-6},
  publisher={Springer}
}

@article{roberts2009examplesMCMC,
  title={Examples of adaptive MCMC},
  author={Roberts, Gareth O and Rosenthal, Jeffrey S},
  journal={Journal of computational and graphical statistics},
  volume={18},
  number={2},
  pages={349--367},
  year={2009},
  publisher={Taylor \& Francis}
}

@article{greenwood2021stable,
title = {On the evolution of thermally stratified layers at the top of Earth's core},
journal = {Phys. Earth Planet. Inter.},
volume = {318},
pages = {106763},
year = {2021},
issn = {0031-9201},
doi = {https://doi.org/10.1016/j.pepi.2021.106763},
url = {https://www.sciencedirect.com/science/article/pii/S0031920121001217},
author = {Sam Greenwood and Christopher J. Davies and Jon E. Mound}
}

@article{sambridge2013transdimensional,
  title={Transdimensional inference in the geosciences},
  author={Sambridge, Malcolm and Bodin, Thomas and Gallagher, Kerry and Tkal{\v{c}}i{\'c}, H},
  journal={Philos. Trans. R. Soc.},
  volume={371},
  number={1984},
  pages={20110547},
  year={2013},
  publisher={The Royal Society Publishing}
}

@article{driscoll2023paradox,
author = {Driscoll, P. and Davies, C.},
title = {The “New Core Paradox”: Challenges and Potential Solutions},
journal = {J. Geophys. Res. Solid Earth},
volume = {128},
number = {1},
pages = {e2022JB025355},
doi = {https://doi.org/10.1029/2022JB025355},
year = {2023}
}

@article{edmund2024thermal,
author = {Edmund, Eric and Chuvashova, Irina and Konôpková, Zuzana and Husband, Rachel and Strohm, Cornelius and Appel, Karen and Bähtz, Carsten and Ball, Orianna and Bouffetier, Victorien and Brugman, Kara and Buakor, Khachiwan and Chantel, Julien and Chariton, Stella and Duff, Matthew and Dwivedi, Anand and Glazyrin, Konstantin and Hosseini-Saber, S. M. A. and Jaisle, Nicolas and Laurus, Torsten and Li, Xiang and Masani, Bernhard and McHardy, James and McMahon, Malcolm and Merkel, Sébastien and Mohrbach, Katharina and Mondal, Anshuman and Morard, Guillaume and Prakapenka, Vitali B. and Prescher, Clemens and Ryu, Young-Jay and Schwinkendorf, Jan-Patrick and Tang, Minxue and Younes, Zena and Sanchez-Valle, Carmen and Liermann, Hanns-Peter and Badro, James and Lin, Jung-Fu and McWilliams, R. Stewart and Goncharov, Alexander F.},
title = {The Thermal Conductivity of Bridgmanite at Lower Mantle Conditions Using a Multi-Technique Approach},
journal = {J. Geophys. Res. Solid Earth},
volume = {129},
number = {6},
pages = {e2024JB028823},
doi = {https://doi.org/10.1029/2024JB028823},
url = {https://agupubs.onlinelibrary.wiley.com/doi/abs/10.1029/2024JB028823},
eprint = {https://agupubs.onlinelibrary.wiley.com/doi/pdf/10.1029/2024JB028823},
note = {e2024JB028823 2024JB028823},
year = {2024}
}

@article{labrosse2015thermal,
title = {Thermal evolution of the core with a high thermal conductivity},
journal = {Phys. Earth Planet. Inter.},
volume = {247},
pages = {36-55},
year = {2015},
note = {Transport Properties of the Earth’s Core},
issn = {0031-9201},
doi = {https://doi.org/10.1016/j.pepi.2015.02.002},
url = {https://www.sciencedirect.com/science/article/pii/S0031920115000175},
author = {Stéphane Labrosse}
}

@article{olson2013new,
author = {Peter Olson },
title = {The New Core Paradox},
journal = {Science},
volume = {342},
number = {6157},
pages = {431-432},
year = {2013},
doi = {10.1126/science.1243477},
URL = {https://www.science.org/doi/abs/10.1126/science.1243477},
eprint = {https://www.science.org/doi/pdf/10.1126/science.1243477}}

@article{alasad2024coupled,
author = {Manar {Al Asad}  and Harriet C. P. Lau },
title = {Coupled fates of Earth’s mantle and core: Early sluggish-lid tectonics and a long-lived geodynamo},
journal = {Sci. Adv.},
volume = {10},
number = {31},
pages = {eadp1991},
year = {2024},
doi = {10.1126/sciadv.adp1991},
URL = {https://www.science.org/doi/abs/10.1126/sciadv.adp1991},
eprint = {https://www.science.org/doi/pdf/10.1126/sciadv.adp1991}}

@article{forst2022multidisciplinary,
author = {Frost, Daniel A. and Avery, Margaret S. and Buffett, Bruce A. and Chidester, Bethany A. and Deng, Jie and Dorfman, Susannah M. and Li, Zhi and Liu, Lijun and Lv, Mingda and Martin, Joshua F.},
title = {Multidisciplinary Constraints on the Thermal-Chemical Boundary Between Earth's Core and Mantle},
journal = {Geochem. Geophys. Geosyst.},
volume = {23},
number = {3},
pages = {e2021GC009764},
doi = {https://doi.org/10.1029/2021GC009764},
url = {https://agupubs.onlinelibrary.wiley.com/doi/abs/10.1029/2021GC009764},
eprint = {https://agupubs.onlinelibrary.wiley.com/doi/pdf/10.1029/2021GC009764},
note = {e2021GC009764 2021GC009764},
year = {2022}
}

@article{stern2023orosirian,
title = {The Orosirian (1800–2050 Ma) plate tectonic episode: Key for reconstructing the Proterozoic tectonic record},
journal = {Geosci. Front.},
volume = {14},
number = {3},
pages = {101553},
year = {2023},
issn = {1674-9871},
doi = {https://doi.org/10.1016/j.gsf.2023.101553},
url = {https://www.sciencedirect.com/science/article/pii/S1674987123000208},
author = {Robert J. Stern}
}

@article{harrison2024review,
author = {T. Mark Harrison },
title = {We don't know when plate tectonics began},
journal = {J. Geol. Soc.},
volume = {181},
number = {4},
pages = {jgs2023-212},
year = {2024},
doi = {10.1144/jgs2023-212},
URL = {https://www.lyellcollection.org/doi/abs/10.1144/jgs2023-212},
eprint = {https://www.lyellcollection.org/doi/pdf/10.1144/jgs2023-212},
}

@article{driscoll2014thermal,
title = {On the thermal and magnetic histories of Earth and Venus: Influences of melting, radioactivity, and conductivity},
journal = {Phys. Earth Planet. Inter.},
volume = {236},
pages = {36-51},
year = {2014},
issn = {0031-9201},
doi = {https://doi.org/10.1016/j.pepi.2014.08.004},
url = {https://www.sciencedirect.com/science/article/pii/S0031920114001903},
author = {P. Driscoll and D. Bercovici},
}

@article{davies2015cooling,
title = {Cooling history of Earth’s core with high thermal conductivity},
journal = {Phys. Earth Planet. Inter.},
volume = {247},
pages = {65-79},
year = {2015},
note = {Transport Properties of the Earth’s Core},
issn = {0031-9201},
doi = {https://doi.org/10.1016/j.pepi.2015.03.007},
url = {https://www.sciencedirect.com/science/article/pii/S0031920115000473},
author = {Christopher J. Davies},
}

@article{christensen2006scaling,
    author = {Christensen, U. R. and Aubert, J.},
    title = {Scaling properties of convection-driven dynamos in rotating spherical shells and application to planetary magnetic fields},
    journal = {Geophys. J. Int.},
    volume = {166},
    number = {1},
    pages = {97-114},
    year = {2006},
    month = {07},
    issn = {0956-540X},
    doi = {10.1111/j.1365-246X.2006.03009.x},
    url = {https://doi.org/10.1111/j.1365-246X.2006.03009.x},
}

@article{hirose2021light,
  title={Light elements in the Earth’s core},
  author={Hirose, Kei and Wood, Bernard and Vo{\v{c}}adlo, Lidunka},
  journal={Nat. Rev. Earth Environ.},
  volume={2},
  number={9},
  pages={645--658},
  year={2021},
  doi = {10.1038/s43017-021-00203-6},
  publisher={Nature Publishing Group UK London}
}

@article{korenaga2025tectonics,
author = {Korenaga, Jun},
title = {Tectonics and Surface Environments on Early Earth},
journal = {Astrobiology},
year = {2025},
doi = {10.1089/ast.2024.0093},
URL = {https://doi.org/10.1089/ast.2024.0093}
}

@article{greber2017titanium,
author = {Nicolas D. Greber  and Nicolas Dauphas  and Andrey Bekker  and Matouš P. Ptáček  and Ilya N. Bindeman  and Axel Hofmann },
title = {Titanium isotopic evidence for felsic crust and plate tectonics 3.5 billion years ago},
journal = {Science},
volume = {357},
number = {6357},
pages = {1271-1274},
year = {2017},
doi = {10.1126/science.aan8086},
URL = {https://www.science.org/doi/abs/10.1126/science.aan8086}}

@article{lipp2021composition,
  title={The composition and weathering of the continents over geologic time},
  author={Lipp, AG and Shorttle, O and Sperling, EA and Brocks, JJ and Cole, DB and Crockford, PW and Del Mouro, L and Dewing, K and Dornbos, SQ and Emmings, JF and others},
  journal={Geochem. Perspect. Lett.},
  volume={17},
  pages={21--26},
  year={2021},
  doi = {10.7185/geochemlet.2109},
  publisher={European Association of Geochemistry}
}

@article{stern2007and,
  title={When and how did plate tectonics begin? Theoretical and empirical considerations},
  author={Stern, RJ},
  journal={Chin. Sci. Bull.},
  volume={52},
  number={5},
  pages={578--591},
  year={2007},
  doi = {10.1007/s11434-007-0073-8},
  publisher={Springer}
}

@article{stern2018understanding,
title = {Subduction initiation in nature and models: A review},
journal = {Tectonophysics},
volume = {746},
pages = {173-198},
year = {2018},
note = {Understanding geological processes through modelling - A Memorial Volume honouring Evgenii Burov},
issn = {0040-1951},
doi = {https://doi.org/10.1016/j.tecto.2017.10.014},
author = {Robert J. Stern and Taras Gerya}
}

@article{condie2008did,
 author  = {Condie, Kent C and Kröner, Al...},
  title={When did plate tectonics begin? Evidence from the geologic record},
  author={Condie, Kent C and Kr{\"o}ner, Alfred},
  doi     = {10.1130/2008.2440(14)},
  journal = {Geol. Soc. Am. Bull.},
  year={2008}
}

@article{condie2018zircon,
    author = {Condie, Kent C. and Arndt, Nicholas and Davaille, Anne and Puetz, Stephen J.},
    title = {Zircon age peaks: Production or preservation of continental crust?},
    journal = {Geosphere},
    volume = {13},
    number = {2},
    pages = {227-234},
    year = {2017},
    month = {04},
    issn = {1553-040X},
    doi = {10.1130/GES01361.1},
    url = {https://doi.org/10.1130/GES01361.1}
}

@article{brenner2020paleomagnetic,
author = {Alec R. Brenner  and Roger R. Fu  and David A.D. Evans  and Aleksey V. Smirnov  and Raisa Trubko  and Ian R. Rose },
title = {Paleomagnetic evidence for modern-like plate motion velocities at 3.2 Ga},
journal = {Sci. Adv.},
volume = {6},
number = {17},
pages = {eaaz8670},
year = {2020},
doi = {10.1126/sciadv.aaz8670}}

@article{McDonough1995,
  title = {The composition of the Earth},
  volume = {120},
  ISSN = {0009-2541},
  url = {http://dx.doi.org/10.1016/0009-2541(94)00140-4},
  DOI = {10.1016/0009-2541(94)00140-4},
  number = {3-4},
  journal = {Chem. Geol.},
  publisher = {Elsevier BV},
  author = {McDonough,  W.F. and Sun,  S.-s.},
  year = {1995},
  month = Mar,
  pages = {223–253}
}

@article{farquhar2002mass,
author = {J. Farquhar  and B. A. Wing  and K. D. McKeegan  and J. W. Harris  and P. Cartigny  and M. H. Thiemens },
title = {Mass-Independent Sulfur of Inclusions in Diamond and Sulfur Recycling on Early Earth},
journal = {Science},
volume = {298},
number = {5602},
pages = {2369-2372},
year = {2002},
doi = {10.1126/science.1078617}}

@article{cawood2018geological,
    author = {Cawood, Peter A. and Hawkesworth, Chris J. and Pisarevsky, Sergei A. and Dhuime, Bruno and Capitanio, Fabio A. and Nebel, Oliver},
    title = {Geological archive of the onset of plate tectonics},
    journal = {Philos. Trans. R. Soc.},
    volume = {376},
    number = {2132},
    pages = {20170405},
    year = {2018},
    month = {10},
    issn = {1364-503X},
    doi = {10.1098/rsta.2017.0405}
}

@article{debaille2013stagnant,
  title={Stagnant-lid tectonics in early Earth revealed by 142Nd variations in late Archean rocks},
  author={Debaille, Vinciane and O'Neill, Craig and Brandon, Alan D and Haenecour, Pierre and Yin, Qing-Zhu and Mattielli, Nadine and Treiman, Allan H},
  journal={Earth Planet. Sci. Lett.},
  volume={373},
  pages={83--92},
  year={2013},
  publisher={Elsevier},
  doi = {10.1016/j.epsl.2013.04.016}
}

@article{lenardic2018diversity,
  title={The diversity of tectonic modes and thoughts about transitions between them},
  author={Lenardic, A},
  journal={Philos. Trans. R. Soc.},
  volume={376},
  number={2132},
  pages={20170416},
  year={2018},
  doi={10.1098/rsta.2017.0416},
  publisher={The Royal Society Publishing}
}

@article{lourencco2020plutonic,
  title={Plutonic-squishy lid: A new global tectonic regime generated by intrusive magmatism on Earth-like planets},
  author={Louren{\c{c}}o, Diogo L and Rozel, Antoine B and Ballmer, Maxim D and Tackley, Paul J},
  journal={Geochem. Geophys. Geosyst.},
  volume={21},
  number={4},
  pages={e2019GC008756},
  year={2020},
  doi={10.1029/2019GC008756},
  publisher={Wiley Online Library}
}

@article{arndt2013episodic,
  title={Episodic earth evolution},
  author={Arndt, Nicholas and Davaille, Anne},
  journal={Tectonophysics},
  volume={609},
  pages={661--674},
  year={2013},
  doi = {10.1016/j.tecto.2013.07.002},
  publisher={Elsevier}
}

@article{herzberg2009petrological,
  title={Petrological evidence for secular cooling in mantle plumes},
  author={Herzberg, Claude and Gazel, Esteban},
  journal={Nature},
  volume={458},
  number={7238},
  pages={619--622},
  year={2009}, 
  doi={doi.org/10.1038/nature07857},
  publisher={Nature Publishing Group UK London}
}

@article{ganne2017mantle,
author = {Ganne, Jérôme and Feng, Xiaojun},
title = {Primary magmas and mantle temperatures through time},
journal = {Geochem. Geophys. Geosyst.},
volume = {18},
number = {3},
pages = {872-888},
doi = {https://doi.org/10.1002/2016GC006787},
year = {2017}
}

@article{herzberg2010thermal,
title = {Thermal history of the Earth and its petrological expression},
journal = {Earth Planet. Sci. Lett.},
volume = {292},
number = {1},
pages = {79-88},
year = {2010},
issn = {0012-821X},
doi = {https://doi.org/10.1016/j.epsl.2010.01.022},
author = {Claude Herzberg and Kent Condie and Jun Korenaga}
}

@incollection{foley2020heat,
author = {Foley, Bradford J and Houser, Christine and Noack, Lena and Tosi, Nicola},
title = {The Heat Budget of Rocky Planets},
booktitle = {Planetary Diversity},
publisher = {IOP Publishing},
year = {2020},
series = {2514-3433},
type = {Book Chapter},
pages = {4-1 to 4-60},
doi = {10.1088/2514-3433/abb4d9ch4},
isbn = {978-0-7503-2140-2}
}

@article{pozzo2022toward,
title = {Towards reconciling experimental and computational determinations of Earth's core thermal conductivity},
journal = {Earth Planet. Sci. Lett.},
volume = {584},
pages = {117466},
year = {2022},
issn = {0012-821X},
doi = {https://doi.org/10.1016/j.epsl.2022.117466},
url = {https://www.sciencedirect.com/science/article/pii/S0012821X22001029},
author = {Monica Pozzo and Christopher J. Davies and Dario Alfè}
}

@article{konopkova2016direct,
  title={Direct measurement of thermal conductivity in solid iron at planetary core conditions},
  author={Kon{\^o}pkov{\'a}, Zuzana and McWilliams, R Stewart and G{\'o}mez-P{\'e}rez, Natalia and Goncharov, Alexander F},
  journal={Nature},
  volume={534},
  number={7605},
  pages={99--101},
  year={2016},
  doi = {10.1038/nature18009},
  publisher={Nature Publishing Group UK London}
}

@article{andrault2025experimental,
title = {Long-lived magnetic field in earth-like terrestrial planets},
journal = {Phys. Earth Planet. Inter.},
volume = {360},
pages = {107315},
year = {2025},
issn = {0031-9201},
doi = {https://doi.org/10.1016/j.pepi.2025.107315},
author = {D. Andrault and L. {Pison Pacynski} and J. Monteux and E. Gardés and A. Mathieu}
}

@article{bono2021pint,
  title = {The PINT database: a definitive compilation of absolute palaeomagnetic intensity determinations since 4 billion years ago},
  volume = {229},
  ISSN = {1365-246X},
  url = {http://dx.doi.org/10.1093/gji/ggab490},
  DOI = {10.1093/gji/ggab490},
  number = {1},
  journal = {Geophys. J. Int.},
  publisher = {Oxford University Press (OUP)},
  author = {Bono,  Richard K and Paterson,  Greig A and van der Boon,  Annique and Engbers,  Yael A and Michael Grappone,  J and Handford,  Benjamin and Hawkins,  Louise M A and Lloyd,  Simon J and Sprain,  Courtney J and Thallner,  Daniele and Biggin,  Andrew J},
  year = {2021},
  month = dec,
  pages = {522–545}
}

@inbook{jaupart2015temperature,
  title = {Temperatures,  Heat,  and Energy in the Mantle of the Earth},
  ISBN = {9780444538031},
  url = {http://dx.doi.org/10.1016/B978-0-444-53802-4.00126-3},
  DOI = {10.1016/b978-0-444-53802-4.00126-3},
  booktitle = {Treatise on Geophysics},
  publisher = {Elsevier},
  author = {Jaupart,  C. and Labrosse,  S. and Lucazeau,  F. and Mareschal,  J.-C.},
  year = {2015},
  pages = {223–270}
}

@inbook{korenaga2006archean,
  title = {Archean geodynamics and the thermal evolution of Earth},
  ISBN = {9780875904290},
  ISSN = {0065-8448},
  url = {http://dx.doi.org/10.1029/164GM03},
  DOI = {10.1029/164gm03},
  booktitle = {Archean Geodynamics and Environments},
  publisher = {American Geophysical Union},
  author = {Korenaga,  Jun},
  year = {2006},
  pages = {7–32}
}

@article{koker2012electrical,
author = {{De Koker}, Nico  and Steinle-Neumann, Gerd  and  Vlček, Vojtěch },
title = {Electrical resistivity and thermal conductivity of liquid Fe alloys at high <i>P</i> and <i>T</i>, and heat flux in Earth\&\#x2019;s core},
journal = {Proc. Natl. Acad. Sci. U.S.A.},
volume = {109},
number = {11},
pages = {4070-4073},
year = {2012},
doi = {10.1073/pnas.1111841109}}

@article{bono2019young,
  title={Young inner core inferred from Ediacaran ultra-low geomagnetic field intensity},
  author={Bono, Richard K and Tarduno, John A and Nimmo, Francis and Cottrell, Rory D},
  journal={Nat. Geosci.},
  volume={12},
  number={2},
  pages={143--147},
  year={2019},
  doi={10.1038/s41561-018-0288-0},
  publisher={Nature Publishing Group UK London}
}

@article{lloyd2021first,
    author = {Lloyd, Simon J and Biggin, Andrew J and Halls, Henry and Hill, Mimi J},
    title = {First palaeointensity data from the cryogenian and their potential implications for inner core nucleation age},
    journal = {Geophys. J. Int.},
    volume = {226},
    number = {1},
    pages = {66-77},
    year = {2021},
    month = {03},
    issn = {0956-540X},
    doi = {10.1093/gji/ggab090}
}

@article{badro2016early,
  title={An early geodynamo driven by exsolution of mantle components from Earth’s core},
  author={Badro, James and Siebert, Julien and Nimmo, Francis},
  journal={Nature},
  volume={536},
  number={7616},
  pages={326--328},
  year={2016},
  doi={https://doi.org/10.1038/nature18594},
  publisher={Nature Publishing Group UK London}
}

@article{orourke2016powering,
  title={Powering Earth’s dynamo with magnesium precipitation from the core},
  author={O’Rourke, Joseph G and Stevenson, David J},
  journal={Nature},
  volume={529},
  number={7586},
  pages={387--389},
  year={2016},
  doi={https://doi.org/10.1038/nature16495},
  publisher={Nature Publishing Group UK London}
}

@article{du2019experimental,
author = {Du, Zhixue and Boujibar, Asmaa and Driscoll, Peter and Fei, Yingwei},
title = {Experimental Constraints on an MgO Exsolution-Driven Geodynamo},
journal = {Geophys. Res. Lett.},
volume = {46},
number = {13},
pages = {7379-7385},
doi = {https://doi.org/10.1029/2019GL083017},
year = {2019}
}

@article{andrault2016deep,
title = {The deep Earth may not be cooling down},
journal = {Earth Planet. Sci. Lett.},
volume = {443},
pages = {195-203},
year = {2016},
issn = {0012-821X},
doi = {https://doi.org/10.1016/j.epsl.2016.03.020},
author = {Denis Andrault and Julien Monteux and Michael {Le Bars} and Henri Samuel},
}

@article{landeau2022sustaining,
  title = {Sustaining Earth’s magnetic dynamo},
  volume = {3},
  ISSN = {2662-138X},
  url = {http://dx.doi.org/10.1038/s43017-022-00264-1},
  DOI = {10.1038/s43017-022-00264-1},
  number = {4},
  journal = {Nat. Rev. Earth Environ.},
  publisher = {Springer Science and Business Media LLC},
  author = {Landeau,  Maylis and Fournier,  Alexandre and Nataf,  Henri-Claude and Cébron,  David and Schaeffer,  Nathanaël},
  year = {2022},
  month = mar,
  pages = {255–269}
}

@article{fu2024,
  title = {Statistical reanalysis of Archean zircon paleointensities: No evidence for stagnant-lid tectonics},
  volume = {634},
  ISSN = {0012-821X},
  url = {http://dx.doi.org/10.1016/j.epsl.2024.118679},
  DOI = {10.1016/j.epsl.2024.118679},
  journal = {Earth Planet. Sci. Lett.},
  publisher = {Elsevier BV},
  author = {Fu,  Roger R. and Drabon,  Nadja and Weiss,  Benjamin P. and Borlina,  Cau\^e and Kirkpatrick,  Heather},
  year = {2024},
  month = may,
  pages = {118679}
}

@article{tarduno2023hadaean,
  title = {Hadaean to Palaeoarchaean stagnant-lid tectonics revealed by zircon magnetism},
  volume = {618},
  ISSN = {1476-4687},
  url = {http://dx.doi.org/10.1038/s41586-023-06024-5},
  DOI = {10.1038/s41586-023-06024-5},
  number = {7965},
  journal = {Nature},
  publisher = {Springer Science and Business Media LLC},
  author = {Tarduno,  John A. and Cottrell,  Rory D. and Bono,  Richard K. and Rayner,  Nicole and Davis,  William J. and Zhou,  Tinghong and Nimmo,  Francis and Hofmann,  Axel and Jodder,  Jaganmoy and Ibañez-Mejia,  Mauricio and Watkeys,  Michael K. and Oda,  Hirokuni and Mitra,  Gautam},
  year = {2023},
  month = jun,
  pages = {531–536}
}

@article{lin2016precession,
  title = {Precession-driven dynamos in a full sphere and the role of large scale cyclonic vortices},
  volume = {28},
  ISSN = {1089-7666},
  url = {http://dx.doi.org/10.1063/1.4954295},
  DOI = {10.1063/1.4954295},
  number = {6},
  journal = {Phys. Fluids},
  publisher = {AIP Publishing},
  author = {Lin,  Yufeng and Marti,  Philippe and Noir,  Jerome and Jackson,  Andrew},
  year = {2016},
  month = jun 
}

@article{bouffard2019chemical,
  title = {Chemical Convection and Stratification in the Earth’s Outer Core},
  volume = {7},
  ISSN = {2296-6463},
  url = {http://dx.doi.org/10.3389/feart.2019.00099},
  DOI = {10.3389/feart.2019.00099},
  journal = {Front. Earth Sci.},
  publisher = {Frontiers Media SA},
  author = {Bouffard,  Mathieu and Choblet,  Gaël and Labrosse,  Stéphane and Wicht,  Johannes},
  year = {2019},
  month = may 
}

@article{gubbins2013stratified,
  title = {The stratified layer at the core–mantle boundary caused by barodiffusion of oxygen,  sulphur and silicon},
  volume = {215},
  ISSN = {0031-9201},
  url = {http://dx.doi.org/10.1016/j.pepi.2012.11.001},
  DOI = {10.1016/j.pepi.2012.11.001},
  journal = {Phys. Earth Planet. Inter.},
  publisher = {Elsevier BV},
  author = {Gubbins,  D. and Davies,  C.J.},
  year = {2013},
  month = feb,
  pages = {21–28}
}

@article{irving2018seismically,
  title = {Seismically determined elastic parameters for Earth’s outer core},
  volume = {4},
  ISSN = {2375-2548},
  url = {http://dx.doi.org/10.1126/sciadv.aar2538},
  DOI = {10.1126/sciadv.aar2538},
  number = {6},
  journal = {Sci. Adv.},
  publisher = {American Association for the Advancement of Science (AAAS)},
  author = {Irving,  Jessica C. E. and Cottaar,  Sanne and Lekić,  Vedran},
  year = {2018},
  month = jun 
}

@article{davies2020transfer,
  title = {Transfer of oxygen to Earth’s core from a long-lived magma ocean},
  volume = {538},
  ISSN = {0012-821X},
  url = {http://dx.doi.org/10.1016/j.epsl.2020.116208},
  DOI = {10.1016/j.epsl.2020.116208},
  journal = {Earth Planet. Sci. Lett.},
  publisher = {Elsevier BV},
  author = {Davies,  Christopher J. and Pozzo,  Monica and Gubbins,  David and Alfè,  Dario},
  year = {2020},
  month = may,
  pages = {116208}
}

@article{kaneshima2018array,
  title = {Array analyses of SmKS waves and the stratification of Earth’s outermost core},
  volume = {276},
  ISSN = {0031-9201},
  url = {http://dx.doi.org/10.1016/j.pepi.2017.03.006},
  DOI = {10.1016/j.pepi.2017.03.006},
  journal = {Phys. Earth Planet. Inter.},
  publisher = {Elsevier BV},
  author = {Kaneshima,  Satoshi},
  year = {2018},
  month = mar,
  pages = {234–246}
}

@article{tanaka2007possibility,
  title = {Possibility of a low P-wave velocity layer in the outermost core from global SmKS waveforms},
  volume = {259},
  ISSN = {0012-821X},
  url = {http://dx.doi.org/10.1016/j.epsl.2007.05.007},
  DOI = {10.1016/j.epsl.2007.05.007},
  number = {3–4},
  journal = {Earth Planet. Sci. Lett.},
  publisher = {Elsevier BV},
  author = {Tanaka,  Satoru},
  year = {2007},
  month = jul,
  pages = {486–499}
}

@article{stern2018evolution,
  title = {The evolution of plate tectonics},
  volume = {376},
  ISSN = {1471-2962},
  url = {http://dx.doi.org/10.1098/rsta.2017.0406},
  DOI = {10.1098/rsta.2017.0406},
  number = {2132},
  journal = {Philos. Trans. R. Soc. A-Math. Phys. Eng. Sci.},
  publisher = {The Royal Society},
  author = {Stern,  Robert J.},
  year = {2018},
  month = oct,
  pages = {20170406}
}

@article{labrosse2007thermal,
  title = {Thermal evolution of the Earth: Secular changes and fluctuations of plate characteristics},
  volume = {260},
  ISSN = {0012-821X},
  url = {http://dx.doi.org/10.1016/j.epsl.2007.05.046},
  DOI = {10.1016/j.epsl.2007.05.046},
  number = {3–4},
  journal = {Earth Planet. Sci. Lett.},
  publisher = {Elsevier BV},
  author = {Labrosse,  S. and Jaupart,  C.},
  year = {2007},
  month = aug,
  pages = {465–481}
}

@article{bedard2024gradual,
  title = {A gradual Proterozoic transition from an unstable stagnant lid to the modern plate tectonic system},
  volume = {181},
  ISSN = {2041-479X},
  url = {http://dx.doi.org/10.1144/jgs2024-023},
  DOI = {10.1144/jgs2024-023},
  number = {4},
  journal = {J. Geol. Soc.},
  publisher = {Geological Society of London},
  author = {Bédard,  Jean H.},
  year = {2024},
  month = jul 
}

@article{lyu2025dissecting,
  title = {Dissecting the puzzle of tectonic lid regimes in terrestrial planets},
  volume = {16},
  ISSN = {2041-1723},
  url = {http://dx.doi.org/10.1038/s41467-025-65943-1},
  DOI = {10.1038/s41467-025-65943-1},
  number = {1},
  journal = {Nat. Commun.},
  publisher = {Springer Science and Business Media LLC},
  author = {Lyu,  Tianyang and Ballmer,  Maxim D. and Li,  Zhong-Hai and Lee,  Man Hoi and Yan,  Jun and Wu,  Benjun and Zhao,  Guochun},
  year = {2025},
  month = nov 
}

@article{gunawardana2024correlating,
  title = {Correlating mantle cooling with tectonic transitions on early Earth},
  volume = {52},
  ISSN = {1943-2682},
  url = {http://dx.doi.org/10.1130/G51874.1},
  DOI = {10.1130/g51874.1},
  number = {4},
  journal = {Geology},
  publisher = {Geol. Soc. Am. Bull.},
  author = {Gunawardana,  Prasanna M. and Chowdhury,  Priyadarshi and Morra,  Gabriele and Cawood,  Peter A.},
  year = {2024},
  month = jan,
  pages = {230–234}
}

@article{oneill2018inception,
  title = {The inception of plate tectonics: a record of failure},
  volume = {376},
  ISSN = {1471-2962},
  url = {http://dx.doi.org/10.1098/rsta.2017.0414},
  DOI = {10.1098/rsta.2017.0414},
  number = {2132},
  journal = {Philos. Trans. R. Soc. A-Math. Phys. Eng. Sci.},
  publisher = {The Royal Society},
  author = {O’Neill,  Craig and Turner,  Simon and Rushmer,  Tracy},
  year = {2018},
  month = oct,
  pages = {20170414}
}

@article{sudholz2025xenolith,
  title = {Xenolith Constraints on the Mantle Potential Temperature and Thickness of Cratonic Roots Through Time},
  volume = {52},
  ISSN = {1944-8007},
  url = {http://dx.doi.org/10.1029/2024GL112851},
  DOI = {10.1029/2024gl112851},
  number = {2},
  journal = {Geophys. Res. Lett.},
  publisher = {American Geophysical Union (AGU)},
  author = {Sudholz,  Z. J. and Copley,  A.},
  year = {2025},
  month = jan 
}

@article{schaeffer2025energetically,
  title = {Energetically expensive dynamo action in Earth’s basal magma ocean},
  volume = {122},
  ISSN = {1091-6490},
  url = {http://dx.doi.org/10.1073/pnas.2507575122},
  DOI = {10.1073/pnas.2507575122},
  number = {45},
  journal = {Proc. Natl. Acad. Sci. U.S.A.},
  publisher = {Proc. Natl. Acad. Sci. U.S.A.},
  author = {Schaeffer,  Nathanaël and Labrosse,  Stéphane and Aurnou,  Jonathan M.},
  year = {2025},
  month = nov 
}

@article{dragulet2025electrical,
  title = {Electrical and thermal conductivity of Earth’s iron-enriched basal magma ocean},
  volume = {122},
  ISSN = {1091-6490},
  url = {http://dx.doi.org/10.1073/pnas.2509771122},
  DOI = {10.1073/pnas.2509771122},
  number = {42},
  journal = {Proc. Natl. Acad. Sci. U.S.A.},
  publisher = {Proc. Natl. Acad. Sci. U.S.A.},
  author = {Dragulet,  Francis and Stixrude,  Lars},
  year = {2025},
  month = oct 
}

@article{ziegler2013implications,
  title = {Implications of a long‐lived basal magma ocean in generating Earth’s ancient magnetic field},
  volume = {14},
  ISSN = {1525-2027},
  url = {http://dx.doi.org/10.1002/2013GC005001},
  DOI = {10.1002/2013gc005001},
  number = {11},
  journal = {Geochem. Geophys. Geosyst.},
  publisher = {American Geophysical Union (AGU)},
  author = {Ziegler,  L. B. and Stegman,  D. R.},
  year = {2013},
  month = nov,
  pages = {4735–4742}
}

@article{lherm2024thermal,
  title = {Thermal and magnetic evolution of an Earth-like planet with a basal magma ocean},
  volume = {356},
  ISSN = {0031-9201},
  url = {http://dx.doi.org/10.1016/j.pepi.2024.107267},
  DOI = {10.1016/j.pepi.2024.107267},
  journal = {Phys. Earth Planet. Inter.},
  publisher = {Elsevier BV},
  author = {Lherm,  Victor and Nakajima,  Miki and Blackman,  Eric G.},
  year = {2024},
  month = nov,
  pages = {107267}
}

@article{stixrude2020silicate,
  title = {A silicate dynamo in the early Earth},
  volume = {11},
  ISSN = {2041-1723},
  url = {http://dx.doi.org/10.1038/s41467-020-14773-4},
  DOI = {10.1038/s41467-020-14773-4},
  number = {1},
  journal = {Nat. Commun.},
  publisher = {Springer Science and Business Media LLC},
  author = {Stixrude,  Lars and Scipioni,  Roberto and Desjarlais,  Michael P.},
  year = {2020},
  month = feb 
}

@article{labrosse2007crystallizing,
  title = {A crystallizing dense magma ocean at the base of the Earth’s mantle},
  volume = {450},
  ISSN = {1476-4687},
  url = {http://dx.doi.org/10.1038/nature06355},
  DOI = {10.1038/nature06355},
  number = {7171},
  journal = {Nature},
  publisher = {Springer Science and Business Media LLC},
  author = {Labrosse,  S. and Hernlund,  J. W. and Coltice,  N.},
  year = {2007},
  month = dec,
  pages = {866–869}
}

@article{boukare2025solidification,
  title = {Solidification of Earth’s mantle led inevitably to a basal magma ocean},
  volume = {640},
  ISSN = {1476-4687},
  url = {http://dx.doi.org/10.1038/s41586-025-08701-z},
  DOI = {10.1038/s41586-025-08701-z},
  number = {8057},
  journal = {Nature},
  publisher = {Springer Science and Business Media LLC},
  author = {Boukaré,  Charles-Édouard and Badro,  James and Samuel,  Henri},
  year = {2025},
  month = mar,
  pages = {114–119}
}

@article{karato2010rheology,
  title = {Rheology of the Earth’s mantle: A historical review},
  volume = {18},
  ISSN = {1342-937X},
  url = {http://dx.doi.org/10.1016/j.gr.2010.03.004},
  DOI = {10.1016/j.gr.2010.03.004},
  number = {1},
  journal = {Gondwana Res.},
  publisher = {Elsevier BV},
  author = {Karato,  Shun-ichiro},
  year = {2010},
  month = jul,
  pages = {17–45}
}

@article{parmentier1992chemical,
  title = {Chemical dieferentiation of a convecting planetary interior: Consequences for a one plate planet such as Venus},
  volume = {19},
  ISSN = {1944-8007},
  url = {http://dx.doi.org/10.1029/92GL01862},
  DOI = {10.1029/92gl01862},
  number = {20},
  journal = {Geophys. Res. Lett.},
  publisher = {American Geophysical Union (AGU)},
  author = {Parmentier,  E. M. and Hess,  P. C.},
  year = {1992},
  month = oct,
  pages = {2015–2018}
}

@article{hirth1996water,
  title = {Water in the oceanic upper mantle: implications for rheology,  melt extraction and the evolution of the lithosphere},
  volume = {144},
  ISSN = {0012-821X},
  url = {http://dx.doi.org/10.1016/0012-821X(96)00154-9},
  DOI = {10.1016/0012-821x(96)00154-9},
  number = {1–2},
  journal = {Earth Planet. Sci. Lett.},
  publisher = {Elsevier BV},
  author = {Hirth,  Greg and Kohlstedt,  David L.},
  year = {1996},
  month = oct,
  pages = {93–108}
}

@article{korenaga2009scaling,
  title = {Scaling of stagnant-lid convection with Arrhenius rheology and the effects of mantle melting},
  volume = {179},
  ISSN = {1365-246X},
  url = {http://dx.doi.org/10.1111/j.1365-246X.2009.04272.x},
  DOI = {10.1111/j.1365-246x.2009.04272.x},
  number = {1},
  journal = {Geophys. J. Int.},
  publisher = {Oxford University Press (OUP)},
  author = {Korenaga,  Jun},
  year = {2009},
  month = oct,
  pages = {154–170}
}

@article{korenaga2011thermal,
  title = {Thermal evolution with a hydrating mantle and the initiation of plate tectonics in the early Earth},
  volume = {116},
  ISSN = {0148-0227},
  url = {http://dx.doi.org/10.1029/2011JB008410},
  DOI = {10.1029/2011jb008410},
  number = {B12},
  journal = {J. Geophys. Res.},
  publisher = {American Geophysical Union (AGU)},
  author = {Korenaga,  J.},
  year = {2011},
  month = dec 
}

@article{bercovici2014plate,
  title = {Plate tectonics,  damage and inheritance},
  volume = {508},
  ISSN = {1476-4687},
  url = {http://dx.doi.org/10.1038/nature13072},
  DOI = {10.1038/nature13072},
  number = {7497},
  journal = {Nature},
  publisher = {Springer Science and Business Media LLC},
  author = {Bercovici,  David and Ricard,  Yanick},
  year = {2014},
  month = apr,
  pages = {513–516}
}

@article{foley2014scaling,
  title = {Scaling laws for convection with temperature-dependent viscosity and grain-damage},
  volume = {199},
  ISSN = {0956-540X},
  url = {http://dx.doi.org/10.1093/gji/ggu275},
  DOI = {10.1093/gji/ggu275},
  number = {1},
  journal = {Geophys. J. Int.},
  publisher = {Oxford University Press (OUP)},
  author = {Foley,  Bradford J. and Bercovici,  David},
  year = {2014},
  month = aug,
  pages = {580–603}
}

@article{alfe2002composition,
  title = {Composition and temperature of the Earth’s core constrained by combining ab initio calculations and seismic data},
  volume = {195},
  ISSN = {0012-821X},
  url = {http://dx.doi.org/10.1016/S0012-821X(01)00568-4},
  DOI = {10.1016/s0012-821x(01)00568-4},
  number = {1–2},
  journal = {Earth Planet. Sci. Lett.},
  publisher = {Elsevier BV},
  author = {Alfè,  D and Gillan,  M.J and Price,  G.D},
  year = {2002},
  month = jan,
  pages = {91–98}
}

@article{stacey2007,
  title={A revised estimate of the conductivity of iron alloy at high pressure and implications for the core energy balance},
  author={Stacey, FD and Loper, DE},
  journal={Phys. Earth Planet. Inter.},
  volume={161},
  number={1-2},
  pages={13--18},
  year={2007},
  doi={10.1016/j.pepi.2006.12.001}
}

@article{labrosse1997cooling,
  title={On cooling of the Earth's core},
  author={Labrosse, St{\'e}phane and Poirier, Jean-Paul and Le Mou{\"e}l, Jean-Louis},
  journal={Phys. Earth Planet. Inter.},
  volume={99},
  number={1-2},
  pages={1--17},
  year={1997},
  doi={10.1016/S0031-9201(96)03207-4}
}

@article{moresi1995,
  title={Numerical investigation of 2D convection with extremely large viscosity variations},
  author={Moresi, L-N and Solomatov, VS},
  journal={Phys. Fluids},
  volume={7},
  number={9},
  pages={2154--2162},
  year={1995},
  doi={10.1063/1.868465}
}

@article{patocka2020,
  title={Minimum heat flow from the core and thermal evolution of the Earth},
  author={Pato{\v{c}}ka, V and {\v{S}}r{\'a}mek, O and Tosi, Nicola},
  journal={Phys. Earth Planet. Inter.},
  volume={305},
  pages={106457},
  year={2020},
  doi={10.1016/j.pepi.2020.106457}
}

@incollection{labrosse2016,
    author = {Labrosse, St{\'e}phane},
    title = {Thermal state and evolution of the Earth core and deep mantle},
    booktitle = {Deep Earth: Physics and Chemistry of the Lower Mantle and Core},
    publisher = {American Geophysical Union},
    editors = {H. Terasaki and R. Fischer},
    pages = {43-56},
    year = {2016}
}

@article{spohn1991,
  title={Mantle differentiation and thermal evolution of Mars, Mercury, and Venus},
  author={Spohn, Tilman},
  journal={Icarus},
  volume={90},
  number={2},
  pages={222--236},
  year={1991},
  doi={/10.1016/0019-1035(91)90103-Z}
}

@article{morschhauser2011,
  title={Crustal recycling, mantle dehydration, and the thermal evolution of Mars},
  author={Morschhauser, Achim and Grott, M and Breuer, D},
  journal={Icarus},
  volume={212},
  number={2},
  pages={541--558},
  year={2011},
  doi={10.1016/j.icarus.2010.12.028}
}

@article{tosi2017,
  title={The habitability of a stagnant-lid Earth},
  author={Tosi, Nicola and Godolt, Mareike and Stracke, Barbara and Ruedas, Thomas and Grenfell, John Lee and H{\"o}ning, Dennis and Nikolaou, Athanasia and Plesa, A-C and Breuer, Doris and Spohn, Tilman},
  journal={Astron. Astrophys},
  volume={605},
  pages={A71},
  year={2017},
  doi={10.1051/0004-6361/201730728}
}

@ARTICLE{scipy2025,
  author  = {Virtanen, Pauli and Gommers, Ralf and Oliphant, Travis E. and
            Haberland, Matt and Reddy, Tyler and Cournapeau, David and
            Burovski, Evgeni and Peterson, Pearu and Weckesser, Warren and
            Bright, Jonathan and {van der Walt}, St{\'e}fan J. and
            Brett, Matthew and Wilson, Joshua and Millman, K. Jarrod and
            Mayorov, Nikolay and Nelson, Andrew R. J. and Jones, Eric and
            Kern, Robert and Larson, Eric and Carey, C J and
            Polat, {\.I}lhan and Feng, Yu and Moore, Eric W. and
            {VanderPlas}, Jake and Laxalde, Denis and Perktold, Josef and
            Cimrman, Robert and Henriksen, Ian and Quintero, E. A. and
            Harris, Charles R. and Archibald, Anne M. and
            Ribeiro, Ant{\^o}nio H. and Pedregosa, Fabian and
            {van Mulbregt}, Paul and {SciPy 1.0 Contributors}},
  title   = {{{SciPy} 1.0: Fundamental Algorithms for Scientific
            Computing in Python}},
  journal = {Nat. Methods},
  year    = {2020},
  volume  = {17},
  pages   = {261--272},
  adsurl  = {https://rdcu.be/b08Wh},
  doi     = {10.1038/s41592-019-0686-2},
}

@article{tosi2013mantle,
  title = {Mantle dynamics with pressure- and temperature-dependent thermal expansivity and conductivity},
  volume = {217},
  ISSN = {0031-9201},
  url = {http://dx.doi.org/10.1016/j.pepi.2013.02.004},
  DOI = {10.1016/j.pepi.2013.02.004},
  journal = {Phys. Earth Planet. Inter.},
  publisher = {Elsevier BV},
  author = {Tosi,  Nicola and Yuen,  David A. and de Koker,  Nico and Wentzcovitch,  Renata M.},
  year = {2013},
  month = apr,
  pages = {48–58}
}

@article{deschamps2001thermal,
  title={Thermal convection in the outer shell of large icy satellites},
  author={Deschamps, Fr{\'e}d{\'e}ric and Sotin, Christophe},
  journal={J. Geophys. Res. Planets},
  volume={106},
  number={E3},
  pages={5107--5121},
  year={2001},
  doi = {https://doi.org/10.1029/2000JE001253},
  publisher={Wiley Online Library}
}

@article{Thiriet2019,
  title = {Scaling laws of convection for cooling planets in a stagnant lid regime},
  volume = {286},
  ISSN = {0031-9201},
  url = {http://dx.doi.org/10.1016/j.pepi.2018.11.003},
  DOI = {10.1016/j.pepi.2018.11.003},
  journal = {Phys. Earth Planet. Inter.},
  publisher = {Elsevier BV},
  author = {Thiriet,  Mélanie and Breuer,  Doris and Michaut,  Chloé and Plesa,  Ana-Catalina},
  year = {2019},
  month = jan,
  pages = {138–153}
}

@article{schubert1979subsolidus,
  title = {Subsolidus convective cooling histories of terrestrial planets},
  volume = {38},
  ISSN = {0019-1035},
  url = {http://dx.doi.org/10.1016/0019-1035(79)90178-7},
  DOI = {10.1016/0019-1035(79)90178-7},
  number = {2},
  journal = {Icarus},
  publisher = {Elsevier BV},
  author = {Schubert,  G. and Cassen,  P. and Young,  R.E.},
  year = {1979},
  month = may,
  pages = {192–211}
}

@article{xiong2018ab,
  title = {Ab Initio Prediction of Potassium Partitioning Into Earth’s Core},
  volume = {123},
  ISSN = {2169-9356},
  url = {http://dx.doi.org/10.1029/2018JB015522},
  DOI = {10.1029/2018jb015522},
  number = {8},
  journal = {J. Geophys. Res. Solid Earth},
  publisher = {American Geophysical Union (AGU)},
  author = {Xiong,  Zhihua and Tsuchiya,  Taku and Taniuchi,  Takashi},
  year = {2018},
  month = aug,
  pages = {6451–6458}
}

@article{knibbe2021modelling,
  title = {Modelling of thermal stratification at the top of a planetary core: Application to the cores of Earth and Mercury and the thermal coupling with their mantles},
  volume = {321},
  ISSN = {0031-9201},
  url = {http://dx.doi.org/10.1016/j.pepi.2021.106804},
  DOI = {10.1016/j.pepi.2021.106804},
  journal = {Phys. Earth Planet. Inter.},
  publisher = {Elsevier BV},
  author = {Knibbe,  J.S. and Van Hoolst,  T.},
  year = {2021},
  month = dec,
  pages = {106804}
}

@article{gubbins2004gross,
  title = {Gross thermodynamics of two-component core convection},
  volume = {157},
  ISSN = {1365-246X},
  url = {http://dx.doi.org/10.1111/j.1365-246X.2004.02219.x},
  DOI = {10.1111/j.1365-246x.2004.02219.x},
  number = {3},
  journal = {Geophys. J. Int.},
  publisher = {Oxford University Press (OUP)},
  author = {Gubbins,  David and Alfè,  Dario and Masters,  Guy and Price,  G. David and Gillan,  Michael},
  year = {2004},
  month = jun,
  pages = {1407–1414}
}

@article{ruedas2017radioactive,
  title = {Radioactive heat production of six geologically important nuclides},
  volume = {18},
  ISSN = {1525-2027},
  url = {http://dx.doi.org/10.1002/2017GC006997},
  DOI = {10.1002/2017gc006997},
  number = {9},
  journal = {Geochem. Geophys. Geosyst.},
  publisher = {American Geophysical Union (AGU)},
  author = {Ruedas,  Thomas},
  year = {2017},
  month = sep,
  pages = {3530–3541}
}

@article{condie2014growth,
title={Growth of continental crust: a balance between preservation and recycling}, volume={78},
DOI={10.1180/minmag.2014.078.3.11},
number={3},
journal={Mineralogical Magazine}, 
author={Condie, K. C.},
year={2014},
pages={623–637}}

@article{thiriet2018hemispheric,
  title = {Hemispheric Dichotomy in Lithosphere Thickness on Mars Caused by Differences in Crustal Structure and Composition},
  volume = {123},
  ISSN = {2169-9100},
  url = {http://dx.doi.org/10.1002/2017JE005431},
  DOI = {10.1002/2017je005431},
  number = {4},
  journal = {J. Geophys. Res. Planets},
  publisher = {American Geophysical Union (AGU)},
  author = {Thiriet,  Mélanie and Michaut,  Chloé and Breuer,  Doris and Plesa,  Ana‐Catalina},
  year = {2018},
  month = apr,
  pages = {823–848}
}

@book{turcotte2002geodynamics,
  title={Geodynamics},
  author={Turcotte, Donald L and Schubert, Gerald},
  year={2002},
  publisher={Cambridge university press}
}

@article{solomatov1995scaling,
  title = {Scaling of temperature- and stress-dependent viscosity convection},
  volume = {7},
  ISSN = {1089-7666},
  url = {http://dx.doi.org/10.1063/1.868624},
  DOI = {10.1063/1.868624},
  number = {2},
  journal = {Phys. Fluids},
  publisher = {AIP Publishing},
  author = {Solomatov,  V. S.},
  year = {1995},
  month = feb,
  pages = {266–274}
}

@article{masters1990summary,
  title = {Summary of seismological constraints on the structure of the Earth’s core},
  volume = {95},
  ISSN = {0148-0227},
  url = {http://dx.doi.org/10.1029/JB095iB13p21691},
  DOI = {10.1029/jb095ib13p21691},
  number = {B13},
  journal = {J. Geophys. Res. Solid Earth},
  publisher = {American Geophysical Union (AGU)},
  author = {Masters,  T. G. and Shearer,  P. M.},
  year = {1990},
  month = dec,
  pages = {21691–21695}
}

@article{olson2006dipole,
title = {Dipole moment scaling for convection-driven planetary dynamos},
journal = {Earth Planet. Sci. Lett.},
volume = {250},
number = {3},
pages = {561-571},
year = {2006},
issn = {0012-821X},
doi = {https://doi.org/10.1016/j.epsl.2006.08.008},
author = {Peter Olson and Ulrich R. Christensen},
}

@article{lark2026coupled,
  title = {Coupled thermochemical evolution of the early Earth’s solid mantle and basal magma ocean: The role of melting and melt transport},
  volume = {680},
  ISSN = {0012-821X},
  url = {http://dx.doi.org/10.1016/j.epsl.2026.119880},
  DOI = {10.1016/j.epsl.2026.119880},
  journal = {Earth Planet. Sci. Lett.},
  publisher = {Elsevier BV},
  author = {Lark,  Laura H. and Boukaré,  Charles-Édouard and Badro,  James and Samuel,  Henri},
  year = {2026},
  month = apr,
  pages = {119880}
}

@article{rosenthal2025finger,
  title = {Finger properties in bounded double diffusive finger convection},
  volume = {37},
  ISSN = {1089-7666},
  url = {http://dx.doi.org/10.1063/5.0253859},
  DOI = {10.1063/5.0253859},
  number = {2},
  journal = {Phys. Fluids},
  publisher = {AIP Publishing},
  author = {Rosenthal,  A. and Tilgner,  A.},
  year = {2025},
  month = feb 
}

@article{davies1980thermal,
  title = {Thermal histories of convective Earth models and constraints on radiogenic heat production in the Earth},
  volume = {85},
  ISSN = {0148-0227},
  url = {http://dx.doi.org/10.1029/JB085iB05p02517},
  DOI = {10.1029/jb085ib05p02517},
  number = {B5},
  journal = {J. Geophys. Res. Solid Earth},
  publisher = {American Geophysical Union (AGU)},
  author = {Davies,  Geoffrey F.},
  year = {1980},
  month = may,
  pages = {2517–2530}
}

@article{christensen1985thermal,
  title = {Thermal evolution models for the Earth},
  volume = {90},
  ISSN = {0148-0227},
  url = {http://dx.doi.org/10.1029/JB090iB04p02995},
  DOI = {10.1029/jb090ib04p02995},
  number = {B4},
  journal = {J. Geophys. Res. Solid Earth},
  publisher = {American Geophysical Union (AGU)},
  author = {Christensen,  Ulrich R.},
  year = {1985},
  month = mar,
  pages = {2995–3007}
}

@article{burke2011wilson,
  title = {Plate Tectonics,  the Wilson Cycle,  and Mantle Plumes: Geodynamics from the Top},
  volume = {39},
  ISSN = {1545-4495},
  url = {http://dx.doi.org/10.1146/annurev-earth-040809-152521},
  DOI = {10.1146/annurev-earth-040809-152521},
  number = {1},
  journal = {Ann. Rev. Earth Planet. Sci.},
  publisher = {Annual Reviews},
  author = {Burke,  Kevin},
  year = {2011},
  month = may,
  pages = {1–29}
}

@article{badro2015core,
  title = {Core formation and core composition from coupled geochemical and geophysical constraints},
  volume = {112},
  ISSN = {1091-6490},
  url = {http://dx.doi.org/10.1073/pnas.1505672112},
  DOI = {10.1073/pnas.1505672112},
  number = {40},
  journal = {Proc. Natl. Acad. Sci. U.S.A.},
  publisher = {Proc. Natl. Acad. Sci. U.S.A.},
  author = {Badro,  James and Brodholt,  John P. and Piet,  Hélène and Siebert,  Julien and Ryerson,  Frederick J.},
  year = {2015},
  month = sep,
  pages = {12310–12314}
}

@article{deschamps2022estimating,
  title = {Estimating core-mantle boundary temperature from seismic shear velocity and attenuation},
  volume = {10},
  ISSN = {2296-6463},
  url = {http://dx.doi.org/10.3389/feart.2022.1031507},
  DOI = {10.3389/feart.2022.1031507},
  journal = {Front. Earth Sci.},
  publisher = {Frontiers Media SA},
  author = {Deschamps,  Frédéric and Cobden,  Laura},
  year = {2022},
  month = dec 
}

@article{andrault2011solidus,
  title = {Solidus and liquidus profiles of chondritic mantle: Implication for melting of the Earth across its history},
  volume = {304},
  ISSN = {0012-821X},
  url = {http://dx.doi.org/10.1016/j.epsl.2011.02.006},
  DOI = {10.1016/j.epsl.2011.02.006},
  number = {1–2},
  journal = {Earth Planet. Sci. Lett.},
  publisher = {Elsevier BV},
  author = {Andrault,  Denis and Bolfan-Casanova,  Nathalie and Nigro,  Giacomo Lo and Bouhifd,  Mohamed A. and Garbarino,  Gaston and Mezouar,  Mohamed},
  year = {2011},
  month = apr,
  pages = {251–259}
}

@article{yuan2023hydrogen,
  title = {Hydrogen distribution between the Earth’s inner and outer core},
  volume = {609},
  ISSN = {0012-821X},
  url = {http://dx.doi.org/10.1016/j.epsl.2023.118084},
  DOI = {10.1016/j.epsl.2023.118084},
  journal = {Earth Planet. Sci. Lett.},
  publisher = {Elsevier BV},
  author = {Yuan,  Liang and Steinle-Neumann,  Gerd},
  year = {2023},
  month = may,
  pages = {118084}
}

@article{mooney2023Earth,
	author = {Walter D. Mooney and Carol Barrera-Lopez and Mar{\'\i}a Gabriela Su{\'a}rez and Miguel A. Castelblanco},
	journal = {Earth-Science Reviews},
	pages = {104493},
	title = {Earth Crustal Model 1 (ECM1): A 1$\,^{\circ}$ x 1$\,^{\circ}$ Global Seismic and Density Model},
	volume = {243},
	year = {2023}
}

@article{alfe2002iron,
  title = {Iron under Earth's core conditions: Liquid-state thermodynamics and high-pressure melting curve from ab initio calculations},
  author = {Alf\`e, D. and Price, G. D. and Gillan, M. J.},
  journal = {Phys. Rev. B},
  volume = {65},
  issue = {16},
  pages = {165118},
  numpages = {11},
  year = {2002},
  month = apr,
  publisher = {American Physical Society},
  doi = {10.1103/PhysRevB.65.165118},
  url = {https://link.aps.org/doi/10.1103/PhysRevB.65.165118}
}

@misc{bonnet2026zenodo,
  doi = {10.5281/ZENODO.19211141},
  url = {https://zenodo.org/doi/10.5281/zenodo.19211141},
  author = {Bonnet Gibet,  Valentin and Tosi,  Nicola},
  title = {Core and mantle thermal evolution constraints on the onset of plate tectonics and a long-lived geodynamo -- Data and Software},
  publisher = {Zenodo},
  year = {2026},
  copyright = {Creative Commons Attribution 4.0 International}
}

@article{Blanchard2017,
  title = {The solubility of heat-producing elements in Earth’s core},
  ISSN = {2410-3403},
  url = {http://dx.doi.org/10.7185/geochemlet.1737},
  DOI = {10.7185/geochemlet.1737},
  journal = {Geochem. Perspect. Lett.},
  publisher = {European Association of Geochemistry},
  author = {Blanchard,  I. and Siebert,  J. and Borensztajn,  S. and Badro,  J.},
  year = {2017},
  month = Oct,
  pages = {1–5}
}

@article{Chidester2022,
  title = {The Lithophile Element Budget of Earth’s Core},
  volume = {23},
  ISSN = {1525-2027},
  url = {http://dx.doi.org/10.1029/2021GC009986},
  DOI = {10.1029/2021gc009986},
  number = {2},
  journal = {Geochem. Geophys. Geosyst.},
  publisher = {American Geophysical Union (AGU)},
  author = {Chidester,  B. A. and Lock,  S. J. and Swadba,  K. E. and Rahman,  Z. and Righter,  K. and Campbell,  A. J.},
  year = {2022},
  month = Feb 
}

@article{Jain2019,
  title = {Global Analysis of Experimental Data on the Rheology of Olivine Aggregates},
  volume = {124},
  ISSN = {2169-9356},
  url = {http://dx.doi.org/10.1029/2018JB016558},
  DOI = {10.1029/2018jb016558},
  number = {1},
  journal = {J. Geophys. Res. Solid Earth},
  publisher = {American Geophysical Union (AGU)},
  author = {Jain,  Chhavi and Korenaga,  Jun and Karato,  Shun‐ichiro},
  year = {2019},
  month = Jan,
  pages = {310–334}
}

@article{Lau2016,
  title = {Inferences of mantle viscosity based on ice age data sets: Radial structure},
  volume = {121},
  ISSN = {2169-9356},
  url = {http://dx.doi.org/10.1002/2016JB013043},
  DOI = {10.1002/2016jb013043},
  number = {10},
  journal = {J. Geophys. Res. Solid Earth},
  publisher = {American Geophysical Union (AGU)},
  author = {Lau,  Harriet C. P. and Mitrovica,  Jerry X. and Austermann,  Jacqueline and Crawford,  Ophelia and Al‐Attar,  David and Latychev,  Konstantin},
  year = {2016},
  month = Oct,
  pages = {6991–7012}
}

@article{Zhang2022free,
  title = {Free Energies of Fe‐O‐Si Ternary Liquids at High Temperatures and Pressures: Implications for the Evolution of the Earth’s Core Composition},
  volume = {49},
  ISSN = {1944-8007},
  url = {http://dx.doi.org/10.1029/2021GL096749},
  DOI = {10.1029/2021gl096749},
  number = {4},
  journal = {Geophys. Res. Lett.},
  publisher = {American Geophysical Union (AGU)},
  author = {Zhang,  Zhigang and Csányi,  Gábor and Alfè,  Dario and Zhang,  Yigang and Li,  Juan and Liu,  Jin},
  year = {2022},
  month = Feb 
}

@article{Hirose2017crystallization,
  title = {Crystallization of silicon dioxide and compositional evolution of the Earth’s core},
  volume = {543},
  ISSN = {1476-4687},
  url = {http://dx.doi.org/10.1038/nature21367},
  DOI = {10.1038/nature21367},
  number = {7643},
  journal = {Nature},
  publisher = {Springer Science and Business Media LLC},
  author = {Hirose,  Kei and Morard,  Guillaume and Sinmyo,  Ryosuke and Umemoto,  Koichio and Hernlund,  John and Helffrich,  George and Labrosse,  Stéphane},
  year = {2017},
  month = Feb,
  pages = {99–102}
}

@article{Wilson2022powering,
  title = {Powering Earth’s Ancient Dynamo With Silicon Precipitation},
  volume = {49},
  ISSN = {1944-8007},
  url = {http://dx.doi.org/10.1029/2022GL100692},
  DOI = {10.1029/2022gl100692},
  number = {22},
  journal = {Geophys. Res. Lett.},
  publisher = {American Geophysical Union (AGU)},
  author = {Wilson,  Alfred J. and Pozzo,  Monica and Alfè,  Dario and Walker,  Andrew M. and Greenwood,  Sam and Pommier,  Anne and Davies,  Christopher J.},
  year = {2022},
  month = Nov 
}

@article{Insixiengmay2025,
  title = {MgO miscibility in liquid iron},
  volume = {654},
  ISSN = {0012-821X},
  url = {http://dx.doi.org/10.1016/j.epsl.2025.119242},
  DOI = {10.1016/j.epsl.2025.119242},
  journal = {Earth Planet. Sci. Lett.},
  publisher = {Elsevier BV},
  author = {Insixiengmay,  Leslie and Stixrude,  Lars},
  year = {2025},
  month = Mar,
  pages = {119242}
}

@article{MCKENZIE2005,
  title = {Thermal structure of oceanic and continental lithosphere},
  volume = {233},
  ISSN = {0012-821X},
  url = {http://dx.doi.org/10.1016/j.epsl.2005.02.005},
  DOI = {10.1016/j.epsl.2005.02.005},
  number = {3-4},
  journal = {Earth Planet. Sci. Lett.},
  publisher = {Elsevier BV},
  author = {McKenzie,  D and Jackson,  J and Priestley,  K},
  year = {2005},
  month = May,
  pages = {337–349}
}

@article{Breuer2003,
  title = {Early plate tectonics versus single‐plate tectonics on Mars: Evidence from magnetic field history and crust evolution},
  volume = {108},
  ISSN = {0148-0227},
  url = {http://dx.doi.org/10.1029/2002JE001999},
  DOI = {10.1029/2002je001999},
  number = {E7},
  journal = {J. Geophys. Res. Planets},
  publisher = {American Geophysical Union (AGU)},
  author = {Breuer,  D. and Spohn,  T.},
  year = {2003},
  month = July 
}

@article{Ritsema2020,
  title = {Heterogeneity of Seismic Wave Velocity in Earth’s Mantle},
  volume = {48},
  ISSN = {1545-4495},
  url = {http://dx.doi.org/10.1146/annurev-earth-082119-065909},
  DOI = {10.1146/annurev-earth-082119-065909},
  number = {1},
  journal = {Ann. Rev. Earth Planet. Sci.},
  publisher = {Annual Reviews},
  author = {Ritsema,  Jeroen and Lekić,  Vedran},
  year = {2020},
  month = May,
  pages = {377–401}
}

@article{Frasson2024,
  title = {On the impact of true polar wander on heat  flux patterns at the core–mantle boundary},
  volume = {15},
  ISSN = {1869-9529},
  url = {http://dx.doi.org/10.5194/se-15-617-2024},
  DOI = {10.5194/se-15-617-2024},
  number = {5},
  journal = {Solid Earth},
  publisher = {Copernicus GmbH},
  author = {Frasson,  Thomas and Labrosse,  Stéphane and Nataf,  Henri-Claude and Coltice,  Nicolas and Flament,  Nicolas},
  year = {2024},
  month = May,
  pages = {617–637}
}

@article{TerraNova2024,
  title = {Regionally-triggered geomagnetic reversals},
  volume = {14},
  ISSN = {2045-2322},
  url = {http://dx.doi.org/10.1038/s41598-024-59849-z},
  DOI = {10.1038/s41598-024-59849-z},
  number = {1},
  journal = {Sci. Rep.},
  publisher = {Springer Science and Business Media LLC},
  author = {Terra-Nova,  Filipe and Amit,  Hagay},
  year = {2024},
  month = Apr 
}

@misc{Davies2023,
  title = {Dynamics in Earth’s Core Arising from Thermo‐Chemical Interactions with the Mantle},
  ISBN = {9781119526919},
  ISSN = {2328-8779},
  url = {http://dx.doi.org/10.1002/9781119526919.ch12},
  DOI = {10.1002/9781119526919.ch12},
  journal = {Core‐Mantle Co‐Evolution},
  publisher = {Wiley},
  author = {Davies,  Christopher J. and Greenwood,  Sam},
  year = {2023},
  month = June,
  pages = {219–258}
}

\end{document}